\documentclass[aps, superscriptaddress,reprint,amsmath,amssymb,prl]{revtex4-2}
\usepackage[]{graphicx}
\usepackage{multirow}
\usepackage{dcolumn}
\usepackage[usenames]{color}
\usepackage{booktabs}
\usepackage[version=4]{mhchem}
\usepackage[utf8]{inputenc}
\usepackage{subcaption}
\usepackage{makecell}
\usepackage{array}
\usepackage{bm}
\usepackage[colorlinks=true, allcolors=blue]{hyperref}
\usepackage{caption}
\usepackage{xcolor}
\usepackage[normalem]{ulem}

\definecolor{darkgreen}{RGB}{0,150,0}

\usepackage{siunitx}
\DeclareSIUnit\angstrom{\text{Å}}

\begin{document}

\title{Atoms that Matter: Largest Materials Datasets Are 99\% Redundant, and Local Chemistry Can Guide the Required 1\%}

\author{Evgeny Moerman}
\affiliation{Department of Physics and Materials Science, University of Luxembourg, L-1511 Luxembourg City, Luxembourg}

\author{Igor Poltavsky}
\email{igor.poltavsky@uni.lu}
\affiliation{Department of Physics and Materials Science, University of Luxembourg, L-1511 Luxembourg City, Luxembourg}

\author{Anton Charkin-Gorbulin}
\affiliation{Department of Physics and Materials Science, University of Luxembourg, L-1511 Luxembourg City, Luxembourg}

\author{Alexandre Tkatchenko}
\email{alexandre.tkatchenko@uni.lu}
\affiliation{Department of Physics and Materials Science, University of Luxembourg, L-1511 Luxembourg City, Luxembourg}

\begin{abstract}
\noindent General-purpose machine-learning force fields (MLFFs) for materials are now 
trained on the order of 100 million structures, assuming transferability 
requires such scale. We test this premise 
on OMat24, the largest available materials dataset, by training MACE models on subsets of $6.5$ 
million atoms, less than 1\% of OMat24, selected in three complementary ways: at random, by latent-space clustering of a pretrained MLFF (DIRECT), and by a new model-free method maximizing the diversity of discrete, 
interpretable local environments (EnvTok). All three models approach the full-data 
reference MACE-OMAT-0 model to within a few tens of meV/\AA{} in forces, and remain close to or exceed it on 
formation energies, elastic moduli, heat capacities, Li$^{+}$ migration barriers, and 
molecular-dynamics stability up to $3000\,$K and $500\,$GPa. This near-independence from the 
selection strategy shows that chemical environment diversity, not size, defines OMat24's 
information content: the remaining $99\%$ of structures are largely redundant. Broad 
sampling of chemical space thus remains essential, but reference labels are needed only for 
structurally and chemically distinct configurations. Using atomic positions and 
species, optionally augmented by magnetic moments, EnvTok identifies these before any labelling or training, yielding compact, architecture-independent datasets.
This hundredfold reduction makes hybrid-DFT or higher-level relabelling of general-purpose training sets feasible, and frees computational budget for extending environmental coverage
and multi-fidelity learning.

\end{abstract}

\maketitle

\section{Introduction}
Machine-learning force fields (MLFFs) promise atomistic simulations with \emph{ab initio} accuracy at a computational cost comparable to that of classical force fields~\cite{unke2021machine, deringer2019machine}. The ambition of the field has shifted from system-specific models to general-purpose (GP) MLFFs that can be applied across entire chemical compound spaces directly or after minimal fine-tuning~\cite{deng2023chgnet, batatia2025foundation, park2024sevennet, zhang2024dpa2}. The route to such transferability has so far been data volume~\cite{frey2023neural, schmidt2024improving, yuan2026foundation, yang2024mattersim, zhang2026dpa3}: for inorganic materials this development culminated in the Open Materials 2024 (OMat24) dataset of over 110 million structures~\cite{barroso2024open}, which has since underpinned some of the most accurate GP materials MLFFs to date~\cite{batatia2025cross, fu2025learning, wood2026family, rhodes2025orb, lysogorskiy2026graph}.

Scale, however, comes at a price. Training on $10^{8}$ structures requires substantial computational resources, often limiting training to large batches and few epochs. Reference calculations are also costly. Selecting informative training configurations~\cite{mazitov2025petmad,mazitov2025mad,liang2026nep89} and choosing an appropriate density-functional theory (DFT) reference~\cite{kaplan2025foundational,kim2025sevennetmf,huang2025crossfunctional,kim2026crossdomain} remain central contemporary challenges in developing GP-MLFFs. Because of the computational cost, OMat24 and comparable datasets rely on semilocal DFT with generalized-gradient (GGA) or meta-GGA functionals~\cite{perdew1996generalized, sun2015strongly,furness2020accurate}. Recomputing these labels at a higher level of theory remains prohibitive~\cite{ko2025multifidelity}. More fundamentally, the fraction of data that provides distinct, learnable information remains unknown~\cite{sorscher2022beyond, schwalbe2025model}. The way these datasets are generated suggests substantial redundancy. OMat24 starts from equilibrium structures drawn from existing databases. Each is used to generate many non-equilibrium configurations through rattling, \emph{ab initio} molecular dynamics, and Boltzmann sampling~\cite{barroso2024open}. Configurations derived from the same parent structure share many local atomic environments. Substantial overlap also exists between environments in different parent structures. The number of \emph{distinct} environments therefore grows much more slowly than the number of structures~\cite{karabin2020entropy}.

This observation motivates dataset pruning, and several strategies have been proposed to extract a maximally informative subset. 
For property-prediction models trained on datasets of $10^{4}$–$10^{6}$ materials, a model-guided pruning of $45$--$95\%$ of the data raised errors on formation energies and band gaps 
by less than $10\%$ without severely affecting out-of-distribution robustness~\cite{li2023exploiting}.
Farthest point sampling (FPS)~\cite{imbalzano2018automatic} maximizes distances in a descriptor space and therefore favours outliers over the densely populated near-equilibrium regions; its gradient-guided successor~\cite{trestman2025gradient} corrects this but requires force labels. The DIRECT protocol~\cite{qi2024robust} clusters structures in the latent space of a pre-trained MLFF (M3GNet)~\cite{chen2022universal} after principal component analysis (PCA) and samples from each cluster. It has been scaled to over a million structures and achieved a dataset reduction of 
$85\%$, but its result depends on several tunable parameters and, more importantly, on the pre-trained model whose latent space defines the clusters.

Information-theoretic approaches maximize the entropy of the selected subset~\cite{karabin2020entropy,schwalbe2025model, subramanyam2025entropy}; their state-of-the-art representative, the QUESTS compression algorithm~\cite{schwalbe2025model,yu2026maximizing}, uses an element-agnostic distance-based descriptor and kernel density estimates at a single bandwidth, and has so far been applied only to single-element datasets of moderate size. 
Active-learning schemes instead select configurations by model uncertainty or extrapolation grade~\cite{podryabinkin2017active,kulichenko2023uncertainty}.
 MP-ALOE applies active learning to a pool of roughly $10^{8}$ candidate structures~\cite{kuner2025mpaloe}, but its iterative, model-guided acquisition of new DFT labels in batches of $\sim$10$^{5}$ structures differs from the model-agnostic pruning of an already labelled dataset considered here.
Alternatively, compact datasets can be constructed directly rather than selected from large existing ones. SMAX~\cite{bochkarev2026exploring} is built iteratively by optimizing the atomic positions of randomly initialized candidates to maximize the feature entropy of the growing dataset, yielding $\sim$10$^{6}$ structures that cover far-from-equilibrium regions absent from OMat24. To also represent the near-equilibrium regime, this core dataset was supplemented by a comparable number of low-energy and low-force structures, partly drawn from existing databases. As a generative approach, SMAX does not address how much of an existing dataset is redundant.
Consequently, it remains open whether an MLFF trained on a small fraction of a dataset at the scale of OMat24 can match the full-data model,
since none of these methods is simultaneously label-free, model-free and practical for multi-element datasets of $10^{8}$ structures.

\begin{figure*}
    \includegraphics[width=\textwidth]{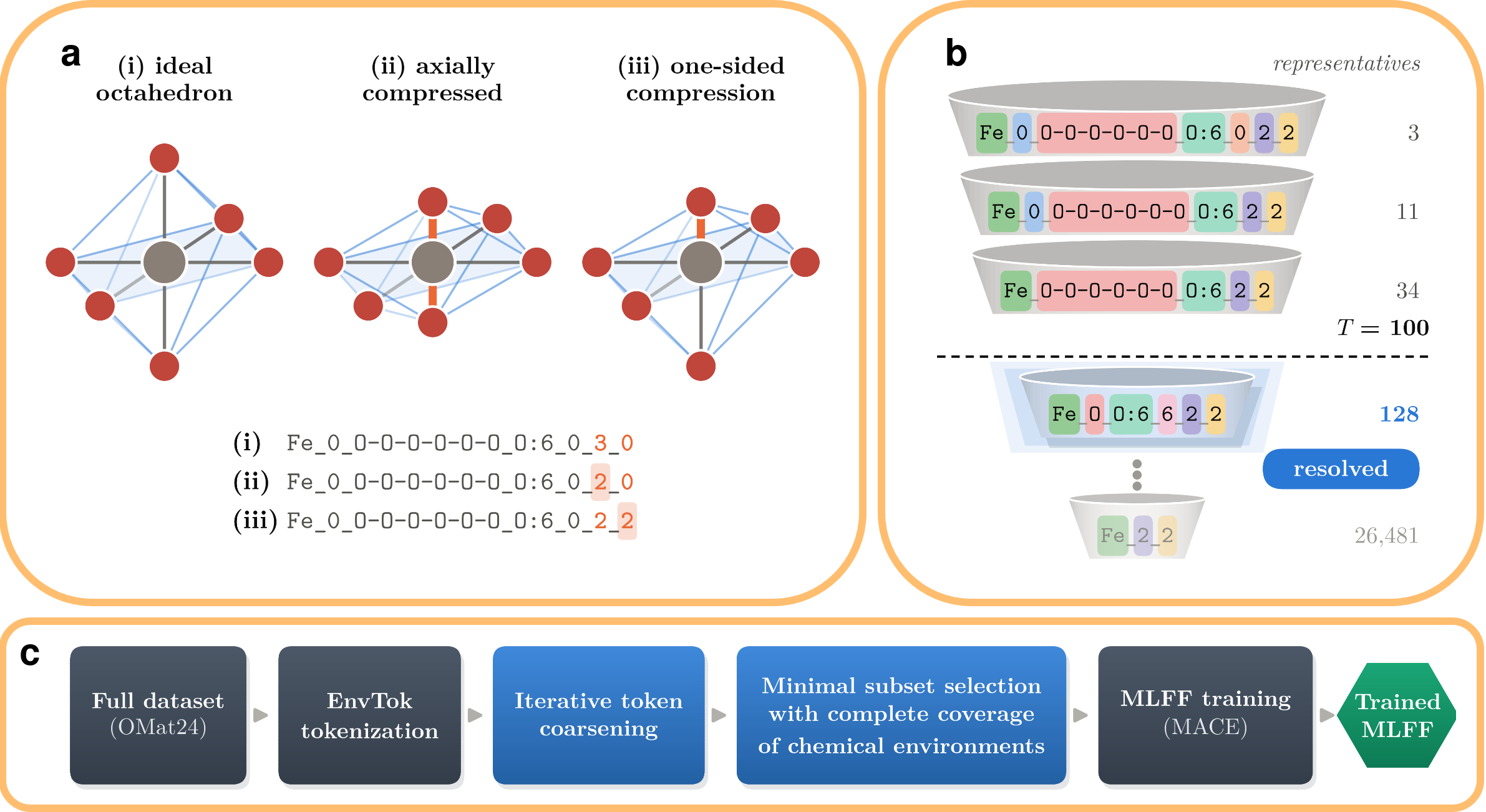}
    \caption{Overview of the EnvTok descriptor definition and subset selection procedure.
    \textbf{a} Examples of the EnvTok token and its sensitivity to shape and bond-length distortions.
    \textbf{b} Construction of the EnvTok coarsening ladder.
    \textbf{c} Workflow of EnvTok-based subset construction and MLFF training}
    \label{fig:overview}
\end{figure*}

Here we address the above-mentioned challenges. Using OMat24 as a case study, we train MLFFs on subsets containing less than $1\%$ of the full dataset, selected by three strategies with distinct objectives: random sampling, DIRECT, and EnvTok, a coverage-based method introduced in this work (see Fig.~\ref{fig:overview}c). We benchmark these models against MACE-OMAT-0~\cite{batatia2025cross}, trained on the full dataset, across atomic forces, near-equilibrium thermodynamic and mechanical properties, molecular-dynamics (MD) stability under extreme conditions, and Li$^{+}$ migration barriers. The subset-trained models perform similarly to one another and, across most of these assessments, to the full-data model, revealing substantial redundancy ($\sim 99 \%$) in OMat24, a compression level and raw dataset size 
well beyond those reported previously.

The EnvTok approach proposed in this work offers a chemically interpretable basis for selecting structures that preserve the diversity of environments within a compact subset. It assigns each atom a discrete coordination-environment descriptor based on the central species, coordinating elements, coordination polyhedron, distortion, and bond compression. It then seeks the smallest subset of structures that collectively covers all environments identified in the dataset. Requiring only atomic positions and species, optionally supplemented by magnetic moments where available, EnvTok can guide structure selection before reference calculations and is independent of the MLFF architecture. 
It can therefore serve both top-down, to prune existing datasets, and bottom-up, to decide which configurations to label in the first place, for instance at a higher level of theory than the original data.
We further show that its environment classes map onto compact, connected regions in the latent space of a trained MLFF, indicating that this model-free descriptor captures chemically meaningful distinctions also learned by the model.

\section{Methods}
\label{sec:subsets}

We start by excluding structures containing $f$-block elements (lanthanides, $Z=57$--$71$, and actinides, $Z\geq89$). Their strongly localized $f$ electrons exhibit pronounced static correlation and self-interaction error that semilocal functionals describe poorly~\cite{cohen2008insights, caucci2025performance}, and unlike the $3d$--$5d$ transition metals in oxides and fluorides, for which OMat24 inherits the Materials Project Hubbard-$U$ correction~\cite{dudarev1998electron, wang2006oxidation,jain2011formation}, the $f$-block has no such correction. Their PBE labels are therefore the least reliable in the dataset, and retaining them would conflate label errors with the data-selection effects under study. The resulting $f$-block-free OMat24 comprises approximately 65 million structures and $1.2$ billion atoms.

From this pool we constructed six subsets of approximately $6.5$ million atoms each, i.e.\ less than $1\%$ of the complete $f$-block-free dataset: one with the EnvTok algorithm, one with DIRECT~\cite{qi2024robust}, and four random draws with different seeds that set the scale of run-to-run noise. We deliberately fix the budget in atoms rather than structures, because the number of atoms governs both the cost of DFT labels and the cost of training.

\paragraph*{Random selection} provides the baseline. If a dataset is highly redundant, a random $1\%$ already contains almost every common environment, and the only information it can miss is the rare environments in the tails of the distribution.

\paragraph*{DIRECT}~\cite{qi2024robust} performs PCA on the per-structure features of a pre-trained M3GNet model, clusters the result and draws a fixed number of structures from each cluster. It thus optimizes coverage of a model's latent space at the level of whole structures, and is explicitly designed to ``ensure coverage of the extrema of the configuration space''~\cite{qi2024robust}.

\paragraph*{EnvTok} is an atom-centred representation of local chemical environments proposed in this work and defines the discrete descriptor space in which our subset-selection algorithm operates. Each atom receives a seven-field token encoding its elemental identity, magnetic state, coordination, geometry, and bond strain (Fig.~\ref{fig:overview}a). Atoms with identical tokens belong to the same environment class. The first two fields specify the central element and the integer-rounded site magnetic moment reported in OMat24, with the spin field set to zero when magnetic labels are unavailable.
The third field lists the coordinating elements, including repeated entries, thereby encoding both coordination number and local composition. Neighbors are identified using the \texttt{ChemEnv} package~\cite{waroquiers2020chemenv} in \texttt{pymatgen}: a Voronoi analysis selects candidate neighbors according to their distances relative to the shortest bond and their shares of the coordination sphere, without a fixed radial cutoff. The fourth field identifies the best-fitting ideal coordination polyhedron, such as an octahedron (\texttt{O:6}). The fifth discretizes the continuous symmetry measure of this fit into four distortion levels, from 0 (ideal coordination) to 3 (severely distorted). The sixth quantifies bond compression through the minimum normalized bond length, $\min_j[d_{ij}/(r_i^{\mathrm{cov}}+r_j^{\mathrm{cov}})]$, where $d_{ij}$ is the distance from the central atom $i$ to neighbor $j$ and $r^{\mathrm{cov}}$ denotes a covalent radius. This minimum identifies the most compressed bond relative to its element-dependent reference length, which need not be the shortest bond in absolute distance. It is discretized into five bins: 0 denotes ratios below 0.60, bins 1 and 2 represent intermediate compression, bin 3 spans the reference-length range of 0.85--1.15, and bin 4 denotes ratios above 1.15. The seventh field measures compression asymmetry using the magnitude of a sum of unit bond vectors weighted towards shorter bonds. It is discretized into three levels, from 0, where opposing contributions essentially cancel, through 1 to 2, indicating strongly one-sided compression. Exact neighbor-selection parameters, weighting choices, and bin boundaries are provided in Sec.~S1 of the Supplementary. As an illustrative example, a sodium site in ideal rock-salt NaCl with bonds in the reference-length bin is encoded as \texttt{Na\_0\_Cl-Cl-Cl-Cl-Cl-Cl\_O:6\_0\_3\_0}: a nonmagnetic sodium center with six chlorine neighbors, ideal octahedral coordination and no compression asymmetry. In this symmetric environment, hydrostatic compression changes the sixth field when a bin boundary is crossed, provided the neighbor set remains unchanged; an off-center displacement can also change the distortion and asymmetry fields. Every field has a direct chemical or crystallographic meaning, yielding an element-resolved, interpretable representation whose resolution is controlled by the discretization boundaries.
We emphasize that the spin field is the only component not derived from the atomic 
structure alone.
With only $1.6\%$ of spin-polarized atoms in the entire OMat24 dataset,
it plays a subordinate role in the present study and can alternatively be omitted entirely
or estimated
from the structure itself via bond-valence analysis~\cite{o1991atom} (also implemented in \texttt{pymatgen}),
which in combination with the coordination polyhedron already assigned by \texttt{ChemEnv} yields crystal-field high- or low-spin moments.

EnvTok subset selection is formulated as a weighted set-cover problem: find the set of structures with the smallest total number of atoms that covers every sufficiently frequent environment class. Roughly half of the distinct OMat24 tokens occur at a single atomic site. We therefore map each token onto a fixed hierarchy of coarser descriptions, successively removing the distortion bin, magnetic moment, and explicit neighbor list, and retain the first level shared by at least $T$ structures as illustrated in Fig.~\ref{fig:overview}b. These classes define the coverage targets. A greedy algorithm constructs the cover, followed by pruning of redundant structures, subject to presence constraints for rare element--coordination combinations. The single threshold $T$ controls the subset resolution (see Sec.~S2). This construction is related to entropy-based selection as motivated in Sec.~S2.3. Set cover enforces class coverage at minimal atom count, whereas entropy maximization also favors balanced occupations. EnvTok thus replaces continuous similarity with discrete, chemically defined equivalence and requires neither reference energy and force labels nor a trained model. We note that simplified tokenization schemes were previously used to construct predictive and interpretable multilinear models for molecular and material properties~\cite{allen2022machine}.

At a fixed budget of approximately $6.5$ million atoms, the three selection strategies yield markedly different subsets of OMat24 (see Sec.~S2.5 and Fig.~S1). EnvTok achieves environment coverage primarily through small cells, selecting approximately $796$ thousand structures with an average of $8$ atoms per structure. DIRECT and random selection instead largely preserve the OMat24 cell-size distribution, selecting $459$ and $356$ thousand structures with averages of $14$ and $18$ atoms, respectively. Random selection reproduces the parent atomic-force distribution, whereas DIRECT shifts it toward higher forces, particularly in the high-force tail, consistent with its preference for extrema. EnvTok produces the broadest force distribution, reflecting coverage of compression and asymmetry classes across distortion levels. These differences provide a basis for testing how training-set composition affects model performance at fixed atom count.

Finally, for each training dataset, we trained a MACE model~\cite{batatia2022mace} using the hyperparameters of the MACE-OMAT-0 (medium) reference model~\cite{batatia2025cross}, adjusting only the effective batch size (see Sec.~S3). We compared these models with MACE-OMAT-0, trained on the full OMat24 dataset. Unless stated otherwise, random-subset results are reported as the mean $\pm$ standard deviation over four seeds.

\section{Results and Discussion}

\begin{table*}[htbp]
	\centering
	\small
	\setlength{\tabcolsep}{6pt}
	\renewcommand{\arraystretch}{1.08}
	\sisetup{
		table-format = 4.1,
		group-minimum-digits = 4,
		group-separator = {\,},
		detect-weight = true,
		mode = text,
	}
	\caption{
    Per-atom force-error profile on MatPES compared to MACE-OMAT-0.
      Atoms are binned by the per-atom force-error magnitude
      $|\Delta\mathbf{F}_i|$ of the full-data reference model MACE-OMAT-0.
      Every entry is the median, over the atoms of that bin, of
      $|\Delta\mathbf{F}_i|$ against the DFT reference.
      All errors are in meV/\AA.
      The \emph{Random} column is the mean $\pm$ standard deviation
      over the four random subsets.
      Bold marks the lowest of the three candidate columns in a row, and only
      where its margin over the next-lowest exceeds that standard deviation}
	\label{tab:performance-matpes}
	\begin{tabular}{@{}
			l
			S[table-format=2.2]
			S
			S
			S
			S[table-alignment-mode=none, table-number-alignment=right]@{\;$\pm$\;}S[table-alignment-mode=none, table-number-alignment=left]
			@{}}
		\toprule
		& & \multicolumn{5}{c@{}}{Median per-atom force error $|\Delta\mathbf{F}_i|$ (meV/\AA)} \\
		\cmidrule(l){3-7}
		\makecell[l]{MACE-OMAT-0\\force error (meV/\AA)} &
		\multicolumn{1}{c}{\makecell{Share of\\atoms (\%)}} &
		\multicolumn{1}{c}{\makecell{Reference\\MACE-OMAT-0}} &
		\multicolumn{4}{c@{}}{Subset-trained models} \\
		\cmidrule(l){4-7}
		& & & {EnvTok} & {DIRECT} & \multicolumn{2}{c@{}}{Random} \\
		\midrule
          \multicolumn{7}{@{}l}{\emph{Bulk} ($2\,234\,968$ \emph{atoms})} \\
          \addlinespace[2pt]
          0--25        &  15.72 &     5.9 &   \bfseries{11.6} &            13.9 &            12.1 &     0.2 \\
          25--50       &  10.75 &    37.0 &   \bfseries{67.0} &            78.4 &            69.2 &     0.7 \\
          50--100      &  16.25 &    72.5 &  \bfseries{112.5} &           126.2 &           115.4 &     0.9 \\
          100--200     &  19.55 &   142.4 &  \bfseries{189.9} &           206.0 &           195.4 &     1.1 \\
          200--400     &  18.90 &   278.8 &  \bfseries{326.8} &           343.0 &           334.9 &     1.5 \\
          400--800     &  12.93 &   535.6 &  \bfseries{552.1} &           560.6 &           567.8 &     2.5 \\
          800--1600    &   4.88 &  1002.1 &             943.5 & \bfseries{933.7} &           973.2 &     2.6 \\  
          1600--3200   &   0.65 &  1921.9 &            1799.1 & \bfseries{1760.9} &          1845.4 &    13.4 \\
          $\ge 3200$   &   0.38 &  5527.3 &            3633.3 &          2459.0 &          2550.8 &   260.9 \\
          \addlinespace[3pt]
          \midrule
          \multicolumn{7}{@{}l}{\emph{Non-bulk} ($775\,930$ \emph{atoms})} \\
          \addlinespace[2pt]
          0--25        &   5.74 &     7.2 &            16.6 &            19.0 &            16.7 &     1.2 \\
          25--50       &   5.64 &    38.2 &           114.8 &           115.8 &  \bfseries{112.0} &     2.4 \\
          50--100      &  12.27 &    74.5 &           171.1 & \bfseries{158.6} &           166.0 &     2.8 \\ 
          100--200     &  19.21 &   145.5 &           259.1 & \bfseries{243.9} &           258.2 &     3.8 \\  
          200--400     &  23.23 &   283.6 &           420.0 & \bfseries{391.2} &           421.1 &     6.6 \\  
          400--800     &  19.44 &   546.1 &           697.9 & \bfseries{647.7} &           700.3 &     8.9 \\  
          800--1600    &   9.51 &  1032.6 &          1175.1 & \bfseries{1109.1} &          1180.8 &    16.6 \\
          1600--3200   &   2.35 &  1992.7 &          2197.6 & \bfseries{1965.7} &          2086.1 &    43.1 \\
          $\geq 3200$  &   2.61 &  6429.1 &          4186.6 &          2482.6 &          2550.0 &   341.3 \\
		\bottomrule
	\end{tabular}
\end{table*}

\noindent\textbf{Performance on atomic forces.} First, we assess the models' static accuracy on the $f$-block-free part of MatPES~\cite{kaplan2025foundational}, an independent test set of $323{,}096$ structures computed at the same PBE/PBE+$U$ level as OMat24. Because OMat24 contains only periodic bulk structures~\cite{barroso2024open}, we distinguish three-dimensionally connected bonded networks (``bulk'', $74\%$ of MatPES atoms) from lower-dimensional structures and their combinations (``non-bulk''). For details on the distinction between ``bulk''  and ``non-bulk'' structures see Sec.~S4.1. We bin atoms by the reference-model, MACE-OMAT-0, force error and evaluate the subset-trained models within the same bins (Table~\ref{tab:performance-matpes}).

For bulk structures, training on less than $ 1\%$ of OMat24 largely preserves accuracy. All subset models show a small penalty in the bins of $0$--$800\,\mathrm{meV/\AA}$ covering $94.09\%$ of bulk atoms in MatPES. EnvTok performs best throughout this range, with its advantage over random selection increasing with reference error. DIRECT is least accurate on the $81\%$ of bulk atoms with reference errors below $400\,\mathrm{meV/\AA}$, but performs better than random selection at higher errors.
Subset models also outperform the reference in the remaining $5.91\%$ of bulk atoms ($>800\,\mathrm{meV/\AA}$ bins), but this comparison depends on the reference model's failures and does not establish a benefit of reducing the training set.
EnvTok trails DIRECT in the three highest bins and random selection in the highest bin.
In the two highest bins, this is caused by about 100 structures of highly compressed polymeric nitrogen, which contribute
$11\%$ and $82\%$ of their atoms, respectively. Without them, EnvTok is the most accurate subset-trained model in every bulk bin except $800$--$1600\,\mathrm{meV/\AA}$, where DIRECT remains best.

For non-bulk structures, outside the OMat24 training distribution, subset training leads to larger errors and the preferred selection strategy depends on extrapolation complexity. 
Random selection and EnvTok perform best at the lowest reference errors ($<50\,\mathrm{meV/\AA}$), whereas DIRECT is the most accurate subset-trained model in every bin above, covering $89\%$ of non-bulk atoms.
 Among the more than 3 million atoms in MatPES, extreme force errors ($>10\,\mathrm{eV/\AA}$) are observed in 4150 atoms for MACE-OMAT-0, compared with 1271 for DIRECT, 4428 for EnvTok, and 2924--6226 for random selection. 
 For the subset-trained models, $91$--$96\%$ of these atoms lie in structures containing isolated 0D or 1D fragments. For MACE-OMAT-0, $97\%$ occur in highly compressed polymeric nitrogen phases. Even the largest extreme error count represents around $0.2\%$ of MatPES (see Fig.~S2).

Thus, static accuracy tests indicate that adding the remaining $99\%$ of OMat24 to the selected sub-$1\%$ training sets yields limited gains relative to data and training costs, mainly improving models' typical accuracy rather than their out-of-distribution robustness, where MACE-OMAT-0 falls within the random-subset noise floor and behind the DIRECT-subset model. 
We next assess how these changes in accuracy affect application performance.\newline

\begin{figure*}[ht]
    \centering
    \captionsetup[subfigure]{position=top, justification=raggedright,
                             singlelinecheck=false, skip=0pt}
    \begin{subfigure}{0.3\textwidth}
        \caption{}
        \includegraphics[width=\textwidth]{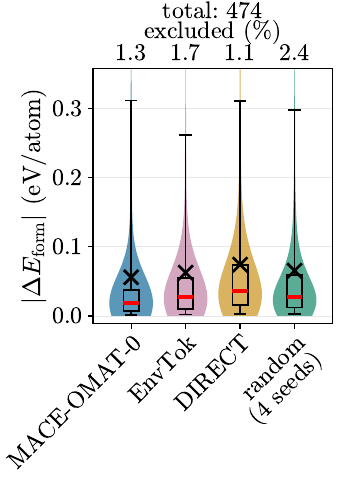}
    \end{subfigure}
    \hfill
    \begin{subfigure}{0.3\textwidth}
        \caption{}
        \includegraphics[width=\textwidth]{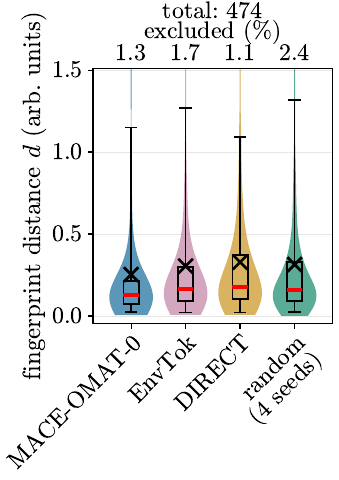}
    \end{subfigure}
    \hfill
    \begin{subfigure}{0.3\textwidth}
        \caption{}
        \includegraphics[width=\textwidth]{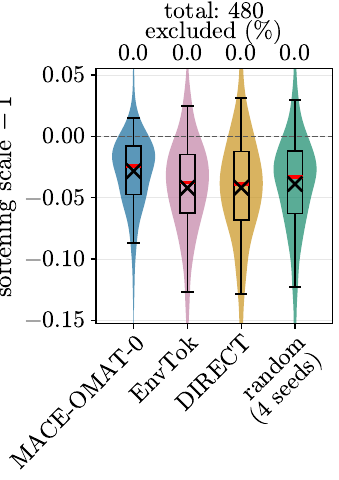}
    \end{subfigure}

    \vskip\baselineskip

    \begin{subfigure}{0.3\textwidth}
        \caption{}
        \includegraphics[width=\textwidth]{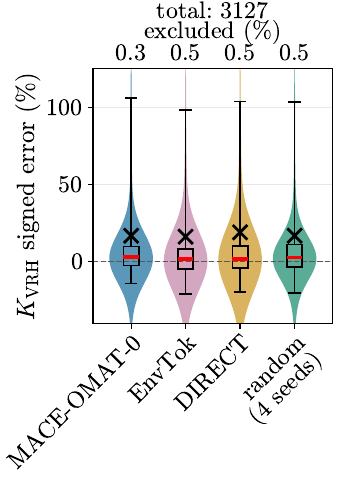}
    \end{subfigure}
    \hfill
    \begin{subfigure}{0.3\textwidth}
        \caption{}
        \includegraphics[width=\textwidth]{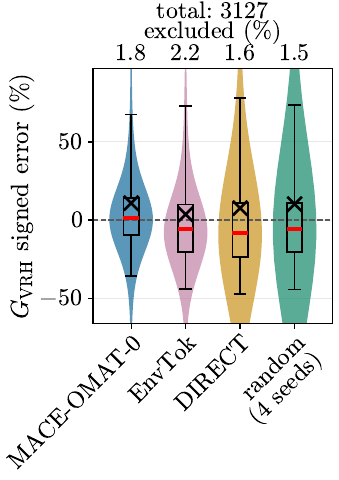}
    \end{subfigure}
    \hfill
    \begin{subfigure}{0.3\textwidth}
        \caption{}
        \label{fig:matcalc-cv}
        \includegraphics[width=\textwidth]{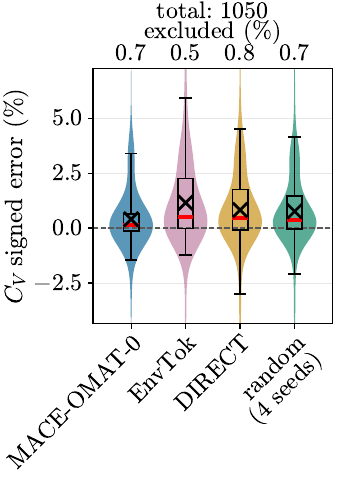}
    \end{subfigure}
    \caption{MatCalc benchmark of MACE-OMAT-0 and the subset-trained models: (a) formation energies, (b) fingerprint distance between MLFF- and DFT-relaxed structures, (c) systematic force softening, (d) bulk modulus, (e) shear modulus and (f) constant-volume heat capacity in the harmonic approximation. Violin plots: cross and red bar, mean and median; box edges, 25th and 75th percentiles; whiskers, 5th and 95th percentiles. 
    Above every violin, the percentage of structures that were excluded due to diverged relaxations
    or unphysical property predictions (i.e non-positive elastic moduli)
    and at the top the total number of data points for the respective benchmark property are given}
    \label{fig:matcalc-bench}
\end{figure*}

\noindent\textbf{Near-equilibrium properties.} We first consider properties dominated by near-equilibrium configurations. The MatCalc suite~\cite{kaplan2025foundational} evaluates formation energies $E_{\mathrm{form}}$, bulk and shear moduli $K$ and $G$, and harmonic heat capacities $C_V$ against the OMat24-consistent PBE/PBE+$U$ reference. Two diagnostic quantities complement these targets: the fingerprint distance $d$ between MLFF- and DFT-relaxed structures and the force-softening scale, which measures systematic under- or overestimation of atomic forces~\cite{deng2025systematic}. Fig.~\ref{fig:matcalc-bench} summarizes the error distributions for all six quantities.

Models trained on less than $1\%$ of OMat24 achieve accuracy comparable to the full model across all targets. For $E_{\mathrm{form}}$ and the fingerprint distance $d$, models trained on EnvTok and randomly selected subsets perform similarly, with only marginal degradation relative to MACE-OMAT-0, while DIRECT shows a somewhat larger formation-energy error (MAE of $74$ vs. $63$--$68\,\mathrm{meV/atom}$, and $56\,\mathrm{meV/atom}$ for MACE-OMAT-0). 
For $K$, EnvTok gives the median signed error closest to zero, almost two percentage points below that of MACE-OMAT-0, while its error distribution remains close to those of MACE-OMAT-0 and the random subsets. 
DIRECT shows a mean signed error exceeding those of all other models by $2$--$3$ percentage points. 
For $G$, subset training shifts the median error from $1.4\%$ to $5.6$--$8.0$ percentage 
points below zero. 
The EnvTok distribution and summary statistics remain within the variation among random subsets. DIRECT produces the broadest distribution for $G$, and the most negative median error ($-8.0\%$).

Dataset reduction most strongly affects force softening. The mean force underestimation increases from $2.8\%$ for MACE-OMAT-0 to $3.9\%$ for random selection, and $4.2\%$ for DIRECT and EnvTok. Random selection thus exhibits the least softening among the subset-trained models despite retaining the smallest fraction of high-force configurations (see Fig.~S1). This suggests that dataset size, rather than selection strategy, primarily controls the residual softening. Denser coverage of chemical environments in the full dataset may reduce interpolation errors and excessive smoothing of the potential energy surface, better preserving its curvature. Nevertheless, even EnvTok's and DIRECT's
$4.2\%$ softening remains within the range reported for state-of-the-art general-purpose MLFFs~\cite{zhou2026matris}.

Force softening lowers phonon frequencies and can therefore overestimate harmonic heat capacities~\cite{deng2025systematic,loew2025universal}. Consistent with this mechanism, the mean $C_V$ errors follow the same ordering as the mean force softening (Fig.~\ref{fig:matcalc-cv}). The degradation remains modest: EnvTok's $95$th-percentile $C_V$ error exceeds that of MACE-OMAT-0 by only about $2.5$ percentage points. 
The small changes under more than $\times$100 reduction in training data indicate substantial redundancy in OMat24 for the near-equilibrium properties examined here and even smaller OMat24 subsets, down to $0.92$ million atoms, degrade these properties only gradually (see Sec.~S4.2). Neither EnvTok nor DIRECT consistently improves on random selection in predicting the properties measured in this
benchmark.\newline

\begin{figure*}[ht]
    \centering
    \begin{subfigure}{0.48\textwidth}
        \includegraphics[width=\textwidth]{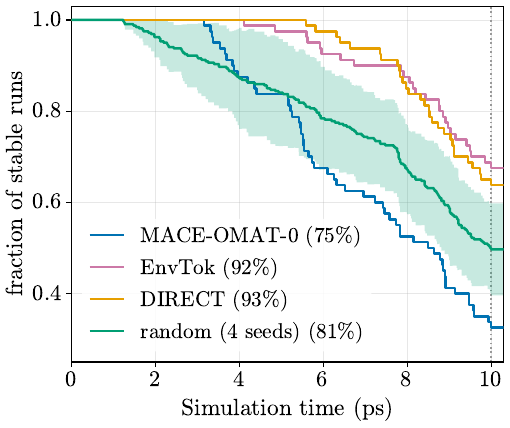}
        \caption{}
        \label{fig:mlip-arena-ecdf}
    \end{subfigure}
    \hfill
    \begin{subfigure}{0.48\textwidth}
        \includegraphics[width=\textwidth]{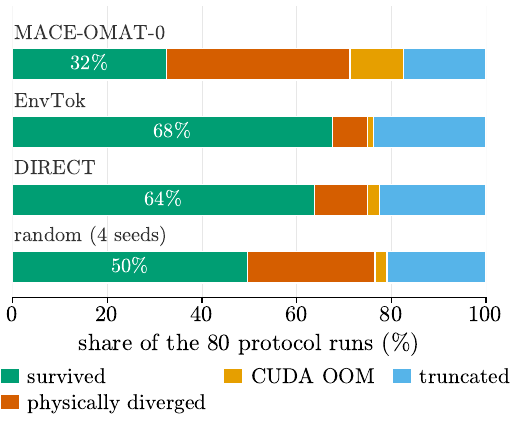}
        \caption{}
        \label{fig:mlip-arena-outcome}
    \end{subfigure}
    \caption{(a) Survival curves of MD runs in the MLIP Arena stability benchmark, in which temperature and pressure are ramped simultaneously from $300$ to $3000\,$K and from $0$ to $500\,$GPa over $10\,$ps. The four random-seed models are shown as mean (solid line) and standard deviation (band). The legend gives the average fraction of MD steps completed. (b) Breakdown of run outcomes}
    \label{fig:mlip-arena-stability}
\end{figure*}

\noindent\textbf{Molecular dynamics stability.} One of the main uses of MLFFs is molecular dynamics at length and time scales beyond the reach of DFT. For inorganic materials, applications ranging from superhard materials to planetary interiors and phase exploration often require high temperatures and pressures. Under these conditions, trajectory instability is a major failure mode~\cite{fu2022forces,poltavsky2025crash1,poltavsky2025crash2,zhang2020pushing,stocker2022robust,bischoff2024hydrogen,loew2026universal,sema2026environment}. This instability is largely decoupled from static or near-equilibrium force accuracy. We therefore evaluated all models using the MLIP Arena stability benchmark~\cite{chiang2026mlip}: $10\,\mathrm{ps}$ NPT trajectories for 80 random mixtures of stable materials, with temperature and pressure ramped linearly from $300$ to $3000\,\mathrm{K}$ and from $0$ to $500\,\mathrm{GPa}$, respectively (see Sec.~S4.3).

We found that the full-data model is the least stable (Fig.~\ref{fig:mlip-arena-stability}). MACE-OMAT-0 sustains all runs until $3.1\,\mathrm{ps}$ ($1147\,\mathrm{K}$, $157\,\mathrm{GPa}$), but failures then accumulate rapidly: only one-third of the runs finish, with a mean trajectory completion of $75\%$. The random models start failing earlier ($1.1$--$2.3\,\mathrm{ps}$), yet complete $81\%$ of the trajectory on average, with half of the runs finishing. EnvTok and DIRECT achieve mean completion rates of $92\%$ and $93\%$, respectively. To distinguish computational limits from physical instability, Fig.~\ref{fig:mlip-arena-outcome} separates runs truncated by the $30\,\mathrm{min}$ wall-clock limit while still integrating normally from those terminated by cell collapse or expansion, force or temperature divergence, or memory exhaustion caused by neighbor-list growth during cell collapse. Truncation rates are similar across models, whereas physical divergence affects $39\%$ of MACE-OMAT-0 runs, $27\%$ for random selection on average, 
$11\%$ for DIRECT, and $8\%$ for EnvTok.

These results show that training on the additional $99\%$ of OMat24 does not ensure greater robustness under extreme conditions: the full-data model is less stable than every subset-trained model tested here. This out-of-distribution test probes extreme thermodynamic conditions, distinct from the changes in system topology encountered for low-dimensional structures in the MatPES dataset, where the full-data model performs best. The stability results also provide the clearest distinction between selection strategies across our benchmarks. Both non-random strategies favor strongly compressed environments --- DIRECT through its preference for extremal structures and EnvTok through the bond-compression and asymmetry fields of its token --- and substantially outperform random selection in stability. This pattern suggests that balanced coverage of relevant environments, rather than training set size, is critical for robust MD.\newline

\begin{table}[t]
\centering
\caption{Migration-barrier errors (in eV) on the nebDFT2k benchmark, split by whether the host material would be computed with or without a Hubbard-$U$ correction in the Materials Project convention used for OMat24. Columns give the MAE, RMSE and median absolute error. The random row is mean $\pm$ one standard deviation over four seeds; bold marks the smallest value in each column}
\label{tab:nebdft2k-abs-errs}
\footnotesize
\setlength{\tabcolsep}{3pt}
\begin{tabular}{lccc}
\toprule
Model & MAE & RMSE & med.\ $|$err$|$ \\
\midrule
\multicolumn{4}{l}{\textit{w/o Hubbard-$U$ (621 hops)} }\\
MACE-OMAT-0 & 0.146 & 0.728 & \textbf{0.038} \\
EnvTok & 0.124 & 0.372 & 0.050 \\
DIRECT & \textbf{0.116} & \textbf{0.325} & 0.059 \\
random & 0.148\,$\pm$\,0.037 & 0.795\,$\pm$\,0.720 & 0.051\,$\pm$\,0.001 \\
\midrule
\multicolumn{4}{l}{\textit{with Hubbard-$U$ (1035 hops)} }\\
MACE-OMAT-0 & 0.270 & 0.823 & 0.136 \\
EnvTok & 0.269 & 0.760 & \textbf{0.134} \\
DIRECT & 0.316 & 1.160 & 0.144 \\
random & \textbf{0.264\,$\pm$\,0.019} & \textbf{0.571\,$\pm$\,0.124} & 0.139\,$\pm$\,0.004 \\
\bottomrule
\end{tabular}
\end{table}

\noindent\textbf{Li$^{+}$ migration.} Finally, we test the strongest extrapolation beyond the training distribution: Li$^{+}$ migration into an adjacent vacancy, a class of defect structures absent from OMat24~\cite{barroso2024open}. The nebDFT2k benchmark~\cite{dembitskiy2025benchmarking} contains 1656 barriers after excluding $f$-block compounds, calculated using non-spin-polarized PBE without Hubbard $U$. Because OMat24 follows the Materials Project convention of applying PBE+$U$ to selected transition-metal oxides and fluorides, we separate hosts according to whether they contain one
of the $U$-corrected transition metals (see Sec.~S4.4). Table~\ref{tab:nebdft2k-abs-errs} reports the errors. The full error distributions are shown in Fig.~S6.

For the 621 hops with matching levels of theory, 
both targeted selections outperform the reference: DIRECT and EnvTok reduce the MAE by $21\%$ and $15\%$ and the RMSE
by $55\%$ and $49\%$ relative to MACE-OMAT-0, respectively, with EnvTok yielding the signed-error distribution most closely centred on zero.
MACE-OMAT-0 retains the lowest median absolute error of $38\,\mathrm{meV}$, but its mean errors are dominated by the largest-error $5\%$ of barriers. Random selection yields the largest seed-averaged MAE and RMSE, with seed-to-seed standard deviations exceeding $24\%$ and $90\%$ of their respective means. Unlike the preceding near-equilibrium benchmarks, defect extrapolation thus depends strongly on which $1\%$ of the training data is selected, with targeted selection outperforming random sampling on average.

For the remaining 1035 hops, every model exhibits a systematic shift in median signed error of $62$--$80\,\mathrm{meV}$, consistent with the PBE/PBE+$U$ mismatch. 
MACE-OMAT-0 and EnvTok differ by only $1\,\mathrm{meV}$ in MAE and $2\,\mathrm{meV}$ in median absolute error, whereas 
DIRECT shows a larger MAE ($0.316\,\mathrm{eV}$) and RMSE ($1.16\,\mathrm{eV}$).
The smaller apparent errors of the random models arise at least partly from cancellation with the level-of-theory mismatch and therefore do not establish higher accuracy.\newline

\begin{figure*}[th]
\centering
      \begin{minipage}[b]{0.293\linewidth}
          \centering
          \begin{subfigure}[b]{\linewidth}
              \centering
              \includegraphics[height=105pt]{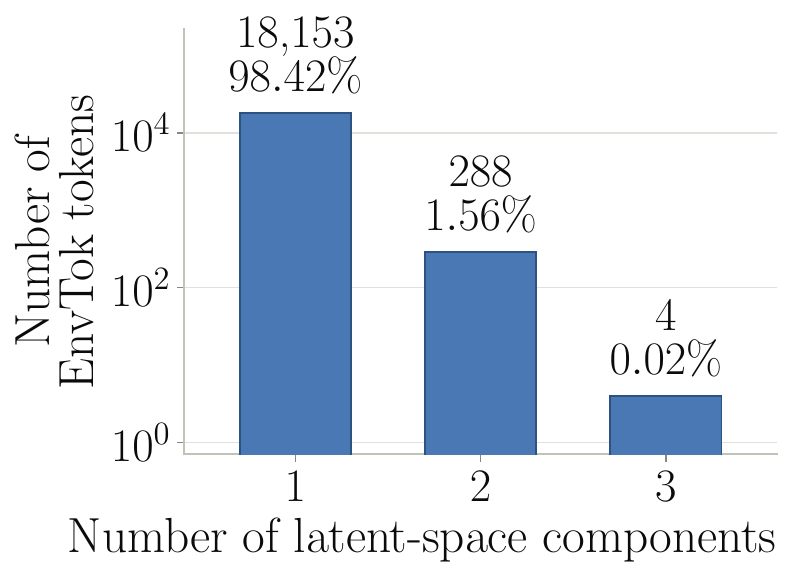}
              \caption{}
              \label{fig:modality}
          \end{subfigure}
          \par\vspace{0.5ex}
          \begin{subfigure}[b]{\linewidth}
              \centering
              \includegraphics[height=105pt]{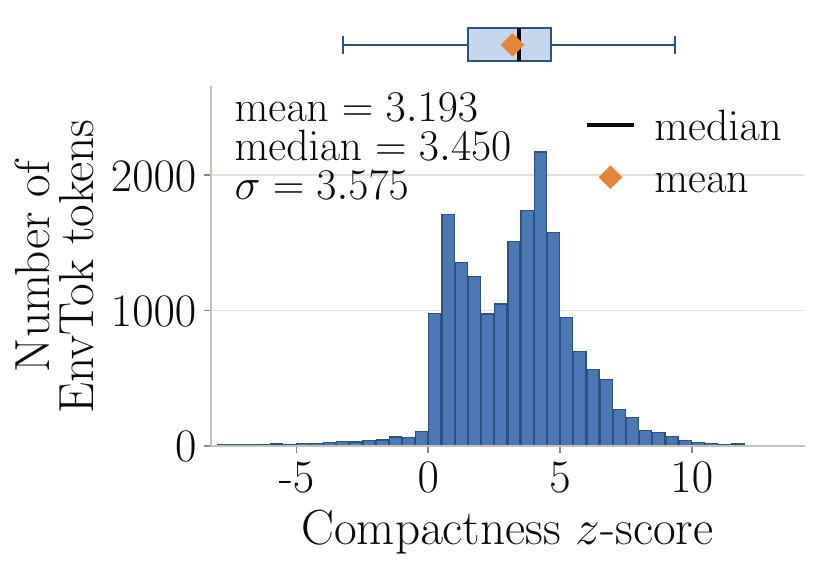}
              \caption{}
              \label{fig:compactness}
          \end{subfigure}
          \par\vspace{0.5ex}
          \begin{subfigure}[b]{\linewidth}
              \centering
              \includegraphics[height=105pt]{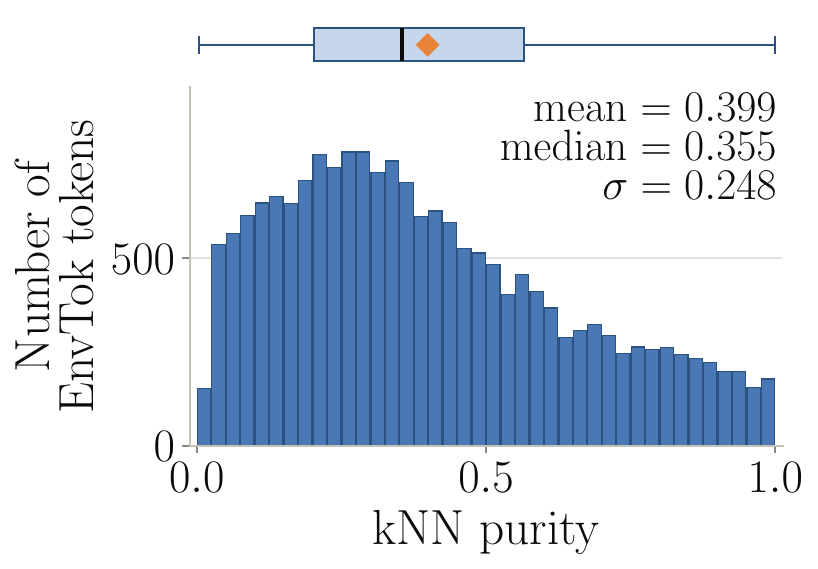}
              \caption{}
              \label{fig:knn_purity}
          \end{subfigure}
      \end{minipage}\hfill
      \begin{subfigure}[b]{0.692\linewidth}
          \centering
          \includegraphics[width=\linewidth]{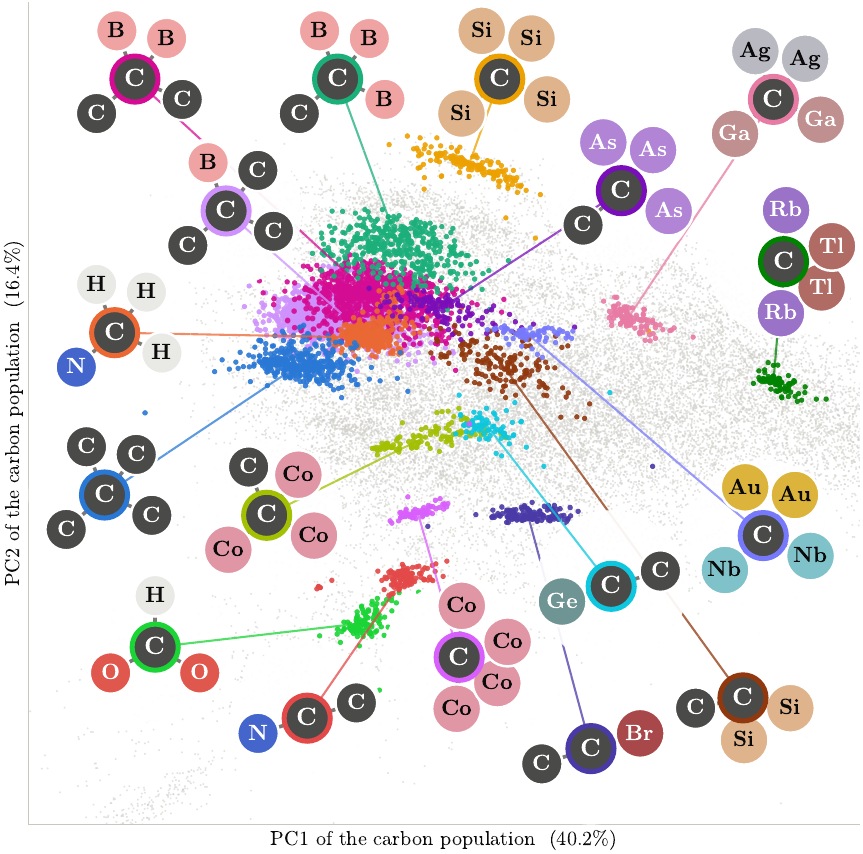}
          \caption{}
          \label{fig:chemenv-scatter-in-latent-space}
      \end{subfigure}
    \caption{Properties of EnvTok descriptors in the latent space of MACE-OMAT-0, evaluated on the $18\,445$ unique descriptors that are represented by at least $30$ atoms in the
    $f$-block-free OMat24 validation set: (\subref{fig:modality}) number of connected regions formed by the atoms of each descriptor, (\subref{fig:compactness}) within-class compactness $z$-score and (\subref{fig:knn_purity}) $k$-nearest-neighbour label purity. (\subref{fig:chemenv-scatter-in-latent-space}) Atoms of selected carbon-centred descriptors in the first two principal components, with representative structures}
    \label{fig:latent_space}
\end{figure*}

\noindent\textbf{Mapping the EnvTok descriptor to MLFF latent space.} The benchmarks above indicate that a suitably selected subset of a few hundred thousand structures captures most of the information in OMat24. EnvTok selects these structures using a predefined, model-free descriptor. To test whether its environment classes reflect the organization learned by an MLFF, we map EnvTok tokens onto the latent space of MACE-OMAT-0 for the 12 million atoms in the $f$-block-free OMat24 validation set. We use PCA-reduced node features and a symmetric $k$-nearest-neighbor graph ($k=15$), restricting the analysis to the 18{\,}445 descriptors represented by at least 30 atoms (for further details on the analysis of the EnvTok tokens in latent space we refer to Sec.~S5).

Three results characterize this correspondence (Fig.~\ref{fig:latent_space}). First, $98.42\%$ of the descriptors form a single connected component in the latent-space graph (see Fig.~\ref{fig:modality}). Only 292 ($1.6\%$) split into two or three components, with no systematic chemical signature. Second, $96.9\%$ form tighter clusters than equally sized random samples of atoms with the same central element, see Fig.~\ref{fig:compactness}. The diffuse minority is dominated by cations coordinated by O or F, consistent with nominally identical polyhedra occurring in distinct host chemistries whose second-shell variation the local token cannot resolve. Third, although the mean nearest-neighbor label purity is $0.4$ (see Fig.~\ref{fig:knn_purity}), differently labelled neighbors share the central element in $99.9\%$ of cases, the ligand-element set in $69.7\%$, and the coordination number in $56.9\%$. Thus, EnvTok primarily partitions chemically similar regions into finer classes. Together, these results support EnvTok diversity as a proxy for latent-space coverage, with some redundancy among neighboring classes.

Earlier studies showed that MLFF representations organize structural phases of elemental solids into connected regions~\cite{cubuk2017representations,smith2021automated} and that this broad organization persists across architectures~\cite{li2026platonic}. Here, we find a finer correspondence within a single model: local environments classified by element, spin, coordination polyhedron, and distortion map predominantly onto connected, compact regions of latent space. This agreement between coordination chemistry and a learned representation supports using EnvTok as a model-free selection criterion, identifying distinct environments without prior model training.

\begin{figure}[t]
    \centering
    \includegraphics[width=\linewidth]{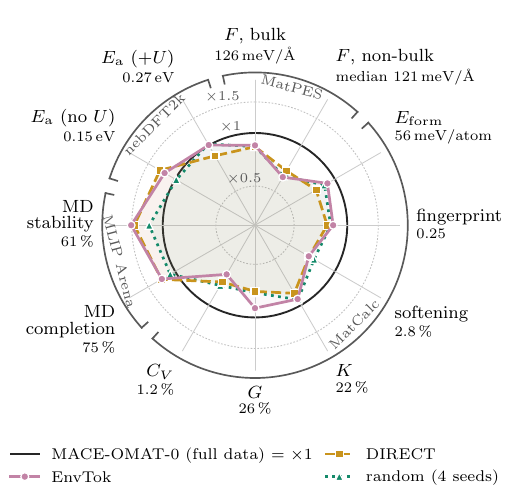}
    \caption{
   Performance of subset-trained models relative to MACE-OMAT-0 (black circle, $\times1$) across all benchmarks. Radial values show the reference-to-model error ratio, or the 
  model-to-reference success-rate ratio for the two MD axes, on a logarithmic scale. Points outside the circle indicate better performance of the $1\%$ model. Absolute MACE-OMAT-0 values 
  are listed below the axis labels and define $\times1$ on each spoke; $K$, $G$ and $C_V$ show mean absolute percentage errors, softening the mean signed deviation of the softening scale 
  from 1, and MD stability the fraction of runs without physical divergence. 
  For example, $\times0.7$ on the softening axis corresponds to $2.8\%$ for MACE-OMAT-0 versus $3.9$ to $4.2\%$
  for the subset-trained models.
  For non-bulk structures, the median per-atom force error is shown instead of the mean, which was
  shown to be dominated by a few atoms with diverging forces}
    \label{fig:radar}
\end{figure}

\section{Conclusions}

We show that MLFFs trained on less than $1\%$ of OMat24, $6.5\times10^{6}$ of $1.2\times10^{9}$ atoms, achieve results comparable to the full-data model across static and dynamic benchmarks. For most properties, performance depends only weakly on the dataset reduction scheme: random sampling, latent-space clustering (DIRECT), or chemical-environment coverage (EnvTok) (Fig.~\ref{fig:radar}).  
Training on the full dataset lowers the bulk force MAE from $149$--$151$ to $126\,\mathrm{meV}/\mathrm{\AA}$ 
and the formation-energy MAE from $63$--$74$ to $56\,\mathrm{meV/atom}$. These gains are modest compared with the hundredfold increase in data. The subset models also reproduce most derived properties with comparable accuracy and outperform the reference in MD stability under extreme conditions.

The weak dependence on selection strategy points to substantial redundancy in OMat24: subsets of this size already sample common environments repeatedly. This is consistent with generating many perturbed configurations from a smaller set of parent structures. Selection nevertheless matters for properties sensitive to rare environments, particularly MD stability and Li$^{+}$ migration barriers. 
EnvTok gives the lowest subset-model errors for bulk forces and shear moduli, whereas DIRECT gives the lowest errors for non-bulk forces but larger formation-energy and elastic-modulus errors. Both outperform random selection in MD stability and in Li$^{+}$ migration barriers without Hubbard $U$ elements. The key to MLFF performance is thus sampling distinct environments rather than labeling repeated instances.

These results motivate separating configuration sampling from reference labeling. Configurational space can be explored broadly, while expensive calculations are reserved for structures that expand environmental coverage. EnvTok provides a practical criterion using atomic positions and species, without reference energy or force labels or a pretrained MLFF. Its correspondence with connected, compact regions of the MACE-OMAT-0 latent space supports this approach. The present benchmarks suggest that much smaller training sets can retain broad predictive accuracy, making reference calculations at higher levels of theory affordable. 

Realizing this potential requires resolving the remaining under-represented environments. Targeted enrichment with low-dimensional structures offers a route to correcting the rare catastrophic failures on MatPES ($<0.3\%$), 
where the full-data model remains reliable and which DIRECT's whole-structure selection already reduces. Extending the Voronoi-based descriptor to handle vacuum would enable coverage-based selection for surfaces, wires, and clusters. Testing the fine-grained correspondence with learned representations across architectures would establish its generality beyond the shared coarse organization reported in Ref.~\cite{li2026platonic}. Together, these steps would advance environmental coverage from an effective data-reduction criterion toward a systematic basis for compact, reliable MLFF training sets.
Preliminary analyses of the large-scale molecular datasets OMol25~\cite{levine2025open} and QCML~\cite{ganscha2025qcml} reveal a similar degree of redundancy. This work is ongoing and will be reported in a forthcoming publication.

\vspace{-2mm}
\section*{Code and Data Availability}
The scripts for the EnvTok descriptor labelling, tokenization, subset selection,
MLFF training and benchmarking are collected in Ref.~\cite{chemenv-scripts}
and additional data including the OMat24 subsets, the trained models, benchmarking results
and MACE latent space analysis has been deposited on Zenodo~\cite{chemenv-data-zenodo}.

\vspace{-2mm}
\section*{Acknowledgements}
All calculations were performed on the Luxembourg national supercomputer MeluXina.

\vspace{-2mm}
\section*{Funding}
E.M., A.C., and A.T. acknowledge the European Research Council under ERC-AdG grant FITMOL (101054629). A.T. acknowledges the Luxembourg National Research Fund under grant FNR-CORE MBD-in-BMD (18093472). I.P. acknowledges the Luxembourg National Research Fund under grant FNR-CORE UMLFF (19563414).

\vspace{-2mm}
\section*{Contributions}
  E.M., I.P. and A.T. conceived the project, and I.P. and A.T. supervised it. 
  E.M. implemented the methods,
  trained the machine-learning force fields, and performed the benchmarks and
  analyses. E.M. and I.P. prepared the figures and wrote the manuscript. A.C.
  contributed to discussions, curated the references, and proofread the manuscript.
  All authors reviewed and approved the final manuscript.

\vspace{-2mm}
\section*{Competing interests}
The authors declare no competing interests.

\bibliography{references}

\begin{thebibliography}{71}%
\makeatletter
\providecommand \@ifxundefined [1]{%
 \@ifx{#1\undefined}
}%
\providecommand \@ifnum [1]{%
 \ifnum #1\expandafter \@firstoftwo
 \else \expandafter \@secondoftwo
 \fi
}%
\providecommand \@ifx [1]{%
 \ifx #1\expandafter \@firstoftwo
 \else \expandafter \@secondoftwo
 \fi
}%
\providecommand \natexlab [1]{#1}%
\providecommand \enquote  [1]{``#1''}%
\providecommand \bibnamefont  [1]{#1}%
\providecommand \bibfnamefont [1]{#1}%
\providecommand \citenamefont [1]{#1}%
\providecommand \href@noop [0]{\@secondoftwo}%
\providecommand \href [0]{\begingroup \@sanitize@url \@href}%
\providecommand \@href[1]{\@@startlink{#1}\@@href}%
\providecommand \@@href[1]{\endgroup#1\@@endlink}%
\providecommand \@sanitize@url [0]{\catcode `\\12\catcode `\$12\catcode
  `\&12\catcode `\#12\catcode `\^12\catcode `\_12\catcode `\%12\relax}%
\providecommand \@@startlink[1]{}%
\providecommand \@@endlink[0]{}%
\providecommand \url  [0]{\begingroup\@sanitize@url \@url }%
\providecommand \@url [1]{\endgroup\@href {#1}{\urlprefix }}%
\providecommand \urlprefix  [0]{URL }%
\providecommand \Eprint [0]{\href }%
\providecommand \doibase [0]{https://doi.org/}%
\providecommand \selectlanguage [0]{\@gobble}%
\providecommand \bibinfo  [0]{\@secondoftwo}%
\providecommand \bibfield  [0]{\@secondoftwo}%
\providecommand \translation [1]{[#1]}%
\providecommand \BibitemOpen [0]{}%
\providecommand \bibitemStop [0]{}%
\providecommand \bibitemNoStop [0]{.\EOS\space}%
\providecommand \EOS [0]{\spacefactor3000\relax}%
\providecommand \BibitemShut  [1]{\csname bibitem#1\endcsname}%
\let\auto@bib@innerbib\@empty
\bibitem [{\citenamefont {Unke}\ \emph {et~al.}(2021)\citenamefont {Unke},
  \citenamefont {Chmiela}, \citenamefont {Sauceda}, \citenamefont {Gastegger},
  \citenamefont {Poltavsky}, \citenamefont {Schutt}, \citenamefont
  {Tkatchenko},\ and\ \citenamefont {Muller}}]{unke2021machine}%
  \BibitemOpen
  \bibfield  {author} {\bibinfo {author} {\bibfnamefont {O.~T.}\ \bibnamefont
  {Unke}}, \bibinfo {author} {\bibfnamefont {S.}~\bibnamefont {Chmiela}},
  \bibinfo {author} {\bibfnamefont {H.~E.}\ \bibnamefont {Sauceda}}, \bibinfo
  {author} {\bibfnamefont {M.}~\bibnamefont {Gastegger}}, \bibinfo {author}
  {\bibfnamefont {I.}~\bibnamefont {Poltavsky}}, \bibinfo {author}
  {\bibfnamefont {K.~T.}\ \bibnamefont {Schutt}}, \bibinfo {author}
  {\bibfnamefont {A.}~\bibnamefont {Tkatchenko}},\ and\ \bibinfo {author}
  {\bibfnamefont {K.-R.}\ \bibnamefont {Muller}},\ }\bibfield  {title}
  {\bibinfo {title} {Machine learning force fields},\ }\href
  {https://doi.org/10.1021/acs.chemrev.0c01111} {\bibfield  {journal} {\bibinfo
   {journal} {Chem. Rev.}\ }\textbf {\bibinfo {volume} {121}},\ \bibinfo
  {pages} {10142} (\bibinfo {year} {2021})}\BibitemShut {NoStop}%
\bibitem [{\citenamefont {Deringer}\ \emph {et~al.}(2019)\citenamefont
  {Deringer}, \citenamefont {Caro},\ and\ \citenamefont
  {Cs{\'a}nyi}}]{deringer2019machine}%
  \BibitemOpen
  \bibfield  {author} {\bibinfo {author} {\bibfnamefont {V.~L.}\ \bibnamefont
  {Deringer}}, \bibinfo {author} {\bibfnamefont {M.~A.}\ \bibnamefont {Caro}},\
  and\ \bibinfo {author} {\bibfnamefont {G.}~\bibnamefont {Cs{\'a}nyi}},\
  }\bibfield  {title} {\bibinfo {title} {Machine learning interatomic
  potentials as emerging tools for materials science},\ }\href
  {https://doi.org/10.1002/adma.201902765} {\bibfield  {journal} {\bibinfo
  {journal} {Adv. Mater.}\ }\textbf {\bibinfo {volume} {31}},\ \bibinfo {pages}
  {1902765} (\bibinfo {year} {2019})}\BibitemShut {NoStop}%
\bibitem [{\citenamefont {Deng}\ \emph {et~al.}(2023)\citenamefont {Deng},
  \citenamefont {Zhong}, \citenamefont {Jun}, \citenamefont {Riebesell},
  \citenamefont {Han}, \citenamefont {Bartel},\ and\ \citenamefont
  {Ceder}}]{deng2023chgnet}%
  \BibitemOpen
  \bibfield  {author} {\bibinfo {author} {\bibfnamefont {B.}~\bibnamefont
  {Deng}}, \bibinfo {author} {\bibfnamefont {P.}~\bibnamefont {Zhong}},
  \bibinfo {author} {\bibfnamefont {K.}~\bibnamefont {Jun}}, \bibinfo {author}
  {\bibfnamefont {J.}~\bibnamefont {Riebesell}}, \bibinfo {author}
  {\bibfnamefont {K.}~\bibnamefont {Han}}, \bibinfo {author} {\bibfnamefont
  {C.~J.}\ \bibnamefont {Bartel}},\ and\ \bibinfo {author} {\bibfnamefont
  {G.}~\bibnamefont {Ceder}},\ }\bibfield  {title} {\bibinfo {title} {Chgnet as
  a pretrained universal neural network potential for charge-informed atomistic
  modelling},\ }\href {https://doi.org/10.1038/s42256-023-00716-3} {\bibfield
  {journal} {\bibinfo  {journal} {Nat. Mach. Intell.}\ }\textbf {\bibinfo
  {volume} {5}},\ \bibinfo {pages} {1031} (\bibinfo {year} {2023})}\BibitemShut
  {NoStop}%
\bibitem [{\citenamefont {Batatia}\ \emph
  {et~al.}(2025{\natexlab{a}})\citenamefont {Batatia} \emph
  {et~al.}}]{batatia2025foundation}%
  \BibitemOpen
  \bibfield  {author} {\bibinfo {author} {\bibfnamefont {I.}~\bibnamefont
  {Batatia}} \emph {et~al.},\ }\bibfield  {title} {\bibinfo {title} {A
  foundation model for atomistic materials chemistry},\ }\href
  {https://doi.org/10.1063/5.0297006} {\bibfield  {journal} {\bibinfo
  {journal} {J. Chem. Phys.}\ }\textbf {\bibinfo {volume} {163}},\ \bibinfo
  {pages} {184110} (\bibinfo {year} {2025}{\natexlab{a}})}\BibitemShut
  {NoStop}%
\bibitem [{\citenamefont {Park}\ \emph {et~al.}(2024)\citenamefont {Park},
  \citenamefont {Kim}, \citenamefont {Hwang},\ and\ \citenamefont
  {Han}}]{park2024sevennet}%
  \BibitemOpen
  \bibfield  {author} {\bibinfo {author} {\bibfnamefont {Y.}~\bibnamefont
  {Park}}, \bibinfo {author} {\bibfnamefont {J.}~\bibnamefont {Kim}}, \bibinfo
  {author} {\bibfnamefont {S.}~\bibnamefont {Hwang}},\ and\ \bibinfo {author}
  {\bibfnamefont {S.}~\bibnamefont {Han}},\ }\bibfield  {title} {\bibinfo
  {title} {Scalable parallel algorithm for graph neural network interatomic
  potentials in molecular dynamics simulations},\ }\href
  {https://doi.org/10.1021/acs.jctc.4c00190} {\bibfield  {journal} {\bibinfo
  {journal} {J. Chem. Theory Comput.}\ }\textbf {\bibinfo {volume} {20}},\
  \bibinfo {pages} {4857} (\bibinfo {year} {2024})}\BibitemShut {NoStop}%
\bibitem [{\citenamefont {Zhang}\ \emph {et~al.}(2024)\citenamefont {Zhang},
  \citenamefont {Liu}, \citenamefont {Zhang} \emph {et~al.}}]{zhang2024dpa2}%
  \BibitemOpen
  \bibfield  {author} {\bibinfo {author} {\bibfnamefont {D.}~\bibnamefont
  {Zhang}}, \bibinfo {author} {\bibfnamefont {X.}~\bibnamefont {Liu}}, \bibinfo
  {author} {\bibfnamefont {X.}~\bibnamefont {Zhang}}, \emph {et~al.},\
  }\bibfield  {title} {\bibinfo {title} {{DPA-2}: a large atomic model as a
  multi-task learner},\ }\href {https://doi.org/10.1038/s41524-024-01493-2}
  {\bibfield  {journal} {\bibinfo  {journal} {npj Comput. Mater.}\ }\textbf
  {\bibinfo {volume} {10}},\ \bibinfo {pages} {293} (\bibinfo {year}
  {2024})}\BibitemShut {NoStop}%
\bibitem [{\citenamefont {Frey}\ \emph {et~al.}(2023)\citenamefont {Frey},
  \citenamefont {Soklaski}, \citenamefont {Axelrod}, \citenamefont {Samsi},
  \citenamefont {Gomez-Bombarelli}, \citenamefont {Coley},\ and\ \citenamefont
  {Gadepally}}]{frey2023neural}%
  \BibitemOpen
  \bibfield  {author} {\bibinfo {author} {\bibfnamefont {N.~C.}\ \bibnamefont
  {Frey}}, \bibinfo {author} {\bibfnamefont {R.}~\bibnamefont {Soklaski}},
  \bibinfo {author} {\bibfnamefont {S.}~\bibnamefont {Axelrod}}, \bibinfo
  {author} {\bibfnamefont {S.}~\bibnamefont {Samsi}}, \bibinfo {author}
  {\bibfnamefont {R.}~\bibnamefont {Gomez-Bombarelli}}, \bibinfo {author}
  {\bibfnamefont {C.~W.}\ \bibnamefont {Coley}},\ and\ \bibinfo {author}
  {\bibfnamefont {V.}~\bibnamefont {Gadepally}},\ }\bibfield  {title} {\bibinfo
  {title} {Neural scaling of deep chemical models},\ }\href
  {https://doi.org/10.1038/s42256-023-00740-3} {\bibfield  {journal} {\bibinfo
  {journal} {Nat. Mach. Intell.}\ }\textbf {\bibinfo {volume} {5}},\ \bibinfo
  {pages} {1297} (\bibinfo {year} {2023})}\BibitemShut {NoStop}%
\bibitem [{\citenamefont {Schmidt}\ \emph {et~al.}(2024)\citenamefont
  {Schmidt}, \citenamefont {Cerqueira}, \citenamefont {Romero}, \citenamefont
  {Loew}, \citenamefont {J{\"a}ger}, \citenamefont {Wang}, \citenamefont
  {Botti},\ and\ \citenamefont {Marques}}]{schmidt2024improving}%
  \BibitemOpen
  \bibfield  {author} {\bibinfo {author} {\bibfnamefont {J.}~\bibnamefont
  {Schmidt}}, \bibinfo {author} {\bibfnamefont {T.~F.}\ \bibnamefont
  {Cerqueira}}, \bibinfo {author} {\bibfnamefont {A.~H.}\ \bibnamefont
  {Romero}}, \bibinfo {author} {\bibfnamefont {A.}~\bibnamefont {Loew}},
  \bibinfo {author} {\bibfnamefont {F.}~\bibnamefont {J{\"a}ger}}, \bibinfo
  {author} {\bibfnamefont {H.-C.}\ \bibnamefont {Wang}}, \bibinfo {author}
  {\bibfnamefont {S.}~\bibnamefont {Botti}},\ and\ \bibinfo {author}
  {\bibfnamefont {M.~A.}\ \bibnamefont {Marques}},\ }\bibfield  {title}
  {\bibinfo {title} {Improving machine-learning models in materials science
  through large datasets},\ }\href
  {https://doi.org/10.1016/j.mtphys.2024.101560} {\bibfield  {journal}
  {\bibinfo  {journal} {Mater. Today Phys.}\ }\textbf {\bibinfo {volume}
  {48}},\ \bibinfo {pages} {101560} (\bibinfo {year} {2024})}\BibitemShut
  {NoStop}%
\bibitem [{\citenamefont {Yuan}\ \emph {et~al.}(2026)\citenamefont {Yuan},
  \citenamefont {Liu}, \citenamefont {Chen} \emph
  {et~al.}}]{yuan2026foundation}%
  \BibitemOpen
  \bibfield  {author} {\bibinfo {author} {\bibfnamefont {E.~C.-Y.}\
  \bibnamefont {Yuan}}, \bibinfo {author} {\bibfnamefont {Y.}~\bibnamefont
  {Liu}}, \bibinfo {author} {\bibfnamefont {J.}~\bibnamefont {Chen}}, \emph
  {et~al.},\ }\bibfield  {title} {\bibinfo {title} {Foundation models for
  atomistic simulation of chemistry and materials},\ }\href
  {https://doi.org/10.1038/s41570-025-00793-5} {\bibfield  {journal} {\bibinfo
  {journal} {Nat. Rev. Chem.}\ }\textbf {\bibinfo {volume} {10}},\ \bibinfo
  {pages} {212} (\bibinfo {year} {2026})}\BibitemShut {NoStop}%
\bibitem [{\citenamefont {Yang}\ \emph {et~al.}(2024)\citenamefont {Yang},
  \citenamefont {Hu}, \citenamefont {Zhou}, \citenamefont {Liu}, \citenamefont
  {Shi}, \citenamefont {Li}, \citenamefont {Li}, \citenamefont {Chen},
  \citenamefont {Chen}, \citenamefont {Zeni} \emph
  {et~al.}}]{yang2024mattersim}%
  \BibitemOpen
  \bibfield  {author} {\bibinfo {author} {\bibfnamefont {H.}~\bibnamefont
  {Yang}}, \bibinfo {author} {\bibfnamefont {C.}~\bibnamefont {Hu}}, \bibinfo
  {author} {\bibfnamefont {Y.}~\bibnamefont {Zhou}}, \bibinfo {author}
  {\bibfnamefont {X.}~\bibnamefont {Liu}}, \bibinfo {author} {\bibfnamefont
  {Y.}~\bibnamefont {Shi}}, \bibinfo {author} {\bibfnamefont {J.}~\bibnamefont
  {Li}}, \bibinfo {author} {\bibfnamefont {G.}~\bibnamefont {Li}}, \bibinfo
  {author} {\bibfnamefont {Z.}~\bibnamefont {Chen}}, \bibinfo {author}
  {\bibfnamefont {S.}~\bibnamefont {Chen}}, \bibinfo {author} {\bibfnamefont
  {C.}~\bibnamefont {Zeni}}, \emph {et~al.},\ }\bibfield  {title} {\bibinfo
  {title} {Mattersim: A deep learning atomistic model across elements,
  temperatures and pressures},\ }\href {https://arxiv.org/abs/2405.04967}
  {\bibfield  {journal} {\bibinfo  {journal} {arXiv preprint arXiv:2405.04967}\
  } (\bibinfo {year} {2024})}\BibitemShut {NoStop}%
\bibitem [{\citenamefont {Zhang}\ \emph {et~al.}(2026)\citenamefont {Zhang},
  \citenamefont {Peng}, \citenamefont {Cai} \emph {et~al.}}]{zhang2026dpa3}%
  \BibitemOpen
  \bibfield  {author} {\bibinfo {author} {\bibfnamefont {D.}~\bibnamefont
  {Zhang}}, \bibinfo {author} {\bibfnamefont {A.}~\bibnamefont {Peng}},
  \bibinfo {author} {\bibfnamefont {C.}~\bibnamefont {Cai}}, \emph {et~al.},\
  }\bibfield  {title} {\bibinfo {title} {A graph neural network for the era of
  large atomistic models},\ }\href {https://doi.org/10.1038/s41524-026-02146-2}
  {\bibfield  {journal} {\bibinfo  {journal} {npj Comput. Mater.}\ }\textbf
  {\bibinfo {volume} {12}},\ \bibinfo {pages} {276} (\bibinfo {year}
  {2026})}\BibitemShut {NoStop}%
\bibitem [{\citenamefont {Barroso-Luque}\ \emph {et~al.}(2024)\citenamefont
  {Barroso-Luque}, \citenamefont {Shuaibi}, \citenamefont {Fu}, \citenamefont
  {Wood}, \citenamefont {Dzamba}, \citenamefont {Gao}, \citenamefont {Rizvi},
  \citenamefont {Zitnick},\ and\ \citenamefont {Ulissi}}]{barroso2024open}%
  \BibitemOpen
  \bibfield  {author} {\bibinfo {author} {\bibfnamefont {L.}~\bibnamefont
  {Barroso-Luque}}, \bibinfo {author} {\bibfnamefont {M.}~\bibnamefont
  {Shuaibi}}, \bibinfo {author} {\bibfnamefont {X.}~\bibnamefont {Fu}},
  \bibinfo {author} {\bibfnamefont {B.~M.}\ \bibnamefont {Wood}}, \bibinfo
  {author} {\bibfnamefont {M.}~\bibnamefont {Dzamba}}, \bibinfo {author}
  {\bibfnamefont {M.}~\bibnamefont {Gao}}, \bibinfo {author} {\bibfnamefont
  {A.}~\bibnamefont {Rizvi}}, \bibinfo {author} {\bibfnamefont {C.~L.}\
  \bibnamefont {Zitnick}},\ and\ \bibinfo {author} {\bibfnamefont {Z.~W.}\
  \bibnamefont {Ulissi}},\ }\bibfield  {title} {\bibinfo {title} {Open
  materials 2024 ({OMat24}) inorganic materials dataset and models},\ }\href
  {https://arxiv.org/abs/2410.12771} {\bibfield  {journal} {\bibinfo  {journal}
  {arXiv preprint arXiv:2410.12771}\ } (\bibinfo {year} {2024})}\BibitemShut
  {NoStop}%
\bibitem [{\citenamefont {Batatia}\ \emph
  {et~al.}(2025{\natexlab{b}})\citenamefont {Batatia}, \citenamefont {Lin},
  \citenamefont {Hart}, \citenamefont {Kasoar}, \citenamefont {Elena},
  \citenamefont {Norwood}, \citenamefont {Wolf},\ and\ \citenamefont
  {Cs{\'a}nyi}}]{batatia2025cross}%
  \BibitemOpen
  \bibfield  {author} {\bibinfo {author} {\bibfnamefont {I.}~\bibnamefont
  {Batatia}}, \bibinfo {author} {\bibfnamefont {C.}~\bibnamefont {Lin}},
  \bibinfo {author} {\bibfnamefont {J.}~\bibnamefont {Hart}}, \bibinfo {author}
  {\bibfnamefont {E.}~\bibnamefont {Kasoar}}, \bibinfo {author} {\bibfnamefont
  {A.~M.}\ \bibnamefont {Elena}}, \bibinfo {author} {\bibfnamefont {S.~W.}\
  \bibnamefont {Norwood}}, \bibinfo {author} {\bibfnamefont {T.}~\bibnamefont
  {Wolf}},\ and\ \bibinfo {author} {\bibfnamefont {G.}~\bibnamefont
  {Cs{\'a}nyi}},\ }\bibfield  {title} {\bibinfo {title} {Cross learning between
  electronic structure theories for unifying molecular, surface, and inorganic
  crystal foundation force fields},\ }\href {https://arxiv.org/abs/2510.25380}
  {\bibfield  {journal} {\bibinfo  {journal} {arXiv preprint arXiv:2510.25380}\
  } (\bibinfo {year} {2025}{\natexlab{b}})}\BibitemShut {NoStop}%
\bibitem [{\citenamefont {Fu}\ \emph {et~al.}(2025)\citenamefont {Fu},
  \citenamefont {Wood}, \citenamefont {Barroso-Luque}, \citenamefont {Levine},
  \citenamefont {Gao}, \citenamefont {Dzamba},\ and\ \citenamefont
  {Zitnick}}]{fu2025learning}%
  \BibitemOpen
  \bibfield  {author} {\bibinfo {author} {\bibfnamefont {X.}~\bibnamefont
  {Fu}}, \bibinfo {author} {\bibfnamefont {B.~M.}\ \bibnamefont {Wood}},
  \bibinfo {author} {\bibfnamefont {L.}~\bibnamefont {Barroso-Luque}}, \bibinfo
  {author} {\bibfnamefont {D.~S.}\ \bibnamefont {Levine}}, \bibinfo {author}
  {\bibfnamefont {M.}~\bibnamefont {Gao}}, \bibinfo {author} {\bibfnamefont
  {M.}~\bibnamefont {Dzamba}},\ and\ \bibinfo {author} {\bibfnamefont {C.~L.}\
  \bibnamefont {Zitnick}},\ }\bibfield  {title} {\bibinfo {title} {Learning
  smooth and expressive interatomic potentials for physical property
  prediction},\ }\href {https://arxiv.org/abs/2502.12147} {\bibfield  {journal}
  {\bibinfo  {journal} {arXiv preprint arXiv:2502.12147}\ } (\bibinfo {year}
  {2025})}\BibitemShut {NoStop}%
\bibitem [{\citenamefont {Wood}\ \emph {et~al.}(2025)\citenamefont {Wood},
  \citenamefont {Dzamba}, \citenamefont {Fu}, \citenamefont {Gao},
  \citenamefont {Shuaibi}, \citenamefont {Barroso-Luque}, \citenamefont
  {Abdelmaqsoud}, \citenamefont {Gharakhanyan}, \citenamefont {Kitchin},
  \citenamefont {Levine} \emph {et~al.}}]{wood2026family}%
  \BibitemOpen
  \bibfield  {author} {\bibinfo {author} {\bibfnamefont {B.}~\bibnamefont
  {Wood}}, \bibinfo {author} {\bibfnamefont {M.}~\bibnamefont {Dzamba}},
  \bibinfo {author} {\bibfnamefont {X.}~\bibnamefont {Fu}}, \bibinfo {author}
  {\bibfnamefont {M.}~\bibnamefont {Gao}}, \bibinfo {author} {\bibfnamefont
  {M.}~\bibnamefont {Shuaibi}}, \bibinfo {author} {\bibfnamefont
  {L.}~\bibnamefont {Barroso-Luque}}, \bibinfo {author} {\bibfnamefont
  {K.}~\bibnamefont {Abdelmaqsoud}}, \bibinfo {author} {\bibfnamefont
  {V.}~\bibnamefont {Gharakhanyan}}, \bibinfo {author} {\bibfnamefont
  {J.}~\bibnamefont {Kitchin}}, \bibinfo {author} {\bibfnamefont
  {D.}~\bibnamefont {Levine}}, \emph {et~al.},\ }\bibfield  {title} {\bibinfo
  {title} {Uma: A family of universal models for atoms},\ }\href
  {https://doi.org/10.52202/085713-4310} {\bibfield  {journal} {\bibinfo
  {journal} {Adv. Neural Inf. Process. Syst.}\ }\textbf {\bibinfo {volume}
  {38}},\ \bibinfo {pages} {129391} (\bibinfo {year} {2025})}\BibitemShut
  {NoStop}%
\bibitem [{\citenamefont {Rhodes}\ \emph {et~al.}(2025)\citenamefont {Rhodes},
  \citenamefont {Vandenhaute}, \citenamefont {{\v{S}}imkus}, \citenamefont
  {Gin}, \citenamefont {Godwin}, \citenamefont {Duignan},\ and\ \citenamefont
  {Neumann}}]{rhodes2025orb}%
  \BibitemOpen
  \bibfield  {author} {\bibinfo {author} {\bibfnamefont {B.}~\bibnamefont
  {Rhodes}}, \bibinfo {author} {\bibfnamefont {S.}~\bibnamefont {Vandenhaute}},
  \bibinfo {author} {\bibfnamefont {V.}~\bibnamefont {{\v{S}}imkus}}, \bibinfo
  {author} {\bibfnamefont {J.}~\bibnamefont {Gin}}, \bibinfo {author}
  {\bibfnamefont {J.}~\bibnamefont {Godwin}}, \bibinfo {author} {\bibfnamefont
  {T.}~\bibnamefont {Duignan}},\ and\ \bibinfo {author} {\bibfnamefont
  {M.}~\bibnamefont {Neumann}},\ }\bibfield  {title} {\bibinfo {title} {Orb-v3:
  atomistic simulation at scale},\ }\href {https://arxiv.org/abs/2504.06231}
  {\bibfield  {journal} {\bibinfo  {journal} {arXiv preprint arXiv:2504.06231}\
  } (\bibinfo {year} {2025})}\BibitemShut {NoStop}%
\bibitem [{\citenamefont {Lysogorskiy}\ \emph {et~al.}(2026)\citenamefont
  {Lysogorskiy}, \citenamefont {Bochkarev},\ and\ \citenamefont
  {Drautz}}]{lysogorskiy2026graph}%
  \BibitemOpen
  \bibfield  {author} {\bibinfo {author} {\bibfnamefont {Y.}~\bibnamefont
  {Lysogorskiy}}, \bibinfo {author} {\bibfnamefont {A.}~\bibnamefont
  {Bochkarev}},\ and\ \bibinfo {author} {\bibfnamefont {R.}~\bibnamefont
  {Drautz}},\ }\bibfield  {title} {\bibinfo {title} {Graph atomic cluster
  expansion for foundational machine learning interatomic potentials},\ }\href
  {https://doi.org/10.1038/s41524-026-01979-1} {\bibfield  {journal} {\bibinfo
  {journal} {npj Comput. Mater.}\ }\textbf {\bibinfo {volume} {12}},\ \bibinfo
  {pages} {114} (\bibinfo {year} {2026})}\BibitemShut {NoStop}%
\bibitem [{\citenamefont {Mazitov}\ \emph
  {et~al.}(2025{\natexlab{a}})\citenamefont {Mazitov}, \citenamefont {Bigi},
  \citenamefont {Kellner}, \citenamefont {Pegolo}, \citenamefont {Tisi},
  \citenamefont {Fraux}, \citenamefont {Pozdnyakov}, \citenamefont {Loche},\
  and\ \citenamefont {Ceriotti}}]{mazitov2025petmad}%
  \BibitemOpen
  \bibfield  {author} {\bibinfo {author} {\bibfnamefont {A.}~\bibnamefont
  {Mazitov}}, \bibinfo {author} {\bibfnamefont {F.}~\bibnamefont {Bigi}},
  \bibinfo {author} {\bibfnamefont {M.}~\bibnamefont {Kellner}}, \bibinfo
  {author} {\bibfnamefont {P.}~\bibnamefont {Pegolo}}, \bibinfo {author}
  {\bibfnamefont {D.}~\bibnamefont {Tisi}}, \bibinfo {author} {\bibfnamefont
  {G.}~\bibnamefont {Fraux}}, \bibinfo {author} {\bibfnamefont
  {S.}~\bibnamefont {Pozdnyakov}}, \bibinfo {author} {\bibfnamefont
  {P.}~\bibnamefont {Loche}},\ and\ \bibinfo {author} {\bibfnamefont
  {M.}~\bibnamefont {Ceriotti}},\ }\bibfield  {title} {\bibinfo {title}
  {{PET-MAD} as a lightweight universal interatomic potential for advanced
  materials modeling},\ }\href {https://doi.org/10.1038/s41467-025-65662-7}
  {\bibfield  {journal} {\bibinfo  {journal} {Nat. Commun.}\ }\textbf {\bibinfo
  {volume} {16}},\ \bibinfo {pages} {10653} (\bibinfo {year}
  {2025}{\natexlab{a}})}\BibitemShut {NoStop}%
\bibitem [{\citenamefont {Mazitov}\ \emph
  {et~al.}(2025{\natexlab{b}})\citenamefont {Mazitov}, \citenamefont {Chorna},
  \citenamefont {Fraux}, \citenamefont {Bercx}, \citenamefont {Pizzi},
  \citenamefont {De},\ and\ \citenamefont {Ceriotti}}]{mazitov2025mad}%
  \BibitemOpen
  \bibfield  {author} {\bibinfo {author} {\bibfnamefont {A.}~\bibnamefont
  {Mazitov}}, \bibinfo {author} {\bibfnamefont {S.}~\bibnamefont {Chorna}},
  \bibinfo {author} {\bibfnamefont {G.}~\bibnamefont {Fraux}}, \bibinfo
  {author} {\bibfnamefont {M.}~\bibnamefont {Bercx}}, \bibinfo {author}
  {\bibfnamefont {G.}~\bibnamefont {Pizzi}}, \bibinfo {author} {\bibfnamefont
  {S.}~\bibnamefont {De}},\ and\ \bibinfo {author} {\bibfnamefont
  {M.}~\bibnamefont {Ceriotti}},\ }\bibfield  {title} {\bibinfo {title}
  {{Massive Atomic Diversity}: a compact universal dataset for atomistic
  machine learning},\ }\href {https://doi.org/10.1038/s41597-025-06109-y}
  {\bibfield  {journal} {\bibinfo  {journal} {Sci. Data}\ }\textbf {\bibinfo
  {volume} {12}},\ \bibinfo {pages} {1857} (\bibinfo {year}
  {2025}{\natexlab{b}})}\BibitemShut {NoStop}%
\bibitem [{\citenamefont {Liang}\ \emph {et~al.}(2026)\citenamefont {Liang},
  \citenamefont {Xu}, \citenamefont {Lindgren} \emph
  {et~al.}}]{liang2026nep89}%
  \BibitemOpen
  \bibfield  {author} {\bibinfo {author} {\bibfnamefont {T.}~\bibnamefont
  {Liang}}, \bibinfo {author} {\bibfnamefont {K.}~\bibnamefont {Xu}}, \bibinfo
  {author} {\bibfnamefont {E.}~\bibnamefont {Lindgren}}, \emph {et~al.},\
  }\bibfield  {title} {\bibinfo {title} {{NEP89}: universal neuroevolution
  potential for inorganic and organic materials across 89 elements},\ }\href
  {https://doi.org/10.1038/s43588-026-01009-6} {\bibfield  {journal} {\bibinfo
  {journal} {Nat. Comput. Sci.}\ }\textbf {\bibinfo {volume} {6}},\ \bibinfo
  {pages} {789} (\bibinfo {year} {2026})}\BibitemShut {NoStop}%
\bibitem [{\citenamefont {Kaplan}\ \emph {et~al.}(2025)\citenamefont {Kaplan},
  \citenamefont {Liu}, \citenamefont {Qi}, \citenamefont {Ko}, \citenamefont
  {Deng}, \citenamefont {Riebesell}, \citenamefont {Ceder}, \citenamefont
  {Persson},\ and\ \citenamefont {Ong}}]{kaplan2025foundational}%
  \BibitemOpen
  \bibfield  {author} {\bibinfo {author} {\bibfnamefont {A.~D.}\ \bibnamefont
  {Kaplan}}, \bibinfo {author} {\bibfnamefont {R.}~\bibnamefont {Liu}},
  \bibinfo {author} {\bibfnamefont {J.}~\bibnamefont {Qi}}, \bibinfo {author}
  {\bibfnamefont {T.~W.}\ \bibnamefont {Ko}}, \bibinfo {author} {\bibfnamefont
  {B.}~\bibnamefont {Deng}}, \bibinfo {author} {\bibfnamefont {J.}~\bibnamefont
  {Riebesell}}, \bibinfo {author} {\bibfnamefont {G.}~\bibnamefont {Ceder}},
  \bibinfo {author} {\bibfnamefont {K.~A.}\ \bibnamefont {Persson}},\ and\
  \bibinfo {author} {\bibfnamefont {S.~P.}\ \bibnamefont {Ong}},\ }\bibfield
  {title} {\bibinfo {title} {A foundational potential energy surface dataset
  for materials},\ }\href {https://arxiv.org/abs/2503.04070} {\bibfield
  {journal} {\bibinfo  {journal} {arXiv preprint arXiv:2503.04070}\ } (\bibinfo
  {year} {2025})}\BibitemShut {NoStop}%
\bibitem [{\citenamefont {Kim}\ \emph {et~al.}(2025)\citenamefont {Kim},
  \citenamefont {Kim}, \citenamefont {Kim}, \citenamefont {Lee}, \citenamefont
  {Park}, \citenamefont {Kang},\ and\ \citenamefont {Han}}]{kim2025sevennetmf}%
  \BibitemOpen
  \bibfield  {author} {\bibinfo {author} {\bibfnamefont {J.}~\bibnamefont
  {Kim}}, \bibinfo {author} {\bibfnamefont {J.}~\bibnamefont {Kim}}, \bibinfo
  {author} {\bibfnamefont {J.}~\bibnamefont {Kim}}, \bibinfo {author}
  {\bibfnamefont {J.}~\bibnamefont {Lee}}, \bibinfo {author} {\bibfnamefont
  {Y.}~\bibnamefont {Park}}, \bibinfo {author} {\bibfnamefont {Y.}~\bibnamefont
  {Kang}},\ and\ \bibinfo {author} {\bibfnamefont {S.}~\bibnamefont {Han}},\
  }\bibfield  {title} {\bibinfo {title} {Data-efficient multifidelity training
  for high-fidelity machine learning interatomic potentials},\ }\href
  {https://doi.org/10.1021/jacs.4c14455} {\bibfield  {journal} {\bibinfo
  {journal} {J. Am. Chem. Soc.}\ }\textbf {\bibinfo {volume} {147}},\ \bibinfo
  {pages} {1042} (\bibinfo {year} {2025})}\BibitemShut {NoStop}%
\bibitem [{\citenamefont {Huang}\ \emph {et~al.}(2025)\citenamefont {Huang},
  \citenamefont {Deng}, \citenamefont {Zhong}, \citenamefont {Kaplan},
  \citenamefont {Persson},\ and\ \citenamefont
  {Ceder}}]{huang2025crossfunctional}%
  \BibitemOpen
  \bibfield  {author} {\bibinfo {author} {\bibfnamefont {X.}~\bibnamefont
  {Huang}}, \bibinfo {author} {\bibfnamefont {B.}~\bibnamefont {Deng}},
  \bibinfo {author} {\bibfnamefont {P.}~\bibnamefont {Zhong}}, \bibinfo
  {author} {\bibfnamefont {A.~D.}\ \bibnamefont {Kaplan}}, \bibinfo {author}
  {\bibfnamefont {K.~A.}\ \bibnamefont {Persson}},\ and\ \bibinfo {author}
  {\bibfnamefont {G.}~\bibnamefont {Ceder}},\ }\bibfield  {title} {\bibinfo
  {title} {Cross-functional transferability in foundation machine learning
  interatomic potentials},\ }\href {https://doi.org/10.1038/s41524-025-01796-y}
  {\bibfield  {journal} {\bibinfo  {journal} {npj Comput. Mater.}\ }\textbf
  {\bibinfo {volume} {11}},\ \bibinfo {pages} {313} (\bibinfo {year}
  {2025})}\BibitemShut {NoStop}%
\bibitem [{\citenamefont {Kim}\ \emph {et~al.}(2026)\citenamefont {Kim},
  \citenamefont {You}, \citenamefont {Park} \emph
  {et~al.}}]{kim2026crossdomain}%
  \BibitemOpen
  \bibfield  {author} {\bibinfo {author} {\bibfnamefont {J.}~\bibnamefont
  {Kim}}, \bibinfo {author} {\bibfnamefont {J.}~\bibnamefont {You}}, \bibinfo
  {author} {\bibfnamefont {Y.}~\bibnamefont {Park}}, \emph {et~al.},\
  }\bibfield  {title} {\bibinfo {title} {Optimizing cross-domain transfer for
  universal machine learning interatomic potentials},\ }\href
  {https://doi.org/10.1038/s41467-026-70195-8} {\bibfield  {journal} {\bibinfo
  {journal} {Nat. Commun.}\ }\textbf {\bibinfo {volume} {17}},\ \bibinfo
  {pages} {3432} (\bibinfo {year} {2026})}\BibitemShut {NoStop}%
\bibitem [{\citenamefont {Perdew}\ \emph {et~al.}(1996)\citenamefont {Perdew},
  \citenamefont {Burke},\ and\ \citenamefont
  {Ernzerhof}}]{perdew1996generalized}%
  \BibitemOpen
  \bibfield  {author} {\bibinfo {author} {\bibfnamefont {J.~P.}\ \bibnamefont
  {Perdew}}, \bibinfo {author} {\bibfnamefont {K.}~\bibnamefont {Burke}},\ and\
  \bibinfo {author} {\bibfnamefont {M.}~\bibnamefont {Ernzerhof}},\ }\bibfield
  {title} {\bibinfo {title} {Generalized gradient approximation made simple},\
  }\href {https://doi.org/10.1103/PhysRevLett.77.3865} {\bibfield  {journal}
  {\bibinfo  {journal} {Phys. Rev. Lett.}\ }\textbf {\bibinfo {volume} {77}},\
  \bibinfo {pages} {3865} (\bibinfo {year} {1996})}\BibitemShut {NoStop}%
\bibitem [{\citenamefont {Sun}\ \emph {et~al.}(2015)\citenamefont {Sun},
  \citenamefont {Ruzsinszky},\ and\ \citenamefont {Perdew}}]{sun2015strongly}%
  \BibitemOpen
  \bibfield  {author} {\bibinfo {author} {\bibfnamefont {J.}~\bibnamefont
  {Sun}}, \bibinfo {author} {\bibfnamefont {A.}~\bibnamefont {Ruzsinszky}},\
  and\ \bibinfo {author} {\bibfnamefont {J.~P.}\ \bibnamefont {Perdew}},\
  }\bibfield  {title} {\bibinfo {title} {Strongly constrained and appropriately
  normed semilocal density functional},\ }\href
  {https://doi.org/10.1103/physrevlett.115.036402} {\bibfield  {journal}
  {\bibinfo  {journal} {Phys. Rev. Lett.}\ }\textbf {\bibinfo {volume} {115}},\
  \bibinfo {pages} {036402} (\bibinfo {year} {2015})}\BibitemShut {NoStop}%
\bibitem [{\citenamefont {Furness}\ \emph {et~al.}(2020)\citenamefont
  {Furness}, \citenamefont {Kaplan}, \citenamefont {Ning}, \citenamefont
  {Perdew},\ and\ \citenamefont {Sun}}]{furness2020accurate}%
  \BibitemOpen
  \bibfield  {author} {\bibinfo {author} {\bibfnamefont {J.~W.}\ \bibnamefont
  {Furness}}, \bibinfo {author} {\bibfnamefont {A.~D.}\ \bibnamefont {Kaplan}},
  \bibinfo {author} {\bibfnamefont {J.}~\bibnamefont {Ning}}, \bibinfo {author}
  {\bibfnamefont {J.~P.}\ \bibnamefont {Perdew}},\ and\ \bibinfo {author}
  {\bibfnamefont {J.}~\bibnamefont {Sun}},\ }\bibfield  {title} {\bibinfo
  {title} {Accurate and numerically efficient r2scan meta-generalized gradient
  approximation},\ }\href {https://doi.org/10.1021/acs.jpclett.0c02405}
  {\bibfield  {journal} {\bibinfo  {journal} {J. Phys. Chem. Lett.}\ }\textbf
  {\bibinfo {volume} {11}},\ \bibinfo {pages} {8208} (\bibinfo {year}
  {2020})}\BibitemShut {NoStop}%
\bibitem [{\citenamefont {Ko}\ and\ \citenamefont
  {Ong}(2025)}]{ko2025multifidelity}%
  \BibitemOpen
  \bibfield  {author} {\bibinfo {author} {\bibfnamefont {T.~W.}\ \bibnamefont
  {Ko}}\ and\ \bibinfo {author} {\bibfnamefont {S.~P.}\ \bibnamefont {Ong}},\
  }\bibfield  {title} {\bibinfo {title} {Data-efficient construction of
  high-fidelity graph deep learning interatomic potentials},\ }\href
  {https://doi.org/10.1038/s41524-025-01550-4} {\bibfield  {journal} {\bibinfo
  {journal} {npj Comput. Mater.}\ }\textbf {\bibinfo {volume} {11}},\ \bibinfo
  {pages} {65} (\bibinfo {year} {2025})}\BibitemShut {NoStop}%
\bibitem [{\citenamefont {Sorscher}\ \emph {et~al.}(2022)\citenamefont
  {Sorscher}, \citenamefont {Geirhos}, \citenamefont {Shekhar}, \citenamefont
  {Ganguli},\ and\ \citenamefont {Morcos}}]{sorscher2022beyond}%
  \BibitemOpen
  \bibfield  {author} {\bibinfo {author} {\bibfnamefont {B.}~\bibnamefont
  {Sorscher}}, \bibinfo {author} {\bibfnamefont {R.}~\bibnamefont {Geirhos}},
  \bibinfo {author} {\bibfnamefont {S.}~\bibnamefont {Shekhar}}, \bibinfo
  {author} {\bibfnamefont {S.}~\bibnamefont {Ganguli}},\ and\ \bibinfo {author}
  {\bibfnamefont {A.}~\bibnamefont {Morcos}},\ }\bibfield  {title} {\bibinfo
  {title} {Beyond neural scaling laws: beating power law scaling via data
  pruning},\ }\href {https://doi.org/10.52202/068431-1419} {\bibfield
  {journal} {\bibinfo  {journal} {Adv. Neural Inf. Process. Syst.}\ }\textbf
  {\bibinfo {volume} {35}},\ \bibinfo {pages} {19523} (\bibinfo {year}
  {2022})}\BibitemShut {NoStop}%
\bibitem [{\citenamefont {Schwalbe-Koda}\ \emph {et~al.}(2025)\citenamefont
  {Schwalbe-Koda}, \citenamefont {Hamel}, \citenamefont {Sadigh}, \citenamefont
  {Zhou},\ and\ \citenamefont {Lordi}}]{schwalbe2025model}%
  \BibitemOpen
  \bibfield  {author} {\bibinfo {author} {\bibfnamefont {D.}~\bibnamefont
  {Schwalbe-Koda}}, \bibinfo {author} {\bibfnamefont {S.}~\bibnamefont
  {Hamel}}, \bibinfo {author} {\bibfnamefont {B.}~\bibnamefont {Sadigh}},
  \bibinfo {author} {\bibfnamefont {F.}~\bibnamefont {Zhou}},\ and\ \bibinfo
  {author} {\bibfnamefont {V.}~\bibnamefont {Lordi}},\ }\bibfield  {title}
  {\bibinfo {title} {Model-free estimation of completeness, uncertainties, and
  outliers in atomistic machine learning using information theory},\ }\href
  {https://doi.org/10.1038/s41467-025-59232-0} {\bibfield  {journal} {\bibinfo
  {journal} {Nat. Commun.}\ }\textbf {\bibinfo {volume} {16}},\ \bibinfo
  {pages} {4014} (\bibinfo {year} {2025})}\BibitemShut {NoStop}%
\bibitem [{\citenamefont {Karabin}\ and\ \citenamefont
  {Perez}(2020)}]{karabin2020entropy}%
  \BibitemOpen
  \bibfield  {author} {\bibinfo {author} {\bibfnamefont {M.}~\bibnamefont
  {Karabin}}\ and\ \bibinfo {author} {\bibfnamefont {D.}~\bibnamefont
  {Perez}},\ }\bibfield  {title} {\bibinfo {title} {An entropy-maximization
  approach to automated training set generation for interatomic potentials},\
  }\href {https://doi.org/10.1063/5.0013059} {\bibfield  {journal} {\bibinfo
  {journal} {J. Chem. Phys.}\ }\textbf {\bibinfo {volume} {153}},\ \bibinfo
  {pages} {094110} (\bibinfo {year} {2020})}\BibitemShut {NoStop}%
\bibitem [{\citenamefont {Li}\ \emph {et~al.}(2023)\citenamefont {Li},
  \citenamefont {Persaud}, \citenamefont {Choudhary}, \citenamefont {DeCost},
  \citenamefont {Greenwood},\ and\ \citenamefont
  {Hattrick-Simpers}}]{li2023exploiting}%
  \BibitemOpen
  \bibfield  {author} {\bibinfo {author} {\bibfnamefont {K.}~\bibnamefont
  {Li}}, \bibinfo {author} {\bibfnamefont {D.}~\bibnamefont {Persaud}},
  \bibinfo {author} {\bibfnamefont {K.}~\bibnamefont {Choudhary}}, \bibinfo
  {author} {\bibfnamefont {B.}~\bibnamefont {DeCost}}, \bibinfo {author}
  {\bibfnamefont {M.}~\bibnamefont {Greenwood}},\ and\ \bibinfo {author}
  {\bibfnamefont {J.}~\bibnamefont {Hattrick-Simpers}},\ }\bibfield  {title}
  {\bibinfo {title} {Exploiting redundancy in large materials datasets for
  efficient machine learning with less data},\ }\href
  {https://doi.org/10.1038/s41467-023-42992-y} {\bibfield  {journal} {\bibinfo
  {journal} {Nat. Commun.}\ }\textbf {\bibinfo {volume} {14}},\ \bibinfo
  {pages} {7283} (\bibinfo {year} {2023})}\BibitemShut {NoStop}%
\bibitem [{\citenamefont {Imbalzano}\ \emph {et~al.}(2018)\citenamefont
  {Imbalzano}, \citenamefont {Anelli}, \citenamefont {Giofr{\'e}},
  \citenamefont {Klees}, \citenamefont {Behler},\ and\ \citenamefont
  {Ceriotti}}]{imbalzano2018automatic}%
  \BibitemOpen
  \bibfield  {author} {\bibinfo {author} {\bibfnamefont {G.}~\bibnamefont
  {Imbalzano}}, \bibinfo {author} {\bibfnamefont {A.}~\bibnamefont {Anelli}},
  \bibinfo {author} {\bibfnamefont {D.}~\bibnamefont {Giofr{\'e}}}, \bibinfo
  {author} {\bibfnamefont {S.}~\bibnamefont {Klees}}, \bibinfo {author}
  {\bibfnamefont {J.}~\bibnamefont {Behler}},\ and\ \bibinfo {author}
  {\bibfnamefont {M.}~\bibnamefont {Ceriotti}},\ }\bibfield  {title} {\bibinfo
  {title} {Automatic selection of atomic fingerprints and reference
  configurations for machine-learning potentials},\ }\href
  {https://doi.org/10.1063/1.5024611} {\bibfield  {journal} {\bibinfo
  {journal} {J. Chem. Phys.}\ }\textbf {\bibinfo {volume} {148}},\ \bibinfo
  {pages} {241730} (\bibinfo {year} {2018})}\BibitemShut {NoStop}%
\bibitem [{\citenamefont {Trestman}\ \emph {et~al.}(2026)\citenamefont
  {Trestman}, \citenamefont {Gugler}, \citenamefont {Faber},\ and\
  \citenamefont {von Lilienfeld}}]{trestman2025gradient}%
  \BibitemOpen
  \bibfield  {author} {\bibinfo {author} {\bibfnamefont {M.}~\bibnamefont
  {Trestman}}, \bibinfo {author} {\bibfnamefont {S.}~\bibnamefont {Gugler}},
  \bibinfo {author} {\bibfnamefont {F.~A.}\ \bibnamefont {Faber}},\ and\
  \bibinfo {author} {\bibfnamefont {O.}~\bibnamefont {von Lilienfeld}},\
  }\bibfield  {title} {\bibinfo {title} {Gradient-guided furthest point
  sampling for robust training set selection},\ }\href
  {https://doi.org/10.1088/2632-2153/ae68b8} {\bibfield  {journal} {\bibinfo
  {journal} {Mach. Learn.: Sci. Technol.}\ }\textbf {\bibinfo {volume} {7}},\
  \bibinfo {pages} {035047} (\bibinfo {year} {2026})}\BibitemShut {NoStop}%
\bibitem [{\citenamefont {Qi}\ \emph {et~al.}(2024)\citenamefont {Qi},
  \citenamefont {Ko}, \citenamefont {Wood}, \citenamefont {Pham},\ and\
  \citenamefont {Ong}}]{qi2024robust}%
  \BibitemOpen
  \bibfield  {author} {\bibinfo {author} {\bibfnamefont {J.}~\bibnamefont
  {Qi}}, \bibinfo {author} {\bibfnamefont {T.~W.}\ \bibnamefont {Ko}}, \bibinfo
  {author} {\bibfnamefont {B.~C.}\ \bibnamefont {Wood}}, \bibinfo {author}
  {\bibfnamefont {T.~A.}\ \bibnamefont {Pham}},\ and\ \bibinfo {author}
  {\bibfnamefont {S.~P.}\ \bibnamefont {Ong}},\ }\bibfield  {title} {\bibinfo
  {title} {Robust training of machine learning interatomic potentials with
  dimensionality reduction and stratified sampling},\ }\href
  {https://doi.org/10.1038/s41524-024-01227-4} {\bibfield  {journal} {\bibinfo
  {journal} {npj Comput. Mater.}\ }\textbf {\bibinfo {volume} {10}},\ \bibinfo
  {pages} {43} (\bibinfo {year} {2024})}\BibitemShut {NoStop}%
\bibitem [{\citenamefont {Chen}\ and\ \citenamefont
  {Ong}(2022)}]{chen2022universal}%
  \BibitemOpen
  \bibfield  {author} {\bibinfo {author} {\bibfnamefont {C.}~\bibnamefont
  {Chen}}\ and\ \bibinfo {author} {\bibfnamefont {S.~P.}\ \bibnamefont {Ong}},\
  }\bibfield  {title} {\bibinfo {title} {A universal graph deep learning
  interatomic potential for the periodic table},\ }\href
  {https://doi.org/10.1038/s43588-022-00349-3} {\bibfield  {journal} {\bibinfo
  {journal} {Nat. Comput. Sci.}\ }\textbf {\bibinfo {volume} {2}},\ \bibinfo
  {pages} {718} (\bibinfo {year} {2022})}\BibitemShut {NoStop}%
\bibitem [{\citenamefont {Subramanyam}\ and\ \citenamefont
  {Perez}(2025)}]{subramanyam2025entropy}%
  \BibitemOpen
  \bibfield  {author} {\bibinfo {author} {\bibfnamefont {A.~P.~A.}\
  \bibnamefont {Subramanyam}}\ and\ \bibinfo {author} {\bibfnamefont
  {D.}~\bibnamefont {Perez}},\ }\bibfield  {title} {\bibinfo {title}
  {Information-entropy-driven generation of material-agnostic datasets for
  machine-learning interatomic potentials},\ }\href
  {https://doi.org/10.1038/s41524-025-01602-9} {\bibfield  {journal} {\bibinfo
  {journal} {npj Comput. Mater.}\ }\textbf {\bibinfo {volume} {11}},\ \bibinfo
  {pages} {218} (\bibinfo {year} {2025})}\BibitemShut {NoStop}%
\bibitem [{\citenamefont {Yu}\ \emph {et~al.}(2026)\citenamefont {Yu},
  \citenamefont {Lordi},\ and\ \citenamefont
  {Schwalbe-Koda}}]{yu2026maximizing}%
  \BibitemOpen
  \bibfield  {author} {\bibinfo {author} {\bibfnamefont {B.}~\bibnamefont
  {Yu}}, \bibinfo {author} {\bibfnamefont {V.}~\bibnamefont {Lordi}},\ and\
  \bibinfo {author} {\bibfnamefont {D.}~\bibnamefont {Schwalbe-Koda}},\
  }\bibfield  {title} {\bibinfo {title} {Maximizing efficiency of dataset
  compression for machine learning potentials with information theory},\ }\href
  {https://doi.org/10.1063/5.0312057} {\bibfield  {journal} {\bibinfo
  {journal} {J. Chem. Phys.}\ }\textbf {\bibinfo {volume} {164}},\ \bibinfo
  {pages} {164111} (\bibinfo {year} {2026})}\BibitemShut {NoStop}%
\bibitem [{\citenamefont {Podryabinkin}\ and\ \citenamefont
  {Shapeev}(2017)}]{podryabinkin2017active}%
  \BibitemOpen
  \bibfield  {author} {\bibinfo {author} {\bibfnamefont {E.~V.}\ \bibnamefont
  {Podryabinkin}}\ and\ \bibinfo {author} {\bibfnamefont {A.~V.}\ \bibnamefont
  {Shapeev}},\ }\bibfield  {title} {\bibinfo {title} {Active learning of
  linearly parametrized interatomic potentials},\ }\href
  {https://doi.org/10.1016/j.commatsci.2017.08.031} {\bibfield  {journal}
  {\bibinfo  {journal} {Comput. Mater. Sci.}\ }\textbf {\bibinfo {volume}
  {140}},\ \bibinfo {pages} {171} (\bibinfo {year} {2017})}\BibitemShut
  {NoStop}%
\bibitem [{\citenamefont {Kulichenko}\ \emph {et~al.}(2023)\citenamefont
  {Kulichenko}, \citenamefont {Barros}, \citenamefont {Lubbers}, \citenamefont
  {Li}, \citenamefont {Messerly}, \citenamefont {Tretiak}, \citenamefont
  {Smith},\ and\ \citenamefont {Nebgen}}]{kulichenko2023uncertainty}%
  \BibitemOpen
  \bibfield  {author} {\bibinfo {author} {\bibfnamefont {M.}~\bibnamefont
  {Kulichenko}}, \bibinfo {author} {\bibfnamefont {K.}~\bibnamefont {Barros}},
  \bibinfo {author} {\bibfnamefont {N.}~\bibnamefont {Lubbers}}, \bibinfo
  {author} {\bibfnamefont {Y.~W.}\ \bibnamefont {Li}}, \bibinfo {author}
  {\bibfnamefont {R.}~\bibnamefont {Messerly}}, \bibinfo {author}
  {\bibfnamefont {S.}~\bibnamefont {Tretiak}}, \bibinfo {author} {\bibfnamefont
  {J.~S.}\ \bibnamefont {Smith}},\ and\ \bibinfo {author} {\bibfnamefont
  {B.}~\bibnamefont {Nebgen}},\ }\bibfield  {title} {\bibinfo {title}
  {Uncertainty-driven dynamics for active learning of interatomic potentials},\
  }\href {https://doi.org/10.1038/s43588-023-00406-5} {\bibfield  {journal}
  {\bibinfo  {journal} {Nat. Comput. Sci.}\ }\textbf {\bibinfo {volume} {3}},\
  \bibinfo {pages} {230} (\bibinfo {year} {2023})}\BibitemShut {NoStop}%
\bibitem [{\citenamefont {Kuner}\ \emph {et~al.}(2025)\citenamefont {Kuner},
  \citenamefont {Kaplan}, \citenamefont {Persson}, \citenamefont {Asta},\ and\
  \citenamefont {Chrzan}}]{kuner2025mpaloe}%
  \BibitemOpen
  \bibfield  {author} {\bibinfo {author} {\bibfnamefont {M.~C.}\ \bibnamefont
  {Kuner}}, \bibinfo {author} {\bibfnamefont {A.~D.}\ \bibnamefont {Kaplan}},
  \bibinfo {author} {\bibfnamefont {K.~A.}\ \bibnamefont {Persson}}, \bibinfo
  {author} {\bibfnamefont {M.}~\bibnamefont {Asta}},\ and\ \bibinfo {author}
  {\bibfnamefont {D.~C.}\ \bibnamefont {Chrzan}},\ }\bibfield  {title}
  {\bibinfo {title} {{MP-ALOE}: an {r2SCAN} dataset for universal machine
  learning interatomic potentials},\ }\href
  {https://doi.org/10.1038/s41524-025-01834-9} {\bibfield  {journal} {\bibinfo
  {journal} {npj Comput. Mater.}\ }\textbf {\bibinfo {volume} {11}},\ \bibinfo
  {pages} {352} (\bibinfo {year} {2025})}\BibitemShut {NoStop}%
\bibitem [{\citenamefont {Bochkarev}\ \emph {et~al.}(2026)\citenamefont
  {Bochkarev}, \citenamefont {Lysogorskiy}, \citenamefont {Subramanyam},
  \citenamefont {Drautz},\ and\ \citenamefont
  {Perez}}]{bochkarev2026exploring}%
  \BibitemOpen
  \bibfield  {author} {\bibinfo {author} {\bibfnamefont {A.}~\bibnamefont
  {Bochkarev}}, \bibinfo {author} {\bibfnamefont {Y.}~\bibnamefont
  {Lysogorskiy}}, \bibinfo {author} {\bibfnamefont {A.}~\bibnamefont
  {Subramanyam}}, \bibinfo {author} {\bibfnamefont {R.}~\bibnamefont
  {Drautz}},\ and\ \bibinfo {author} {\bibfnamefont {D.}~\bibnamefont
  {Perez}},\ }\bibfield  {title} {\bibinfo {title} {Exploring the extremes:
  atomic basis for multi-elemental materials science under complex
  thermodynamic conditions},\ }\href {https://arxiv.org/abs/2602.23489}
  {\bibfield  {journal} {\bibinfo  {journal} {arXiv preprint arXiv:2602.23489}\
  } (\bibinfo {year} {2026})}\BibitemShut {NoStop}%
\bibitem [{\citenamefont {Cohen}\ \emph {et~al.}(2008)\citenamefont {Cohen},
  \citenamefont {Mori-S{\'a}nchez},\ and\ \citenamefont
  {Yang}}]{cohen2008insights}%
  \BibitemOpen
  \bibfield  {author} {\bibinfo {author} {\bibfnamefont {A.~J.}\ \bibnamefont
  {Cohen}}, \bibinfo {author} {\bibfnamefont {P.}~\bibnamefont
  {Mori-S{\'a}nchez}},\ and\ \bibinfo {author} {\bibfnamefont {W.}~\bibnamefont
  {Yang}},\ }\bibfield  {title} {\bibinfo {title} {Insights into current
  limitations of density functional theory},\ }\href
  {https://doi.org/10.1126/science.1158722} {\bibfield  {journal} {\bibinfo
  {journal} {Science}\ }\textbf {\bibinfo {volume} {321}},\ \bibinfo {pages}
  {792} (\bibinfo {year} {2008})}\BibitemShut {NoStop}%
\bibitem [{\citenamefont {Caucci}\ \emph {et~al.}(2025)\citenamefont {Caucci},
  \citenamefont {Sivak}, \citenamefont {Almishal}, \citenamefont {Rost},
  \citenamefont {Dabo}, \citenamefont {Maria},\ and\ \citenamefont
  {Sinnott}}]{caucci2025performance}%
  \BibitemOpen
  \bibfield  {author} {\bibinfo {author} {\bibfnamefont {M.~K.}\ \bibnamefont
  {Caucci}}, \bibinfo {author} {\bibfnamefont {J.~T.}\ \bibnamefont {Sivak}},
  \bibinfo {author} {\bibfnamefont {S.~S.}\ \bibnamefont {Almishal}}, \bibinfo
  {author} {\bibfnamefont {C.~M.}\ \bibnamefont {Rost}}, \bibinfo {author}
  {\bibfnamefont {I.}~\bibnamefont {Dabo}}, \bibinfo {author} {\bibfnamefont
  {J.-P.}\ \bibnamefont {Maria}},\ and\ \bibinfo {author} {\bibfnamefont
  {S.~B.}\ \bibnamefont {Sinnott}},\ }\bibfield  {title} {\bibinfo {title}
  {Performance of exchange-correlation approximations to density functional
  theory for rare-earth oxides},\ }\href
  {https://doi.org/10.1016/j.commatsci.2025.113837} {\bibfield  {journal}
  {\bibinfo  {journal} {Comput. Mater. Sci.}\ }\textbf {\bibinfo {volume}
  {253}},\ \bibinfo {pages} {113837} (\bibinfo {year} {2025})}\BibitemShut
  {NoStop}%
\bibitem [{\citenamefont {Dudarev}\ \emph {et~al.}(1998)\citenamefont
  {Dudarev}, \citenamefont {Botton}, \citenamefont {Savrasov}, \citenamefont
  {Humphreys},\ and\ \citenamefont {Sutton}}]{dudarev1998electron}%
  \BibitemOpen
  \bibfield  {author} {\bibinfo {author} {\bibfnamefont {S.~L.}\ \bibnamefont
  {Dudarev}}, \bibinfo {author} {\bibfnamefont {G.~A.}\ \bibnamefont {Botton}},
  \bibinfo {author} {\bibfnamefont {S.~Y.}\ \bibnamefont {Savrasov}}, \bibinfo
  {author} {\bibfnamefont {C.~J.}\ \bibnamefont {Humphreys}},\ and\ \bibinfo
  {author} {\bibfnamefont {A.~P.}\ \bibnamefont {Sutton}},\ }\bibfield  {title}
  {\bibinfo {title} {Electron-energy-loss spectra and the structural stability
  of nickel oxide: An lsda+u study},\ }\href
  {https://doi.org/10.1103/physrevb.57.1505} {\bibfield  {journal} {\bibinfo
  {journal} {Phys. Rev. B}\ }\textbf {\bibinfo {volume} {57}},\ \bibinfo
  {pages} {1505} (\bibinfo {year} {1998})}\BibitemShut {NoStop}%
\bibitem [{\citenamefont {Wang}\ \emph {et~al.}(2006)\citenamefont {Wang},
  \citenamefont {Maxisch},\ and\ \citenamefont {Ceder}}]{wang2006oxidation}%
  \BibitemOpen
  \bibfield  {author} {\bibinfo {author} {\bibfnamefont {L.}~\bibnamefont
  {Wang}}, \bibinfo {author} {\bibfnamefont {T.}~\bibnamefont {Maxisch}},\ and\
  \bibinfo {author} {\bibfnamefont {G.}~\bibnamefont {Ceder}},\ }\bibfield
  {title} {\bibinfo {title} {Oxidation energies of transition metal oxides
  within the gga+u framework},\ }\href
  {https://doi.org/10.1103/physrevb.73.195107} {\bibfield  {journal} {\bibinfo
  {journal} {Phys. Rev. B - Condens. Matter Mater. Phys.}\ }\textbf {\bibinfo
  {volume} {73}},\ \bibinfo {pages} {195107} (\bibinfo {year}
  {2006})}\BibitemShut {NoStop}%
\bibitem [{\citenamefont {Jain}\ \emph {et~al.}(2011)\citenamefont {Jain},
  \citenamefont {Hautier}, \citenamefont {Ong}, \citenamefont {Moore},
  \citenamefont {Fischer}, \citenamefont {Persson},\ and\ \citenamefont
  {Ceder}}]{jain2011formation}%
  \BibitemOpen
  \bibfield  {author} {\bibinfo {author} {\bibfnamefont {A.}~\bibnamefont
  {Jain}}, \bibinfo {author} {\bibfnamefont {G.}~\bibnamefont {Hautier}},
  \bibinfo {author} {\bibfnamefont {S.~P.}\ \bibnamefont {Ong}}, \bibinfo
  {author} {\bibfnamefont {C.~J.}\ \bibnamefont {Moore}}, \bibinfo {author}
  {\bibfnamefont {C.~C.}\ \bibnamefont {Fischer}}, \bibinfo {author}
  {\bibfnamefont {K.~A.}\ \bibnamefont {Persson}},\ and\ \bibinfo {author}
  {\bibfnamefont {G.}~\bibnamefont {Ceder}},\ }\bibfield  {title} {\bibinfo
  {title} {Formation enthalpies by mixing gga and gga+u calculations},\ }\href
  {https://doi.org/10.1103/physrevb.84.045115} {\bibfield  {journal} {\bibinfo
  {journal} {Phys. Rev. B - Condens. Matter Mater. Phys.}\ }\textbf {\bibinfo
  {volume} {84}},\ \bibinfo {pages} {045115} (\bibinfo {year}
  {2011})}\BibitemShut {NoStop}%
\bibitem [{\citenamefont {Waroquiers}\ \emph {et~al.}(2020)\citenamefont
  {Waroquiers}, \citenamefont {George}, \citenamefont {Horton}, \citenamefont
  {Schenk}, \citenamefont {Persson}, \citenamefont {Rignanese}, \citenamefont
  {Gonze},\ and\ \citenamefont {Hautier}}]{waroquiers2020chemenv}%
  \BibitemOpen
  \bibfield  {author} {\bibinfo {author} {\bibfnamefont {D.}~\bibnamefont
  {Waroquiers}}, \bibinfo {author} {\bibfnamefont {J.}~\bibnamefont {George}},
  \bibinfo {author} {\bibfnamefont {M.}~\bibnamefont {Horton}}, \bibinfo
  {author} {\bibfnamefont {S.}~\bibnamefont {Schenk}}, \bibinfo {author}
  {\bibfnamefont {K.~A.}\ \bibnamefont {Persson}}, \bibinfo {author}
  {\bibfnamefont {G.-M.}\ \bibnamefont {Rignanese}}, \bibinfo {author}
  {\bibfnamefont {X.}~\bibnamefont {Gonze}},\ and\ \bibinfo {author}
  {\bibfnamefont {G.}~\bibnamefont {Hautier}},\ }\bibfield  {title} {\bibinfo
  {title} {Chemenv: a fast and robust coordination environment identification
  tool},\ }\href {https://doi.org/10.1107/s2052520620007994} {\bibfield
  {journal} {\bibinfo  {journal} {Acta Crystallogr. B: Struct. Sci. Cryst. Eng.
  Mater.}\ }\textbf {\bibinfo {volume} {76}},\ \bibinfo {pages} {683} (\bibinfo
  {year} {2020})}\BibitemShut {NoStop}%
\bibitem [{\citenamefont {O'Keefe}\ and\ \citenamefont
  {Brese}(1991)}]{o1991atom}%
  \BibitemOpen
  \bibfield  {author} {\bibinfo {author} {\bibfnamefont {M.}~\bibnamefont
  {O'Keefe}}\ and\ \bibinfo {author} {\bibfnamefont {N.}~\bibnamefont
  {Brese}},\ }\bibfield  {title} {\bibinfo {title} {Atom sizes and bond lengths
  in molecules and crystals},\ }\href {https://doi.org/10.1021/ja00009a002}
  {\bibfield  {journal} {\bibinfo  {journal} {J. Am. Chem. Soc.}\ }\textbf
  {\bibinfo {volume} {113}},\ \bibinfo {pages} {3226} (\bibinfo {year}
  {1991})}\BibitemShut {NoStop}%
\bibitem [{\citenamefont {Allen}\ and\ \citenamefont
  {Tkatchenko}(2022)}]{allen2022machine}%
  \BibitemOpen
  \bibfield  {author} {\bibinfo {author} {\bibfnamefont {A.~E.}\ \bibnamefont
  {Allen}}\ and\ \bibinfo {author} {\bibfnamefont {A.}~\bibnamefont
  {Tkatchenko}},\ }\bibfield  {title} {\bibinfo {title} {Machine learning of
  material properties: Predictive and interpretable multilinear models},\
  }\href@noop {} {\bibfield  {journal} {\bibinfo  {journal} {Sci. Adv.}\
  }\textbf {\bibinfo {volume} {8}},\ \bibinfo {pages} {eabm7185} (\bibinfo
  {year} {2022})}\BibitemShut {NoStop}%
\bibitem [{\citenamefont {Batatia}\ \emph {et~al.}(2022)\citenamefont
  {Batatia}, \citenamefont {Kovacs}, \citenamefont {Simm}, \citenamefont
  {Ortner},\ and\ \citenamefont {Cs{\'a}nyi}}]{batatia2022mace}%
  \BibitemOpen
  \bibfield  {author} {\bibinfo {author} {\bibfnamefont {I.}~\bibnamefont
  {Batatia}}, \bibinfo {author} {\bibfnamefont {D.~P.}\ \bibnamefont {Kovacs}},
  \bibinfo {author} {\bibfnamefont {G.}~\bibnamefont {Simm}}, \bibinfo {author}
  {\bibfnamefont {C.}~\bibnamefont {Ortner}},\ and\ \bibinfo {author}
  {\bibfnamefont {G.}~\bibnamefont {Cs{\'a}nyi}},\ }\bibfield  {title}
  {\bibinfo {title} {Mace: Higher order equivariant message passing neural
  networks for fast and accurate force fields},\ }\href
  {https://doi.org/10.52202/068431-0830} {\bibfield  {journal} {\bibinfo
  {journal} {Adv. Neural Inf. Process. Syst.}\ }\textbf {\bibinfo {volume}
  {35}},\ \bibinfo {pages} {11423} (\bibinfo {year} {2022})}\BibitemShut
  {NoStop}%
\bibitem [{\citenamefont {Deng}\ \emph {et~al.}(2025)\citenamefont {Deng},
  \citenamefont {Choi}, \citenamefont {Zhong}, \citenamefont {Riebesell},
  \citenamefont {Anand}, \citenamefont {Li}, \citenamefont {Jun}, \citenamefont
  {Persson},\ and\ \citenamefont {Ceder}}]{deng2025systematic}%
  \BibitemOpen
  \bibfield  {author} {\bibinfo {author} {\bibfnamefont {B.}~\bibnamefont
  {Deng}}, \bibinfo {author} {\bibfnamefont {Y.}~\bibnamefont {Choi}}, \bibinfo
  {author} {\bibfnamefont {P.}~\bibnamefont {Zhong}}, \bibinfo {author}
  {\bibfnamefont {J.}~\bibnamefont {Riebesell}}, \bibinfo {author}
  {\bibfnamefont {S.}~\bibnamefont {Anand}}, \bibinfo {author} {\bibfnamefont
  {Z.}~\bibnamefont {Li}}, \bibinfo {author} {\bibfnamefont {K.}~\bibnamefont
  {Jun}}, \bibinfo {author} {\bibfnamefont {K.~A.}\ \bibnamefont {Persson}},\
  and\ \bibinfo {author} {\bibfnamefont {G.}~\bibnamefont {Ceder}},\ }\bibfield
   {title} {\bibinfo {title} {Systematic softening in universal machine
  learning interatomic potentials},\ }\href
  {https://doi.org/10.1038/s41524-024-01500-6} {\bibfield  {journal} {\bibinfo
  {journal} {npj Comput. Mater.}\ }\textbf {\bibinfo {volume} {11}},\ \bibinfo
  {pages} {9} (\bibinfo {year} {2025})}\BibitemShut {NoStop}%
\bibitem [{\citenamefont {Zhou}\ \emph {et~al.}(2026)\citenamefont {Zhou},
  \citenamefont {Hu}, \citenamefont {Zhang}, \citenamefont {Wang},
  \citenamefont {Tan},\ and\ \citenamefont {Jia}}]{zhou2026matris}%
  \BibitemOpen
  \bibfield  {author} {\bibinfo {author} {\bibfnamefont {Y.}~\bibnamefont
  {Zhou}}, \bibinfo {author} {\bibfnamefont {S.}~\bibnamefont {Hu}}, \bibinfo
  {author} {\bibfnamefont {X.}~\bibnamefont {Zhang}}, \bibinfo {author}
  {\bibfnamefont {H.}~\bibnamefont {Wang}}, \bibinfo {author} {\bibfnamefont
  {G.}~\bibnamefont {Tan}},\ and\ \bibinfo {author} {\bibfnamefont
  {W.}~\bibnamefont {Jia}},\ }\bibfield  {title} {\bibinfo {title} {Matris:
  Toward reliable and efficient pretrained machine learning interatomic
  potentials},\ }\href {https://arxiv.org/abs/2603.02002} {\bibfield  {journal}
  {\bibinfo  {journal} {arXiv preprint arXiv:2603.02002}\ } (\bibinfo {year}
  {2026})}\BibitemShut {NoStop}%
\bibitem [{\citenamefont {Loew}\ \emph {et~al.}(2025)\citenamefont {Loew},
  \citenamefont {Sun}, \citenamefont {Wang}, \citenamefont {Botti},\ and\
  \citenamefont {Marques}}]{loew2025universal}%
  \BibitemOpen
  \bibfield  {author} {\bibinfo {author} {\bibfnamefont {A.}~\bibnamefont
  {Loew}}, \bibinfo {author} {\bibfnamefont {D.}~\bibnamefont {Sun}}, \bibinfo
  {author} {\bibfnamefont {H.-C.}\ \bibnamefont {Wang}}, \bibinfo {author}
  {\bibfnamefont {S.}~\bibnamefont {Botti}},\ and\ \bibinfo {author}
  {\bibfnamefont {M.~A.}\ \bibnamefont {Marques}},\ }\bibfield  {title}
  {\bibinfo {title} {Universal machine learning interatomic potentials are
  ready for phonons},\ }\href {https://doi.org/10.1038/s41524-025-01650-1}
  {\bibfield  {journal} {\bibinfo  {journal} {npj Comput. Mater.}\ }\textbf
  {\bibinfo {volume} {11}},\ \bibinfo {pages} {178} (\bibinfo {year}
  {2025})}\BibitemShut {NoStop}%
\bibitem [{\citenamefont {Fu}\ \emph {et~al.}(2022)\citenamefont {Fu},
  \citenamefont {Wu}, \citenamefont {Wang}, \citenamefont {Xie}, \citenamefont
  {Keten}, \citenamefont {Gomez-Bombarelli},\ and\ \citenamefont
  {Jaakkola}}]{fu2022forces}%
  \BibitemOpen
  \bibfield  {author} {\bibinfo {author} {\bibfnamefont {X.}~\bibnamefont
  {Fu}}, \bibinfo {author} {\bibfnamefont {Z.}~\bibnamefont {Wu}}, \bibinfo
  {author} {\bibfnamefont {W.}~\bibnamefont {Wang}}, \bibinfo {author}
  {\bibfnamefont {T.}~\bibnamefont {Xie}}, \bibinfo {author} {\bibfnamefont
  {S.}~\bibnamefont {Keten}}, \bibinfo {author} {\bibfnamefont
  {R.}~\bibnamefont {Gomez-Bombarelli}},\ and\ \bibinfo {author} {\bibfnamefont
  {T.}~\bibnamefont {Jaakkola}},\ }\bibfield  {title} {\bibinfo {title} {Forces
  are not enough: Benchmark and critical evaluation for machine learning force
  fields with molecular simulations},\ }\href
  {https://arxiv.org/abs/2210.07237} {\bibfield  {journal} {\bibinfo  {journal}
  {arXiv preprint arXiv:2210.07237}\ } (\bibinfo {year} {2022})}\BibitemShut
  {NoStop}%
\bibitem [{\citenamefont {Poltavsky}\ \emph
  {et~al.}(2025{\natexlab{a}})\citenamefont {Poltavsky}, \citenamefont
  {Charkin-Gorbulin}, \citenamefont {Puleva}, \citenamefont {Fonseca},
  \citenamefont {Batatia}, \citenamefont {Browning}, \citenamefont {Chmiela},
  \citenamefont {Cui}, \citenamefont {Frank}, \citenamefont {Heinen} \emph
  {et~al.}}]{poltavsky2025crash1}%
  \BibitemOpen
  \bibfield  {author} {\bibinfo {author} {\bibfnamefont {I.}~\bibnamefont
  {Poltavsky}}, \bibinfo {author} {\bibfnamefont {A.}~\bibnamefont
  {Charkin-Gorbulin}}, \bibinfo {author} {\bibfnamefont {M.}~\bibnamefont
  {Puleva}}, \bibinfo {author} {\bibfnamefont {G.}~\bibnamefont {Fonseca}},
  \bibinfo {author} {\bibfnamefont {I.}~\bibnamefont {Batatia}}, \bibinfo
  {author} {\bibfnamefont {N.~J.}\ \bibnamefont {Browning}}, \bibinfo {author}
  {\bibfnamefont {S.}~\bibnamefont {Chmiela}}, \bibinfo {author} {\bibfnamefont
  {M.}~\bibnamefont {Cui}}, \bibinfo {author} {\bibfnamefont {J.~T.}\
  \bibnamefont {Frank}}, \bibinfo {author} {\bibfnamefont {S.}~\bibnamefont
  {Heinen}}, \emph {et~al.},\ }\bibfield  {title} {\bibinfo {title} {Crash
  testing machine learning force fields for molecules, materials, and
  interfaces: model analysis in the tea challenge 2023},\ }\href
  {https://doi.org/10.1039/D4SC06529H} {\bibfield  {journal} {\bibinfo
  {journal} {Chem. Sci.}\ }\textbf {\bibinfo {volume} {16}},\ \bibinfo {pages}
  {3720} (\bibinfo {year} {2025}{\natexlab{a}})}\BibitemShut {NoStop}%
\bibitem [{\citenamefont {Poltavsky}\ \emph
  {et~al.}(2025{\natexlab{b}})\citenamefont {Poltavsky}, \citenamefont
  {Puleva}, \citenamefont {Charkin-Gorbulin}, \citenamefont {Fonseca},
  \citenamefont {Batatia}, \citenamefont {Browning}, \citenamefont {Chmiela},
  \citenamefont {Cui}, \citenamefont {Frank}, \citenamefont {Heinen} \emph
  {et~al.}}]{poltavsky2025crash2}%
  \BibitemOpen
  \bibfield  {author} {\bibinfo {author} {\bibfnamefont {I.}~\bibnamefont
  {Poltavsky}}, \bibinfo {author} {\bibfnamefont {M.}~\bibnamefont {Puleva}},
  \bibinfo {author} {\bibfnamefont {A.}~\bibnamefont {Charkin-Gorbulin}},
  \bibinfo {author} {\bibfnamefont {G.}~\bibnamefont {Fonseca}}, \bibinfo
  {author} {\bibfnamefont {I.}~\bibnamefont {Batatia}}, \bibinfo {author}
  {\bibfnamefont {N.~J.}\ \bibnamefont {Browning}}, \bibinfo {author}
  {\bibfnamefont {S.}~\bibnamefont {Chmiela}}, \bibinfo {author} {\bibfnamefont
  {M.}~\bibnamefont {Cui}}, \bibinfo {author} {\bibfnamefont {J.~T.}\
  \bibnamefont {Frank}}, \bibinfo {author} {\bibfnamefont {S.}~\bibnamefont
  {Heinen}}, \emph {et~al.},\ }\bibfield  {title} {\bibinfo {title} {Crash
  testing machine learning force fields for molecules, materials, and
  interfaces: molecular dynamics in the tea challenge 2023},\ }\href
  {https://doi.org/10.1039/D4SC06530A} {\bibfield  {journal} {\bibinfo
  {journal} {Chem. Sci.}\ }\textbf {\bibinfo {volume} {16}},\ \bibinfo {pages}
  {3738} (\bibinfo {year} {2025}{\natexlab{b}})}\BibitemShut {NoStop}%
\bibitem [{\citenamefont {Zhang}\ \emph {et~al.}(2020)\citenamefont {Zhang},
  \citenamefont {Lunghi},\ and\ \citenamefont {Sanvito}}]{zhang2020pushing}%
  \BibitemOpen
  \bibfield  {author} {\bibinfo {author} {\bibfnamefont {Y.}~\bibnamefont
  {Zhang}}, \bibinfo {author} {\bibfnamefont {A.}~\bibnamefont {Lunghi}},\ and\
  \bibinfo {author} {\bibfnamefont {S.}~\bibnamefont {Sanvito}},\ }\bibfield
  {title} {\bibinfo {title} {Pushing the limits of atomistic simulations
  towards ultra-high temperature: A machine-learning force field for zrb2},\
  }\href {https://doi.org/10.1016/j.actamat.2019.12.060} {\bibfield  {journal}
  {\bibinfo  {journal} {Acta Materialia}\ }\textbf {\bibinfo {volume} {186}},\
  \bibinfo {pages} {467} (\bibinfo {year} {2020})}\BibitemShut {NoStop}%
\bibitem [{\citenamefont {Stocker}\ \emph {et~al.}(2022)\citenamefont
  {Stocker}, \citenamefont {Gasteiger}, \citenamefont {Becker}, \citenamefont
  {G{\"u}nnemann},\ and\ \citenamefont {Margraf}}]{stocker2022robust}%
  \BibitemOpen
  \bibfield  {author} {\bibinfo {author} {\bibfnamefont {S.}~\bibnamefont
  {Stocker}}, \bibinfo {author} {\bibfnamefont {J.}~\bibnamefont {Gasteiger}},
  \bibinfo {author} {\bibfnamefont {F.}~\bibnamefont {Becker}}, \bibinfo
  {author} {\bibfnamefont {S.}~\bibnamefont {G{\"u}nnemann}},\ and\ \bibinfo
  {author} {\bibfnamefont {J.~T.}\ \bibnamefont {Margraf}},\ }\bibfield
  {title} {\bibinfo {title} {How robust are modern graph neural network
  potentials in long and hot molecular dynamics simulations?},\ }\href
  {https://doi.org/10.1088/2632-2153/ac9955} {\bibfield  {journal} {\bibinfo
  {journal} {Mach. Learn.: Sci. Technol.}\ }\textbf {\bibinfo {volume} {3}},\
  \bibinfo {pages} {045010} (\bibinfo {year} {2022})}\BibitemShut {NoStop}%
\bibitem [{\citenamefont {Bischoff}\ \emph {et~al.}(2024)\citenamefont
  {Bischoff}, \citenamefont {J{\"a}ckl},\ and\ \citenamefont
  {Rupp}}]{bischoff2024hydrogen}%
  \BibitemOpen
  \bibfield  {author} {\bibinfo {author} {\bibfnamefont {T.}~\bibnamefont
  {Bischoff}}, \bibinfo {author} {\bibfnamefont {B.}~\bibnamefont
  {J{\"a}ckl}},\ and\ \bibinfo {author} {\bibfnamefont {M.}~\bibnamefont
  {Rupp}},\ }\bibfield  {title} {\bibinfo {title} {Hydrogen under pressure as a
  benchmark for machine-learning interatomic potentials},\ }\href
  {https://arxiv.org/abs/2409.13390} {\bibfield  {journal} {\bibinfo  {journal}
  {arXiv preprint arXiv:2409.13390}\ } (\bibinfo {year} {2024})}\BibitemShut
  {NoStop}%
\bibitem [{\citenamefont {Loew}\ \emph {et~al.}(2026)\citenamefont {Loew},
  \citenamefont {Schmidt}, \citenamefont {Botti},\ and\ \citenamefont
  {Marques}}]{loew2026universal}%
  \BibitemOpen
  \bibfield  {author} {\bibinfo {author} {\bibfnamefont {A.}~\bibnamefont
  {Loew}}, \bibinfo {author} {\bibfnamefont {J.}~\bibnamefont {Schmidt}},
  \bibinfo {author} {\bibfnamefont {S.}~\bibnamefont {Botti}},\ and\ \bibinfo
  {author} {\bibfnamefont {M.~A.}\ \bibnamefont {Marques}},\ }\bibfield
  {title} {\bibinfo {title} {Universal machine learning potentials under
  pressure},\ }\href {https://doi.org/10.1088/2515-7639/ae2ba8} {\bibfield
  {journal} {\bibinfo  {journal} {JPhys Mater.}\ }\textbf {\bibinfo {volume}
  {9}},\ \bibinfo {pages} {015010} (\bibinfo {year} {2026})}\BibitemShut
  {NoStop}%
\bibitem [{\citenamefont {Sema}\ \emph {et~al.}(2026)\citenamefont {Sema},
  \citenamefont {Nguyen}, \citenamefont {Wyant},\ and\ \citenamefont
  {Hadjiconstantinou}}]{sema2026environment}%
  \BibitemOpen
  \bibfield  {author} {\bibinfo {author} {\bibfnamefont {D.}~\bibnamefont
  {Sema}}, \bibinfo {author} {\bibfnamefont {N.~C.}\ \bibnamefont {Nguyen}},
  \bibinfo {author} {\bibfnamefont {S.}~\bibnamefont {Wyant}},\ and\ \bibinfo
  {author} {\bibfnamefont {N.}~\bibnamefont {Hadjiconstantinou}},\ }\bibfield
  {title} {\bibinfo {title} {Environment-adaptive machine-learned force fields
  for materials under extreme conditions: hafnium and hafnium dioxide
  polymorphs},\ }\href {https://doi.org/10.1038/s41524-026-01984-4} {\bibfield
  {journal} {\bibinfo  {journal} {npj Comput. Mater.}\ }\textbf {\bibinfo
  {volume} {12}},\ \bibinfo {pages} {117} (\bibinfo {year} {2026})}\BibitemShut
  {NoStop}%
\bibitem [{\citenamefont {Chiang}\ \emph {et~al.}(2025)\citenamefont {Chiang},
  \citenamefont {Kreiman}, \citenamefont {Zhang}, \citenamefont {Kuner},
  \citenamefont {Weaver}, \citenamefont {Amin}, \citenamefont {Park},
  \citenamefont {Lim}, \citenamefont {Kim}, \citenamefont {Chrzan} \emph
  {et~al.}}]{chiang2026mlip}%
  \BibitemOpen
  \bibfield  {author} {\bibinfo {author} {\bibfnamefont {Y.}~\bibnamefont
  {Chiang}}, \bibinfo {author} {\bibfnamefont {T.}~\bibnamefont {Kreiman}},
  \bibinfo {author} {\bibfnamefont {C.}~\bibnamefont {Zhang}}, \bibinfo
  {author} {\bibfnamefont {M.}~\bibnamefont {Kuner}}, \bibinfo {author}
  {\bibfnamefont {E.}~\bibnamefont {Weaver}}, \bibinfo {author} {\bibfnamefont
  {I.}~\bibnamefont {Amin}}, \bibinfo {author} {\bibfnamefont {H.}~\bibnamefont
  {Park}}, \bibinfo {author} {\bibfnamefont {Y.}~\bibnamefont {Lim}}, \bibinfo
  {author} {\bibfnamefont {J.}~\bibnamefont {Kim}}, \bibinfo {author}
  {\bibfnamefont {D.}~\bibnamefont {Chrzan}}, \emph {et~al.},\ }\bibfield
  {title} {\bibinfo {title} {Mlip arena: advancing fairness and transparency in
  machine learning interatomic potentials via an open, accessible benchmark
  platform},\ }\href {https://doi.org/10.52202/085713-4410} {\bibfield
  {journal} {\bibinfo  {journal} {Adv. Neural Inf. Process. Syst.}\ }\textbf
  {\bibinfo {volume} {38}},\ \bibinfo {pages} {146756} (\bibinfo {year}
  {2025})}\BibitemShut {NoStop}%
\bibitem [{\citenamefont {Dembitskiy}\ \emph {et~al.}(2025)\citenamefont
  {Dembitskiy}, \citenamefont {Humonen}, \citenamefont {Eremin}, \citenamefont
  {Aksyonov}, \citenamefont {Fedotov},\ and\ \citenamefont
  {Budennyy}}]{dembitskiy2025benchmarking}%
  \BibitemOpen
  \bibfield  {author} {\bibinfo {author} {\bibfnamefont {A.~D.}\ \bibnamefont
  {Dembitskiy}}, \bibinfo {author} {\bibfnamefont {I.~S.}\ \bibnamefont
  {Humonen}}, \bibinfo {author} {\bibfnamefont {R.~A.}\ \bibnamefont {Eremin}},
  \bibinfo {author} {\bibfnamefont {D.~A.}\ \bibnamefont {Aksyonov}}, \bibinfo
  {author} {\bibfnamefont {S.~S.}\ \bibnamefont {Fedotov}},\ and\ \bibinfo
  {author} {\bibfnamefont {S.~A.}\ \bibnamefont {Budennyy}},\ }\bibfield
  {title} {\bibinfo {title} {Benchmarking machine learning models for
  predicting lithium ion migration},\ }\href
  {https://doi.org/10.1038/s41524-025-01571-z} {\bibfield  {journal} {\bibinfo
  {journal} {npj Comput. Mater.}\ }\textbf {\bibinfo {volume} {11}},\ \bibinfo
  {pages} {131} (\bibinfo {year} {2025})}\BibitemShut {NoStop}%
\bibitem [{\citenamefont {Cubuk}\ \emph {et~al.}(2017)\citenamefont {Cubuk},
  \citenamefont {Malone}, \citenamefont {Onat}, \citenamefont {Waterland},\
  and\ \citenamefont {Kaxiras}}]{cubuk2017representations}%
  \BibitemOpen
  \bibfield  {author} {\bibinfo {author} {\bibfnamefont {E.~D.}\ \bibnamefont
  {Cubuk}}, \bibinfo {author} {\bibfnamefont {B.~D.}\ \bibnamefont {Malone}},
  \bibinfo {author} {\bibfnamefont {B.}~\bibnamefont {Onat}}, \bibinfo {author}
  {\bibfnamefont {A.}~\bibnamefont {Waterland}},\ and\ \bibinfo {author}
  {\bibfnamefont {E.}~\bibnamefont {Kaxiras}},\ }\bibfield  {title} {\bibinfo
  {title} {Representations in neural network based empirical potentials},\
  }\href {https://doi.org/10.1063/1.4990503} {\bibfield  {journal} {\bibinfo
  {journal} {J. Chem. Phys.}\ }\textbf {\bibinfo {volume} {147}},\ \bibinfo
  {pages} {024104} (\bibinfo {year} {2017})}\BibitemShut {NoStop}%
\bibitem [{\citenamefont {Smith}\ \emph {et~al.}(2021)\citenamefont {Smith},
  \citenamefont {Nebgen}, \citenamefont {Mathew}, \citenamefont {Chen},
  \citenamefont {Lubbers}, \citenamefont {Burakovsky}, \citenamefont {Tretiak},
  \citenamefont {Nam}, \citenamefont {Germann}, \citenamefont {Fensin} \emph
  {et~al.}}]{smith2021automated}%
  \BibitemOpen
  \bibfield  {author} {\bibinfo {author} {\bibfnamefont {J.~S.}\ \bibnamefont
  {Smith}}, \bibinfo {author} {\bibfnamefont {B.}~\bibnamefont {Nebgen}},
  \bibinfo {author} {\bibfnamefont {N.}~\bibnamefont {Mathew}}, \bibinfo
  {author} {\bibfnamefont {J.}~\bibnamefont {Chen}}, \bibinfo {author}
  {\bibfnamefont {N.}~\bibnamefont {Lubbers}}, \bibinfo {author} {\bibfnamefont
  {L.}~\bibnamefont {Burakovsky}}, \bibinfo {author} {\bibfnamefont
  {S.}~\bibnamefont {Tretiak}}, \bibinfo {author} {\bibfnamefont {H.~A.}\
  \bibnamefont {Nam}}, \bibinfo {author} {\bibfnamefont {T.}~\bibnamefont
  {Germann}}, \bibinfo {author} {\bibfnamefont {S.}~\bibnamefont {Fensin}},
  \emph {et~al.},\ }\bibfield  {title} {\bibinfo {title} {Automated discovery
  of a robust interatomic potential for aluminum},\ }\href
  {https://doi.org/10.1038/s41467-021-21376-0} {\bibfield  {journal} {\bibinfo
  {journal} {Nat. Commun.}\ }\textbf {\bibinfo {volume} {12}},\ \bibinfo
  {pages} {1257} (\bibinfo {year} {2021})}\BibitemShut {NoStop}%
\bibitem [{\citenamefont {Li}\ and\ \citenamefont
  {Walsh}(2026)}]{li2026platonic}%
  \BibitemOpen
  \bibfield  {author} {\bibinfo {author} {\bibfnamefont {Z.}~\bibnamefont
  {Li}}\ and\ \bibinfo {author} {\bibfnamefont {A.}~\bibnamefont {Walsh}},\
  }\bibfield  {title} {\bibinfo {title} {Platonic representation of foundation
  machine learning interatomic potentials},\ }\href
  {https://doi.org/10.1038/s42256-026-01235-7} {\bibfield  {journal} {\bibinfo
  {journal} {Nat. Mach. Intell.}\ }\textbf {\bibinfo {volume} {8}},\ \bibinfo
  {pages} {830} (\bibinfo {year} {2026})}\BibitemShut {NoStop}%
\bibitem [{\citenamefont {Levine}\ \emph {et~al.}(2025)\citenamefont {Levine},
  \citenamefont {Shuaibi}, \citenamefont {Spotte-Smith}, \citenamefont
  {Taylor}, \citenamefont {Hasyim}, \citenamefont {Michel}, \citenamefont
  {Batatia}, \citenamefont {Cs{\'a}nyi}, \citenamefont {Dzamba}, \citenamefont
  {Eastman} \emph {et~al.}}]{levine2025open}%
  \BibitemOpen
  \bibfield  {author} {\bibinfo {author} {\bibfnamefont {D.~S.}\ \bibnamefont
  {Levine}}, \bibinfo {author} {\bibfnamefont {M.}~\bibnamefont {Shuaibi}},
  \bibinfo {author} {\bibfnamefont {E.~W.~C.}\ \bibnamefont {Spotte-Smith}},
  \bibinfo {author} {\bibfnamefont {M.~G.}\ \bibnamefont {Taylor}}, \bibinfo
  {author} {\bibfnamefont {M.~R.}\ \bibnamefont {Hasyim}}, \bibinfo {author}
  {\bibfnamefont {K.}~\bibnamefont {Michel}}, \bibinfo {author} {\bibfnamefont
  {I.}~\bibnamefont {Batatia}}, \bibinfo {author} {\bibfnamefont
  {G.}~\bibnamefont {Cs{\'a}nyi}}, \bibinfo {author} {\bibfnamefont
  {M.}~\bibnamefont {Dzamba}}, \bibinfo {author} {\bibfnamefont
  {P.}~\bibnamefont {Eastman}}, \emph {et~al.},\ }\bibfield  {title} {\bibinfo
  {title} {The open molecules 2025 ({OMol25}) dataset, evaluations, and
  models},\ }\href {https://arxiv.org/abs/2505.08762} {\bibfield  {journal}
  {\bibinfo  {journal} {arXiv preprint arXiv:2505.08762}\ } (\bibinfo {year}
  {2025})}\BibitemShut {NoStop}%
\bibitem [{\citenamefont {Ganscha}\ \emph {et~al.}(2025)\citenamefont
  {Ganscha}, \citenamefont {Unke}, \citenamefont {Ahlin}, \citenamefont
  {Maennel}, \citenamefont {Kashubin},\ and\ \citenamefont
  {M{\"u}ller}}]{ganscha2025qcml}%
  \BibitemOpen
  \bibfield  {author} {\bibinfo {author} {\bibfnamefont {S.}~\bibnamefont
  {Ganscha}}, \bibinfo {author} {\bibfnamefont {O.~T.}\ \bibnamefont {Unke}},
  \bibinfo {author} {\bibfnamefont {D.}~\bibnamefont {Ahlin}}, \bibinfo
  {author} {\bibfnamefont {H.}~\bibnamefont {Maennel}}, \bibinfo {author}
  {\bibfnamefont {S.}~\bibnamefont {Kashubin}},\ and\ \bibinfo {author}
  {\bibfnamefont {K.-R.}\ \bibnamefont {M{\"u}ller}},\ }\bibfield  {title}
  {\bibinfo {title} {The {QCML} dataset, quantum chemistry reference data from
  33.5 {M} {DFT} and 14.7 {B} semi-empirical calculations},\ }\href
  {https://doi.org/10.1038/s41597-025-04720-7} {\bibfield  {journal} {\bibinfo
  {journal} {Sci. Data}\ }\textbf {\bibinfo {volume} {12}},\ \bibinfo {pages}
  {406} (\bibinfo {year} {2025})}\BibitemShut {NoStop}%
\bibitem [{\citenamefont {Moerman}(2026)}]{chemenv-scripts}%
  \BibitemOpen
  \bibfield  {author} {\bibinfo {author} {\bibfnamefont {E.}~\bibnamefont
  {Moerman}},\ }\href@noop {} {\bibinfo {title} {Scripts for the envtok subset
  selection workflow}},\ \bibinfo {howpublished}
  {\url{https://github.com/em819/omat24-chemenv-subset.git}} (\bibinfo {year}
  {2026})\BibitemShut {NoStop}%
\bibitem [{\citenamefont {Moerman}\ \emph {et~al.}(2026)\citenamefont
  {Moerman}, \citenamefont {Poltavsky}, \citenamefont {Charkin-Gorbulin},\ and\
  \citenamefont {Tkatchenko}}]{chemenv-data-zenodo}%
  \BibitemOpen
  \bibfield  {author} {\bibinfo {author} {\bibfnamefont {E.}~\bibnamefont
  {Moerman}}, \bibinfo {author} {\bibfnamefont {I.}~\bibnamefont {Poltavsky}},
  \bibinfo {author} {\bibfnamefont {A.}~\bibnamefont {Charkin-Gorbulin}},\ and\
  \bibinfo {author} {\bibfnamefont {A.}~\bibnamefont {Tkatchenko}},\ }\href
  {https://doi.org/10.5281/zenodo.23137852} {\bibinfo {title} {Zenodo
  repository of atoms that matter: Largest materials datasets are 99\%
  redundant, and local chemistry can guide the required 1\%}},\ \bibinfo
  {howpublished} {\url{https://doi.org/10.5281/zenodo.23137852}} (\bibinfo
  {year} {2026}),\ \bibinfo {note} {version 1.0.0}\BibitemShut {NoStop}%
\end{thebibliography}%


\begin{thebibliography}{11}%
\makeatletter
\providecommand \@ifxundefined [1]{%
 \@ifx{#1\undefined}
}%
\providecommand \@ifnum [1]{%
 \ifnum #1\expandafter \@firstoftwo
 \else \expandafter \@secondoftwo
 \fi
}%
\providecommand \@ifx [1]{%
 \ifx #1\expandafter \@firstoftwo
 \else \expandafter \@secondoftwo
 \fi
}%
\providecommand \natexlab [1]{#1}%
\providecommand \enquote  [1]{``#1''}%
\providecommand \bibnamefont  [1]{#1}%
\providecommand \bibfnamefont [1]{#1}%
\providecommand \citenamefont [1]{#1}%
\providecommand \href@noop [0]{\@secondoftwo}%
\providecommand \href [0]{\begingroup \@sanitize@url \@href}%
\providecommand \@href[1]{\@@startlink{#1}\@@href}%
\providecommand \@@href[1]{\endgroup#1\@@endlink}%
\providecommand \@sanitize@url [0]{\catcode `\\12\catcode `\$12\catcode
  `\&12\catcode `\#12\catcode `\^12\catcode `\_12\catcode `\%12\relax}%
\providecommand \@@startlink[1]{}%
\providecommand \@@endlink[0]{}%
\providecommand \url  [0]{\begingroup\@sanitize@url \@url }%
\providecommand \@url [1]{\endgroup\@href {#1}{\urlprefix }}%
\providecommand \urlprefix  [0]{URL }%
\providecommand \Eprint [0]{\href }%
\providecommand \doibase [0]{https://doi.org/}%
\providecommand \selectlanguage [0]{\@gobble}%
\providecommand \bibinfo  [0]{\@secondoftwo}%
\providecommand \bibfield  [0]{\@secondoftwo}%
\providecommand \translation [1]{[#1]}%
\providecommand \BibitemOpen [0]{}%
\providecommand \bibitemStop [0]{}%
\providecommand \bibitemNoStop [0]{.\EOS\space}%
\providecommand \EOS [0]{\spacefactor3000\relax}%
\providecommand \BibitemShut  [1]{\csname bibitem#1\endcsname}%
\let\auto@bib@innerbib\@empty
\bibitem [{\citenamefont {Waroquiers}\ \emph {et~al.}(2020)\citenamefont
  {Waroquiers}, \citenamefont {George}, \citenamefont {Horton}, \citenamefont
  {Schenk}, \citenamefont {Persson}, \citenamefont {Rignanese}, \citenamefont
  {Gonze},\ and\ \citenamefont {Hautier}}]{waroquiers2020chemenv}%
  \BibitemOpen
  \bibfield  {author} {\bibinfo {author} {\bibfnamefont {D.}~\bibnamefont
  {Waroquiers}}, \bibinfo {author} {\bibfnamefont {J.}~\bibnamefont {George}},
  \bibinfo {author} {\bibfnamefont {M.}~\bibnamefont {Horton}}, \bibinfo
  {author} {\bibfnamefont {S.}~\bibnamefont {Schenk}}, \bibinfo {author}
  {\bibfnamefont {K.~A.}\ \bibnamefont {Persson}}, \bibinfo {author}
  {\bibfnamefont {G.-M.}\ \bibnamefont {Rignanese}}, \bibinfo {author}
  {\bibfnamefont {X.}~\bibnamefont {Gonze}},\ and\ \bibinfo {author}
  {\bibfnamefont {G.}~\bibnamefont {Hautier}},\ }\bibfield  {title} {\bibinfo
  {title} {Chemenv: a fast and robust coordination environment identification
  tool},\ }\href {https://doi.org/10.1107/s2052520620007994} {\bibfield
  {journal} {\bibinfo  {journal} {Acta Crystallogr. B: Struct. Sci. Cryst. Eng.
  Mater.}\ }\textbf {\bibinfo {volume} {76}},\ \bibinfo {pages} {683} (\bibinfo
  {year} {2020})}\BibitemShut {NoStop}%
\bibitem [{\citenamefont {Ong}\ \emph {et~al.}(2013)\citenamefont {Ong},
  \citenamefont {Richards}, \citenamefont {Jain}, \citenamefont {Hautier},
  \citenamefont {Kocher}, \citenamefont {Cholia}, \citenamefont {Gunter},
  \citenamefont {Chevrier}, \citenamefont {Persson},\ and\ \citenamefont
  {Ceder}}]{ong2013python}%
  \BibitemOpen
  \bibfield  {author} {\bibinfo {author} {\bibfnamefont {S.~P.}\ \bibnamefont
  {Ong}}, \bibinfo {author} {\bibfnamefont {W.~D.}\ \bibnamefont {Richards}},
  \bibinfo {author} {\bibfnamefont {A.}~\bibnamefont {Jain}}, \bibinfo {author}
  {\bibfnamefont {G.}~\bibnamefont {Hautier}}, \bibinfo {author} {\bibfnamefont
  {M.}~\bibnamefont {Kocher}}, \bibinfo {author} {\bibfnamefont
  {S.}~\bibnamefont {Cholia}}, \bibinfo {author} {\bibfnamefont
  {D.}~\bibnamefont {Gunter}}, \bibinfo {author} {\bibfnamefont {V.~L.}\
  \bibnamefont {Chevrier}}, \bibinfo {author} {\bibfnamefont {K.~A.}\
  \bibnamefont {Persson}},\ and\ \bibinfo {author} {\bibfnamefont
  {G.}~\bibnamefont {Ceder}},\ }\bibfield  {title} {\bibinfo {title} {Python
  materials genomics (pymatgen): A robust, open-source python library for
  materials analysis},\ }\href
  {https://doi.org/10.1016/j.commatsci.2012.10.028} {\bibfield  {journal}
  {\bibinfo  {journal} {Comput. Mater. Sci.}\ }\textbf {\bibinfo {volume}
  {68}},\ \bibinfo {pages} {314} (\bibinfo {year} {2013})}\BibitemShut
  {NoStop}%
\bibitem [{\citenamefont {Schwalbe-Koda}\ \emph {et~al.}(2025)\citenamefont
  {Schwalbe-Koda}, \citenamefont {Hamel}, \citenamefont {Sadigh}, \citenamefont
  {Zhou},\ and\ \citenamefont {Lordi}}]{schwalbe2025model}%
  \BibitemOpen
  \bibfield  {author} {\bibinfo {author} {\bibfnamefont {D.}~\bibnamefont
  {Schwalbe-Koda}}, \bibinfo {author} {\bibfnamefont {S.}~\bibnamefont
  {Hamel}}, \bibinfo {author} {\bibfnamefont {B.}~\bibnamefont {Sadigh}},
  \bibinfo {author} {\bibfnamefont {F.}~\bibnamefont {Zhou}},\ and\ \bibinfo
  {author} {\bibfnamefont {V.}~\bibnamefont {Lordi}},\ }\bibfield  {title}
  {\bibinfo {title} {Model-free estimation of completeness, uncertainties, and
  outliers in atomistic machine learning using information theory},\ }\href
  {https://doi.org/10.1038/s41467-025-59232-0} {\bibfield  {journal} {\bibinfo
  {journal} {Nat. Commun.}\ }\textbf {\bibinfo {volume} {16}},\ \bibinfo
  {pages} {4014} (\bibinfo {year} {2025})}\BibitemShut {NoStop}%
\bibitem [{\citenamefont {Yu}\ \emph {et~al.}(2026)\citenamefont {Yu},
  \citenamefont {Lordi},\ and\ \citenamefont
  {Schwalbe-Koda}}]{yu2026maximizing}%
  \BibitemOpen
  \bibfield  {author} {\bibinfo {author} {\bibfnamefont {B.}~\bibnamefont
  {Yu}}, \bibinfo {author} {\bibfnamefont {V.}~\bibnamefont {Lordi}},\ and\
  \bibinfo {author} {\bibfnamefont {D.}~\bibnamefont {Schwalbe-Koda}},\
  }\bibfield  {title} {\bibinfo {title} {Maximizing efficiency of dataset
  compression for machine learning potentials with information theory},\ }\href
  {https://doi.org/10.1063/5.0312057} {\bibfield  {journal} {\bibinfo
  {journal} {J. Chem. Phys.}\ }\textbf {\bibinfo {volume} {164}},\ \bibinfo
  {pages} {164111} (\bibinfo {year} {2026})}\BibitemShut {NoStop}%
\bibitem [{\citenamefont {Qi}\ \emph {et~al.}(2024)\citenamefont {Qi},
  \citenamefont {Ko}, \citenamefont {Wood}, \citenamefont {Pham},\ and\
  \citenamefont {Ong}}]{qi2024robust}%
  \BibitemOpen
  \bibfield  {author} {\bibinfo {author} {\bibfnamefont {J.}~\bibnamefont
  {Qi}}, \bibinfo {author} {\bibfnamefont {T.~W.}\ \bibnamefont {Ko}}, \bibinfo
  {author} {\bibfnamefont {B.~C.}\ \bibnamefont {Wood}}, \bibinfo {author}
  {\bibfnamefont {T.~A.}\ \bibnamefont {Pham}},\ and\ \bibinfo {author}
  {\bibfnamefont {S.~P.}\ \bibnamefont {Ong}},\ }\bibfield  {title} {\bibinfo
  {title} {Robust training of machine learning interatomic potentials with
  dimensionality reduction and stratified sampling},\ }\href
  {https://doi.org/10.1038/s41524-024-01227-4} {\bibfield  {journal} {\bibinfo
  {journal} {npj Comput. Mater.}\ }\textbf {\bibinfo {volume} {10}},\ \bibinfo
  {pages} {43} (\bibinfo {year} {2024})}\BibitemShut {NoStop}%
\bibitem [{\citenamefont {Batatia}\ \emph {et~al.}(2025)\citenamefont
  {Batatia}, \citenamefont {Lin}, \citenamefont {Hart}, \citenamefont {Kasoar},
  \citenamefont {Elena}, \citenamefont {Norwood}, \citenamefont {Wolf},\ and\
  \citenamefont {Cs{\'a}nyi}}]{batatia2025cross}%
  \BibitemOpen
  \bibfield  {author} {\bibinfo {author} {\bibfnamefont {I.}~\bibnamefont
  {Batatia}}, \bibinfo {author} {\bibfnamefont {C.}~\bibnamefont {Lin}},
  \bibinfo {author} {\bibfnamefont {J.}~\bibnamefont {Hart}}, \bibinfo {author}
  {\bibfnamefont {E.}~\bibnamefont {Kasoar}}, \bibinfo {author} {\bibfnamefont
  {A.~M.}\ \bibnamefont {Elena}}, \bibinfo {author} {\bibfnamefont {S.~W.}\
  \bibnamefont {Norwood}}, \bibinfo {author} {\bibfnamefont {T.}~\bibnamefont
  {Wolf}},\ and\ \bibinfo {author} {\bibfnamefont {G.}~\bibnamefont
  {Cs{\'a}nyi}},\ }\bibfield  {title} {\bibinfo {title} {Cross learning between
  electronic structure theories for unifying molecular, surface, and inorganic
  crystal foundation force fields},\ }\href {https://arxiv.org/abs/2510.25380}
  {\bibfield  {journal} {\bibinfo  {journal} {arXiv preprint arXiv:2510.25380}\
  } (\bibinfo {year} {2025})}\BibitemShut {NoStop}%
\bibitem [{mac(2026)}]{macefoundation}%
  \BibitemOpen
  \href@noop {} {\bibinfo {title} {Mace foundations}},\ \bibinfo {howpublished}
  {\url{https://github.com/ACEsuit/mace-foundations}} (\bibinfo {year}
  {2026})\BibitemShut {NoStop}%
\bibitem [{\citenamefont {Kaplan}\ \emph {et~al.}(2025)\citenamefont {Kaplan},
  \citenamefont {Liu}, \citenamefont {Qi}, \citenamefont {Ko}, \citenamefont
  {Deng}, \citenamefont {Riebesell}, \citenamefont {Ceder}, \citenamefont
  {Persson},\ and\ \citenamefont {Ong}}]{kaplan2025foundational}%
  \BibitemOpen
  \bibfield  {author} {\bibinfo {author} {\bibfnamefont {A.~D.}\ \bibnamefont
  {Kaplan}}, \bibinfo {author} {\bibfnamefont {R.}~\bibnamefont {Liu}},
  \bibinfo {author} {\bibfnamefont {J.}~\bibnamefont {Qi}}, \bibinfo {author}
  {\bibfnamefont {T.~W.}\ \bibnamefont {Ko}}, \bibinfo {author} {\bibfnamefont
  {B.}~\bibnamefont {Deng}}, \bibinfo {author} {\bibfnamefont {J.}~\bibnamefont
  {Riebesell}}, \bibinfo {author} {\bibfnamefont {G.}~\bibnamefont {Ceder}},
  \bibinfo {author} {\bibfnamefont {K.~A.}\ \bibnamefont {Persson}},\ and\
  \bibinfo {author} {\bibfnamefont {S.~P.}\ \bibnamefont {Ong}},\ }\bibfield
  {title} {\bibinfo {title} {A foundational potential energy surface dataset
  for materials},\ }\href {https://arxiv.org/abs/2503.04070} {\bibfield
  {journal} {\bibinfo  {journal} {arXiv preprint arXiv:2503.04070}\ } (\bibinfo
  {year} {2025})}\BibitemShut {NoStop}%
\bibitem [{\citenamefont {Larsen}\ \emph {et~al.}(2019)\citenamefont {Larsen},
  \citenamefont {Pandey}, \citenamefont {Strange},\ and\ \citenamefont
  {Jacobsen}}]{larsen2019definition}%
  \BibitemOpen
  \bibfield  {author} {\bibinfo {author} {\bibfnamefont {P.~M.}\ \bibnamefont
  {Larsen}}, \bibinfo {author} {\bibfnamefont {M.}~\bibnamefont {Pandey}},
  \bibinfo {author} {\bibfnamefont {M.}~\bibnamefont {Strange}},\ and\ \bibinfo
  {author} {\bibfnamefont {K.~W.}\ \bibnamefont {Jacobsen}},\ }\bibfield
  {title} {\bibinfo {title} {Definition of a scoring parameter to identify
  low-dimensional materials components},\ }\href
  {https://doi.org/10.1103/PhysRevMaterials.3.034003} {\bibfield  {journal}
  {\bibinfo  {journal} {Phys. Rev. Mater.}\ }\textbf {\bibinfo {volume} {3}},\
  \bibinfo {pages} {034003} (\bibinfo {year} {2019})}\BibitemShut {NoStop}%
\bibitem [{\citenamefont {Chiang}\ \emph {et~al.}(2025)\citenamefont {Chiang},
  \citenamefont {Kreiman}, \citenamefont {Zhang}, \citenamefont {Kuner},
  \citenamefont {Weaver}, \citenamefont {Amin}, \citenamefont {Park},
  \citenamefont {Lim}, \citenamefont {Kim}, \citenamefont {Chrzan} \emph
  {et~al.}}]{chiang2026mlip}%
  \BibitemOpen
  \bibfield  {author} {\bibinfo {author} {\bibfnamefont {Y.}~\bibnamefont
  {Chiang}}, \bibinfo {author} {\bibfnamefont {T.}~\bibnamefont {Kreiman}},
  \bibinfo {author} {\bibfnamefont {C.}~\bibnamefont {Zhang}}, \bibinfo
  {author} {\bibfnamefont {M.}~\bibnamefont {Kuner}}, \bibinfo {author}
  {\bibfnamefont {E.}~\bibnamefont {Weaver}}, \bibinfo {author} {\bibfnamefont
  {I.}~\bibnamefont {Amin}}, \bibinfo {author} {\bibfnamefont {H.}~\bibnamefont
  {Park}}, \bibinfo {author} {\bibfnamefont {Y.}~\bibnamefont {Lim}}, \bibinfo
  {author} {\bibfnamefont {J.}~\bibnamefont {Kim}}, \bibinfo {author}
  {\bibfnamefont {D.}~\bibnamefont {Chrzan}}, \emph {et~al.},\ }\bibfield
  {title} {\bibinfo {title} {Mlip arena: advancing fairness and transparency in
  machine learning interatomic potentials via an open, accessible benchmark
  platform},\ }\href {https://doi.org/10.52202/085713-4410} {\bibfield
  {journal} {\bibinfo  {journal} {Adv. Neural Inf. Process. Syst.}\ }\textbf
  {\bibinfo {volume} {38}},\ \bibinfo {pages} {146756} (\bibinfo {year}
  {2025})}\BibitemShut {NoStop}%
\bibitem [{\citenamefont {Dembitskiy}\ \emph {et~al.}(2025)\citenamefont
  {Dembitskiy}, \citenamefont {Humonen}, \citenamefont {Eremin}, \citenamefont
  {Aksyonov}, \citenamefont {Fedotov},\ and\ \citenamefont
  {Budennyy}}]{dembitskiy2025benchmarking}%
  \BibitemOpen
  \bibfield  {author} {\bibinfo {author} {\bibfnamefont {A.~D.}\ \bibnamefont
  {Dembitskiy}}, \bibinfo {author} {\bibfnamefont {I.~S.}\ \bibnamefont
  {Humonen}}, \bibinfo {author} {\bibfnamefont {R.~A.}\ \bibnamefont {Eremin}},
  \bibinfo {author} {\bibfnamefont {D.~A.}\ \bibnamefont {Aksyonov}}, \bibinfo
  {author} {\bibfnamefont {S.~S.}\ \bibnamefont {Fedotov}},\ and\ \bibinfo
  {author} {\bibfnamefont {S.~A.}\ \bibnamefont {Budennyy}},\ }\bibfield
  {title} {\bibinfo {title} {Benchmarking machine learning models for
  predicting lithium ion migration},\ }\href
  {https://doi.org/10.1038/s41524-025-01571-z} {\bibfield  {journal} {\bibinfo
  {journal} {npj Comput. Mater.}\ }\textbf {\bibinfo {volume} {11}},\ \bibinfo
  {pages} {131} (\bibinfo {year} {2025})}\BibitemShut {NoStop}%
\end{thebibliography}%

\end{document}


\title{Supplementary Information for: \\ Atoms that Matter: Largest Materials Datasets Are 99\% Redundant, and Local Chemistry Can Guide the Required 1\%}

\author{Evgeny Moerman}
\affiliation{Department of Physics and Materials Science, University of Luxembourg, L-1511 Luxembourg City, Luxembourg}
\author{Igor Poltavsky}
\affiliation{Department of Physics and Materials Science, University of Luxembourg, L-1511 Luxembourg City, Luxembourg}
\author{Anton Charkin-Gorbulin}
\affiliation{Department of Physics and Materials Science, University of Luxembourg, L-1511 Luxembourg City, Luxembourg}
\author{Alexandre Tkatchenko}
\affiliation{Department of Physics and Materials Science, University of Luxembourg, L-1511 Luxembourg City, Luxembourg}

\maketitle

\section{Generation of the EnvTok descriptor}
\label{sec:si-chemenv}
The EnvTok descriptors used in this work were generated with the
\texttt{ChemEnv} package~\cite{waroquiers2020chemenv} integrated in
\texttt{pymatgen}~\cite{ong2013python}. For a given material,
\texttt{ChemEnv} performs a Voronoi analysis of each atomic site of the unit
cell to determine the coordination number and the nearest neighbours, and
subsequently assigns the best fitting ideal coordination polyhedron from the
over 60 tabulated ones, covering coordination numbers $1$--$13$ and $20$.

\subsection{Neighbour determination}
Whether a Voronoi neighbour is admitted is governed by two dimensionless
criteria: its distance to the central atom, normalized by the distance to
the closest atom, must not exceed a distance cutoff $r_{\mathrm{max}}$, and
the solid angle it subtends at the central atom, normalized by the largest
such solid angle at that site, must be at least an angle cutoff
$a_{\mathrm{min}}$. We employ the \texttt{SimplestChemenvStrategy} with
$a_{\mathrm{min}}=0.3$ and the settings
\texttt{maximum\_distance\_factor}$\,=2.0$,
\texttt{only\_cations}$\,=$\texttt{False},
\texttt{centering\_type}$\,=$\texttt{"centroid"} with
\texttt{include\_central\_site\_in\_centroid}$\,=$\texttt{True}, and the
symmetry measure type \texttt{csm\_wcs\_ctwcc}. We further set
\texttt{structure\_refinement}$\,=$\texttt{"none"}, which is essential:
with the default setting \texttt{ChemEnv} internally replaces the input
structure by its conventional standard cell, so that the returned site
indices no longer correspond to the atoms of the structure passed to it.
Environments are computed for one representative site per symmetry orbit
(determined with \texttt{spglib}) and broadcast to all equivalent atoms.

A shortcoming of the default \texttt{ChemEnv} settings for a dataset as
diverse as OMat24 is the fixed distance cutoff. The default
$r_{\mathrm{max}}=1.4$ only admits neighbours at most $40\%$ further away
than the closest atom. This gives chemically sensible assignments in the
vast majority of cases, but fails when the shortest distance is
substantially shorter than all other plausible neighbour distances. A
phosphate oxygen is a typical example: in \ce{Ca3(PO4)2} the short P--O bond
of $1.52\,$\AA{} sets the normalization, so that the genuine Ca--O bonds at
$2.28$--$2.59\,$\AA{} fall at normalized distances of $1.50$--$1.70$ and are
discarded, collapsing the site to a single-neighbour environment. Simply
raising $r_{\mathrm{max}}$ risks over-coordinating sensible environments. We
therefore adopt a two-tiered cutoff. For each site the neighbours are
determined with both $r_{\mathrm{max}}=1.4$ and $r_{\mathrm{max}}=1.7$, and
the larger cutoff is adopted only if at least one of the additionally
admitted neighbours has a normalized Voronoi solid-angle weight of at least
$0.45$. This weight is obtained from \texttt{pymatgen}'s \texttt{VoronoiNN}
(with \texttt{tol}$\,=0$) as the solid angle of the Voronoi facet shared with
that neighbour, normalized by the largest facet solid angle at the site; it
measures whether the additional contact occupies a substantial fraction of
the coordination sphere rather than a screened sliver. The choice is made per
atom on the basis of geometry and composition only, so it is applied
identically to every dataset.

\subsection{Distortion field}
In addition to the neighbouring atoms, the coordination number and the
coordination polyhedron, \texttt{ChemEnv} reports the \emph{continuous
symmetry measure} (CSM), which quantifies the quality of the polyhedron fit
and is bounded between $0$ (perfect match) and $100$ (maximal deviation).
The CSM is a measure of local distortion and is included in the descriptor.
Since the CSM is continuous while the descriptor is discrete, it is assigned
to one of four bins, with a fifth value reserved as a sentinel:
\begin{equation}
      p(\mathrm{CSM}) = \begin{cases}
          0 & \mathrm{CSM}\leq1\\
          1 & 1<\mathrm{CSM}\leq5\\
          2 & 5<\mathrm{CSM}<10\\
          3 & \mathrm{CSM}\geq10\\
          999 & \text{CSM undefined.}
      \end{cases}
\end{equation}
The sentinel $999$ is used whenever the CSM is undefined, which occurs in
two situations. First, if the coordination number determined for a site
lies outside the tabulated range, no model polyhedron exists and
\texttt{ChemEnv} returns the label \texttt{UNKNOWN:CN}. Second, if the
Voronoi construction fails for the entire structure, as happens for cells
containing vacuum layers, no environment is obtained at all; in this case we
recover the neighbour list and coordination number with \texttt{CrystalNN}
and label the site \texttt{UNKNOWN:CN} as well. We stress that $p=999$ is not
a signature of strong distortion: because \texttt{SimplestChemenvStrategy}
applies no CSM cutoff, even poorly matching polyhedra are assigned and fall
into bin~$3$.

\subsection{Bond-compression and asymmetry fields}
The token constructed so far is blind to stretching and compression of the
bonds between the central atom and its neighbours, and does not encode the
asymmetry of the coordination independently of the best-fitting polyhedron.
Two further fields are therefore added. The first is the binned ratio of the
most compressed interatomic distance $r_{ij}$ between central atom $i$ and
neighbour $j$ to the equilibrium distance $R_{i,0}+R_{j,0}$ obtained from
Cordero covalent radii:
\begin{equation}
      p\!\left(\frac{r_{ij}}{R_{i,0}+R_{j,0}}\right) = \begin{cases}
          0 & \rho < 0.6\\
          1 & 0.6\leq\rho<0.7\\
          2 & 0.7\leq\rho<0.85\\
          3 & 0.85\leq\rho<1.15\\
          4 & \rho\geq1.15\\
          999 & \text{neighbours undetermined,}
      \end{cases}
\end{equation}

with $\rho\equiv\min_{j}\frac{r_{ij}}{R_{i,0}+R_{j,0}}$.
The bin edges are finer towards short distances to make the descriptor more
sensitive in the compressed limit, while distances close to equilibrium
($0.85\leq\rho<1.15$) are grouped in a single bin.

The asymmetry metric $A_i$ is the magnitude of a bond-length-weighted sum of
the unit vectors from the central atom to its neighbours,
\begin{equation}
    A_i = \left|\sum_{j\in\text{neighbours}}
    \exp\!\left[-\frac{1}{\lambda}\left(\frac{r_{ij}}{R_{i,0}+R_{j,0}} - 1\right)\right]
    \bm{\hat{r}}_{ij}\right|,\, \lambda=0.1,
\end{equation}
which vanishes for a centrosymmetric coordination and grows when the
shortest contacts are concentrated on one side of the central atom. It is
binned as
\begin{equation}
      p(A_i) = \begin{cases}
          0 & A_i<1.0\\
          1 & 1.0\leq A_i<3.0\\
          2 & A_i\geq3.0\\
          999 & \text{neighbours undetermined.}
      \end{cases}
\end{equation}

The complete token thus has the form
\texttt{element\_spin\_neighbours\_polyhedron\_p(CSM)\_p($\rho$)\_p($A$)}; for
the sodium site of rock-salt NaCl it reads
\texttt{Na\_0\_Cl-Cl-Cl-Cl-Cl-Cl\_O:6\_0\_3\_0}. The spin field is the
integer-rounded magnetic moment of the site as provided by OMat24; for
structures without magnetic labels it can be set to zero or dropped, and it
is in any case the second field to be removed by the coarsening ladder
(Sec.~\ref{sec:si-selection}).

\section{EnvTok subset selection}
\label{sec:si-selection}
\subsection{Coarsening ladder}
The selection algorithm is controlled by a single external parameter $T$,
the support threshold, which sets the resolution at which chemical
environments are distinguished. In a pre-processing step the algorithm
constructs a ladder of increasingly coarse environment definitions whose
rungs are defined in Table~\ref{tab:chemenv-ladder}. Each unique EnvTok
token in the full dataset is coarsened to the first rung at which 
the coarsened token occurs in at least $T$ distinct structures, counted 
over the tokens not yet assigned at a finer rung. The assignment is a 
strict partition: every atom belongs to exactly one class, and no token is discarded.
For the selection of the EnvTok subset in the present work $T=100$ was chosen, yielding
a subset of approximately $6.5$ million ($6\,566\,253$) atoms.

\begin{table*}[htbp]
\centering
\footnotesize
\begin{tabular}{ll}
\toprule
\textbf{Coarsening applied to the full token} & \textbf{Example class signature} \\
\midrule
\multicolumn{2}{l}{\itshape The last two fields of the token (compression of the most compressed bond, asymmetry) are always retained} \\
\texttt{R0\_full}: the complete token               & \texttt{Fe|0|Cl-O-O-O|T:4|3|1|2} \\
\texttt{R1dA}: drop the polyhedron-fit (CSM) bin        & \texttt{Fe|0|Cl-O-O-O|T:4|1|2}   \\
\texttt{R2dA}: drop the spin                        & \texttt{Fe|Cl-O-O-O|T:4|1|2}     \\
\texttt{R2adA}: mark unreliable polyhedron fits (CSM bin $\geq2$) as \texttt{UNCL} & \texttt{Fe|Cl-O-O-O|UNCL|1|2} \\
\texttt{R2bdA}: replace neighbour list by binned composition, keep coordination number & \texttt{Fe|Cl:1-O:34|UNCL|4|1|2} \\
\texttt{R3dA}: replace neighbour list by set of unique elements       & \texttt{Fe|Cl-O|UNCL|4|1|2}      \\
\texttt{R4dA}: also drop the polyhedron label       & \texttt{Fe|Cl-O|4|1|2}           \\
\texttt{R5dA}: also drop the coordination number    & \texttt{Fe|Cl-O|1|2}             \\
\texttt{R6dA}: drop neighbour set; keep polyhedron and CN & \texttt{Fe|UNCL|4|1|2}           \\
\texttt{R7dA}: drop polyhedron label; element and CN      & \texttt{Fe|4|1|2}                \\
\texttt{R8dA}: central element only                 & \texttt{Fe|1|2}                  \\
\midrule
\texttt{R8A}: drop bond compression; element and asymmetry only & \texttt{Fe|2}                   \\
\texttt{R8}: element alone     & \texttt{Fe}                      \\
\midrule
\multicolumn{2}{l}{\itshape Not a rung: terminal residue} \\
catch-all, for tokens unresolved after \texttt{R8}  & \texttt{\_\_CATCHALL\_\_|Fe}     \\
\bottomrule
\end{tabular}
\caption{The coarsening ladder, finest first. A token is assigned to the first
rung, read top to bottom, whose signature reaches support $T$, counted over the
still-unresolved token pool. Column~2 traces one atom,
\texttt{Fe\_0\_Cl-O-O-O\_T:4\_3\_1\_2}, through every level
}
\label{tab:chemenv-ladder}
\end{table*}

\subsection{Greedy cover, pruning and presence constraints}
The subsequent greedy algorithm iteratively selects structures such that
every class defined above is included at least once, minimizing the total
number of atoms rather than the number of structures. Exact minimization of
a weighted set cover is NP-hard; the greedy solution is near-minimal.

Because the greedy selection never revisits an earlier choice, a structure
taken early for one or two environments can become entirely redundant once
later selections supply those same environments as surplus. After the
selection terminates, the algorithm therefore removes, largest structure
first, every selected structure whose environments all remain covered
without it; by construction this cannot violate the coverage guarantee.

A second correction addresses environments that the ladder cannot reach. 
Because sufficiently rare tokens are merged into coarser ladder rungs, covering
every coarsened rung does not guarantee that every combination of element, coordination
number and local distortion appears in the final subset.
The algorithm
therefore enforces four presence constraints, on (element, polyhedron),
(element, coordination number), (element, coordination number,
shortest-contact bin) and (element, coordination number, asymmetry bin): a
small number (about a thousand for $T=100$) of further structures are added so that every
such combination occurring in the full dataset occurs at least once in the
subset. These constraints carry no support threshold, which is what allows
them to reach environments too rare to survive the ladder, and they are
applied after pruning so that the pruning pass cannot remove the structures
that satisfy them.

\subsection{Relation to entropy maximization}
The EnvTok objective is closely related to the information-entropy
criterion of QUESTS~\cite{schwalbe2025model,yu2026maximizing}. For a set of
$N$ atomic descriptors $\bm{X}_i$, Ref.~\cite{yu2026maximizing} defines
\begin{equation}\label{eq:quests-entropy}
    \mathcal{H}(\{\bm{X}\}) = -\frac{1}{N}\sum_{i=1}^{N}
    \log\left[\frac{1}{N}\sum_{j=1}^{N}K(\bm{X}_i,\bm{X}_j)\right],
\end{equation}
with a Gaussian kernel $K$ of fixed bandwidth. If the kernel is replaced by a
delta function that is unity for two atoms assigned to the same class of the
coarsening ladder and zero otherwise,
\begin{equation}\label{eq:quests-entropy-delta}
    \mathcal{H}(\{\bm{X}\}) = -\frac{1}{N}\sum_{i=1}^{N}
    \log\left[\frac{1}{N}\sum_{j=1}^{N}\delta(\bm{X}_i,\bm{X}_j)\right]
    = -\sum_{c} p_c \log p_c ,
\end{equation}
where $p_c$ is the fraction of atoms of the subset in class $c$.
Eq.~\eqref{eq:quests-entropy-delta} is the Shannon entropy of the class
occupations, which is maximal when all classes are populated equally.
Covering every class with one representative would, therefore, maximize
Eq.~\eqref{eq:quests-entropy} under this kernel. 
Because selection operates on whole structures, this limit cannot be reached in practice.
Compared with the
continuous kernel, the discrete comparison sacrifices resolution within a
class but gains an element-resolved, chemically interpretable definition of
what counts as a distinct environment, and, once the tokens are computed, the cover step scales to
$10^{8}$ structures because class membership is a hash lookup rather than a
kernel evaluation against all other atoms.

\subsection{DIRECT and random subsets}
The DIRECT subset selection was applied to OMat24 as follows. The structures were
featurized with the M3GNet-Eform-MP-2018.6.1 model, and the feature space was
reduced, following Kaiser's rule as originally proposed, to 21 principal components
retaining $82\%$ of the variance. This space was then partitioned by BIRCH into
$n=116\,500$ clusters, in line with the recommendation of the authors of DIRECT to
choose $n$ as close as possible to the target size of the subset, so that ideally
about one structure per cluster is sampled; this is more than five times the number of
clusters used in the original study~\cite{qi2024robust} and a value
that still safely fits into the available memory. For a dataset of this size,
the iterative rule by which the \texttt{maml} implementation adjusts the maximal
cluster radius was found to be unstable and was replaced by a bisection search,
which determined a radius threshold of 0.1092 instead of the default value of 0.5.
This modification only ensures that BIRCH reliably partitions the feature space into
as many regions as requested; the featurization, the clustering itself and all
further steps of the selection remain unaffected. Finally, up to 4 structures were
picked from each cluster ($k=4$), yielding a DIRECT subset of $458\,754$ OMat24 structures,
or approximately $6.5$ million ($6\,532\,238$) atoms.

The four random subsets were drawn uniformly over structures of the
$f$-block-free OMat24 dataset with four different seeds until the budget of
approximately $6.5$ million ($6\,496\,886$ -- $6\,513\,135$) atoms was reached. To guarantee that each of the
68 allowed elements is represented, a minimal set of 35 structures were reserved beforehand, one for each element not yet covered.

\subsection{Composition of the training subsets}
\label{sec:si-subset-composition}
\begin{figure}[ht]
    \centering
    \includegraphics[width=\linewidth]{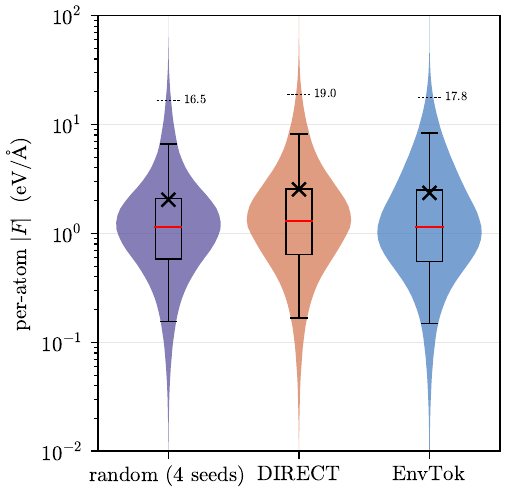}
    \caption{Distribution of atomic forces in the random, DIRECT and EnvTok
    subsets of OMat24 (violin plots; red bar and cross: median and mean; box
    edges: 25th and 75th percentiles; whiskers: 5th and 95th percentiles;
    dashed line: 99th percentile)}
    \label{fig:force-distribution}
\end{figure}

The three selection strategies spend the same budget of approximately $6.5$ million atoms
very differently (Fig.~\ref{fig:force-distribution}). The EnvTok subset
contains about 796 thousand structures (8 atoms per structure on average),
the DIRECT and random subsets 459 and 356 thousand (14 and 18 atoms per
structure, respectively). Covering many distinct environments is most economically
achieved with many small cells, whereas selecting by structure, as DIRECT
and random sampling do, follows the size distribution of OMat24 for the most part, 
with DIRECT noticeably under-sampling cells with more than 40 atoms. 
The atomic-force distributions also differ. 
The random subsets show a narrow
distribution peaked at a median of $1.1\,$eV/\AA, reproducing that of the
parent dataset. DIRECT yields a similarly shaped distribution shifted to
higher forces (median $+0.17\,$eV/\AA, high-force tail shifted by over
$1.5\,$eV/\AA), consistent with its bias towards extreme outliers. The EnvTok subset
has a qualitatively wider distribution: its median is comparable to random,
but its $5$th--$95$th percentile range is the largest of the three, because
the compression and asymmetry fields of the token enforce the inclusion of
environments at all levels of distortion; only in the top $1\%$ does DIRECT
contains more higher-force atoms.

\section{Training of MACE MLFFs}
\label{sec:si-training}
The subset-trained models use the model hyperparameters and training
settings of the MACE-OMAT-0 (medium) reference model~\cite{batatia2025cross}
with a few adjustments due to the substantially smaller training set
(Table~\ref{tab:training-parameters}). Establishing the reference settings
was complicated by disagreements between Ref.~\cite{batatia2025cross}, the
MACE foundation-model repository~\cite{macefoundation} and the metadata of
the released MACE-OMAT-0 medium model: the number of radial basis functions
is quoted as $8$ in the paper and the repository but is $10$ in the model
itself, and we follow the model. The effective batch size was reduced to
$1024$, whereas MACE-OMAT-0 was pre-trained with a batch size of $256$ on
$48$ GPUs, i.e.\ an effective batch size of $12\,288$. Each subset model was
trained until convergence with a patience of 40, i.e 
until the validation loss did not improve for 40 subsequent epochs, 
on $8$ NVIDIA A100 (40\,GB) GPUs. The model precision was reduced from float64 to float32 after
observing no significant difference in model performance.
The ReduceLROnPlateau learning rate schedule was used with a scheduler patience of 5 epochs,
after which the learning rate is reduced by a factor of $0.8$.
Other than that, the training settings were chosen consistently with Ref.~\cite{batatia2025cross}
including the utilization of a weighted Huber loss for the energy, force and stress with $\delta=0.01$,
the AMSGrad variant of the Adam optimizer with default $\beta_1=0.9$, $\beta_2=0.999$, $\epsilon=10^{-8}$, a learning rate of $0.001$, an EMA decay of 0.999 and gradient clipping of $100.0$.

\begin{table}[htbp]
    \centering
    \begin{tabular*}{\columnwidth}{@{\extracolsep{\fill}}llll}
        \toprule
        & \multicolumn{3}{c}{Models}\\
        \cmidrule(lr){2-4}
        Hyperparameter & EnvTok & DIRECT & random\\
        \midrule
        max\_ell &3 &3 &3\\
        correlation &3&3&3\\
        max\_L &1&1&1\\
        num\_channels\_edge &128&128&128\\
        num\_channels\_node &128&128&128\\
        num\_interactions &2&2&2\\
        num\_radial\_basis &10&10&10\\
        r\_max &6.0\,\AA&6.0\,\AA&6.0\,\AA\\
        interactions\_class &linear&linear&linear\\
        irreps &16x0e&16x0e&16x0e\\
        effective batch size &1024&1024&1024\\
        energy coefficient &1&1&1\\
        force coefficient &10&10&10\\
        stress coefficient &10&10&10\\
        \bottomrule
    \end{tabular*}
    \caption{Hyperparameters of the subset-trained MACE models.
    Hyperparameter names follow Table XVII of Ref.~\cite{batatia2025cross}; linear refers to the
    \texttt{RealAgnosticDensityInteractionBlock}
    and \texttt{RealAgnosticDensityResidualInteractionBlock} interaction blocks and irreps to the readout MLP irreps}
    \label{tab:training-parameters}
\end{table}

\section{Benchmark protocols}
\label{sec:si-benchmarks}
Consistent with the restriction of OMat24 to non-$f$-block elements, all
benchmarks exclude unit cells containing lanthanides ($Z=57$--$71$) or
actinides ($Z\geq89$).

\subsection{Atomic forces: MatPES}
\begin{figure*}[ht]
    \centering
    \begin{subfigure}{0.48\textwidth}
        \includegraphics[width=\textwidth]{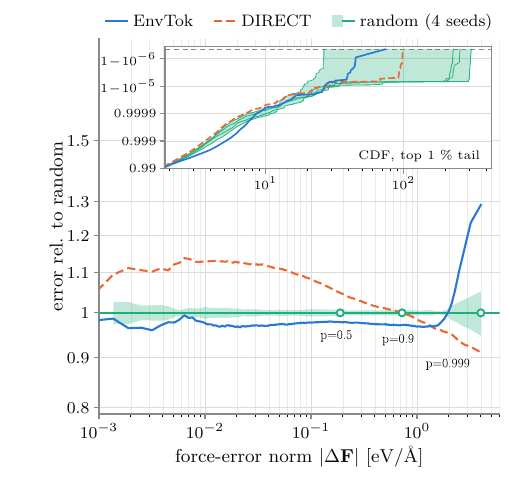}
        \caption{}
        \label{fig:matpes-cdf-bulk}
    \end{subfigure}
    \hfill
    \begin{subfigure}{0.48\textwidth}
        \includegraphics[width=\textwidth]{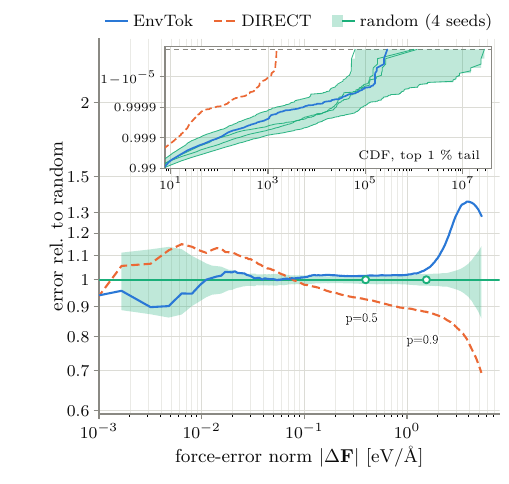}
        \caption{}
        \label{fig:matpes-cdf-non-bulk}
    \end{subfigure}
    \caption{Cumulative distribution function (CDF) of the per-atom
    absolute force error  for (a) bulk and (b) non-bulk MatPES structures. 
    For easier visual distinction the main panels show, at each percentile, 
    the force error of each model relative to the mean over the four random models,
    with the standard deviation as shaded region.
    The insets show the upper 1\% of the per-atom force error CDFs, 
    with the random models as a min–max envelope}
    \label{fig:matpes-cdf}
\end{figure*}

\begin{figure*}[ht]
    \centering
    \begin{subfigure}{0.48\textwidth}
        \includegraphics[width=\textwidth]{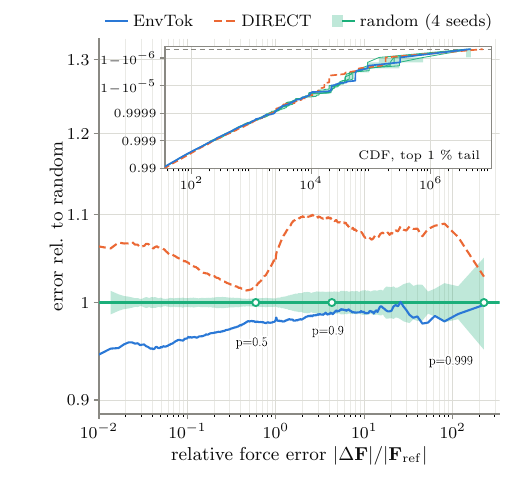}
        \caption{}
        \label{fig:matpes-cdf-bulk-rel}
    \end{subfigure}
    \hfill
    \begin{subfigure}{0.48\textwidth}
        \includegraphics[width=\textwidth]{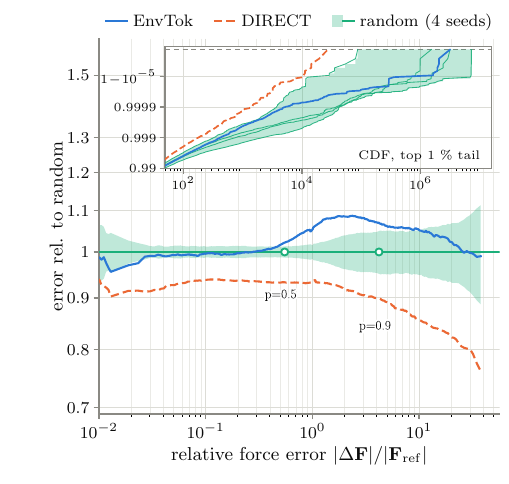}
        \caption{}
        \label{fig:matpes-cdf-non-bulk-rel}
    \end{subfigure}
    \caption{
    Cumulative distribution function (CDF) of the per-atom
    relative force error normalized by the DFT force magnitude
    for (a) bulk and (b) non-bulk MatPES structures. 
    For easier visual distinction the main panels show, at each percentile, 
    the force error of each model relative to the mean over the four random models,
    with the standard deviation as a shaded region.
    The insets show the upper 1\% of the per-atom relative force error CDFs, 
    with the random models as a min–max envelope}
    \label{fig:matpes-cdf-rel}
\end{figure*}

Atomic forces were evaluated on the MatPES PBE v2025.2
dataset~\cite{kaplan2025foundational}, which after the element cut contains
$323\,096$ structures. These were split into three groups:
\begin{itemize}
    \item \textbf{Overcompressed}: unit cells with at least one interatomic
    distance below $40\%$ of the sum of the Cordero covalent radii of the two
    atoms; 11 such cells were identified, considered unphysical and excluded.
    \item \textbf{Non-bulk}: unit cells consisting of topologically
    disconnected components of reduced dimensionality (2D slabs, 1D wires,
    0D clusters and combinations thereof). The structure is represented as a
    graph with atoms as nodes and edges between atoms at a distance of at
    most $1.25$ times the sum of their Cordero covalent radii; the graph is
    evaluated for low-dimensional components with an optimized
    implementation of \texttt{pymatgen}'s
    \texttt{get\_dimensionality\_larsen}~\cite{larsen2019definition}, verified
    against the \texttt{pymatgen} function. $80\,766$ structures.
    \item \textbf{Bulk}: everything else; $242\,319$ structures.
\end{itemize}
For Table~I of the main text, atoms were binned by the force error norm 
of MACE-OMAT-0, and within each bin the same force error metric for every model was computed.
Fig.~\ref{fig:matpes-cdf} shows the same data in binning-free form as
cumulative distribution functions.
The same analysis has also been performed for the
relative force error, which is the absolute force error norm as discussed in Table~I and
Fig.~\ref{fig:matpes-cdf} normalized by the magnitude of the DFT force vector. The analogous
findings are presented in Table~\ref{tab:performance-matpes-rel} and Fig.~\ref{fig:matpes-cdf-rel}
and demonstrate that the main findings on the relative performance of the subset-trained MLFFs
are not an artifact of increased force errors being overrepresented in atoms with large atomic forces.

\begin{table*}[htbp]
    \centering
    \small
    \caption{%
      Per-atom relative force-error profile on MatPES compared to MACE-OMAT-0.
      Atoms are binned by the per-atom force-error normalized by the reference DFT force norm
      $|\Delta\mathbf{F}_i| / |\mathbf{F}_{\mathrm{ref},i}|$ of the full-data reference model MACE-OMAT-0.
      Every entry is the median, over the atoms of that bin of $\rho_i=|\Delta\mathbf{F}_i|/|\mathbf{F}_{\mathrm{ref},i}|$ (dimensionless).
      The \emph{Random} column is the mean $\pm$ standard deviation
      over the four random subsets.
      Bold marks the lowest of the three candidate columns in a row, and only
      where its margin over the next-lowest exceeds that standard deviation. Atoms with a reference DFT force of 0 were
      excluded.}
    \begin{tabular}{@{}lrrrr r@{\;$\pm$\;}l@{}}
      \toprule
      & &\multicolumn{5}{c}{Median relative force error $\rho_i$ (dimensionless)} \\
      \cmidrule(l){3-7}
      \makecell{MACE-OMAT-0\\relative\\force error}& \makecell{\% of\\atoms}& MACE-OMAT-0 & EnvTok & DIRECT & \multicolumn{2}{c@{}}{Random} \\
      \midrule
      \multicolumn{7}{@{}l}{\textit{Bulk} ($2\,110\,829$ atoms)} \\
      $0$--$0.025$      &   1.52 &  0.0174 & \textbf{0.0505} &          0.0603 &          0.0543 &  0.0005 \\
      $0.025$--$0.05$   &   3.47 &  0.0385 & \textbf{0.0796} &          0.0912 &          0.0842 &  0.0008 \\
      $0.05$--$0.1$     &   8.36 &  0.0753 & \textbf{0.1278} &          0.1424 &          0.1345 &  0.0012 \\
      $0.1$--$0.2$      &  14.64 &  0.1465 & \textbf{0.2123} &          0.2292 &          0.2208 &  0.0015 \\
      $0.2$--$0.4$      &  17.86 &  0.2838 & \textbf{0.3670} &          0.3877 &          0.3790 &  0.0022 \\
      $0.4$--$0.8$      &  18.48 &  0.5726 & \textbf{0.6675} &          0.7035 &          0.6812 &  0.0020 \\
      $0.8$--$1.6$      &  18.69 &  1.0755 & \textbf{1.1564} &          1.2406 &          1.1872 &  0.0051 \\
      $1.6$--$3.2$      &   8.76 &  2.1293 & \textbf{2.5333} &          2.7480 &          2.5994 &  0.0258 \\
      $3.2$--$6.4$      &   4.44 &  4.2747 & \textbf{5.3478} &          5.7586 &          5.4433 &  0.0374 \\
      $6.4$--$12.8$     &   2.16 &  8.4784 & \textbf{10.3306} &        10.9511 &         10.5202 &  0.0751 \\
      $\geq 12.8$       &   1.63 & 22.2554 &         28.8894 &         30.9206 &         29.0365 &  0.6177 \\
      \midrule
      \multicolumn{7}{@{}l}{\textit{Non-bulk} ($760\,838$ atoms)} \\
      $0$--$0.025$      &   1.13 &  0.0175 &          0.1004 & \textbf{0.0890} &          0.0976 &  0.0032 \\
      $0.025$--$0.05$   &   2.93 &  0.0390 &          0.1364 & \textbf{0.1259} &          0.1353 &  0.0034 \\
      $0.05$--$0.1$     &   8.44 &  0.0762 &          0.1852 & \textbf{0.1730} &          0.1867 &  0.0037 \\
      $0.1$--$0.2$      &  17.15 &  0.1477 &          0.2668 & \textbf{0.2505} &          0.2687 &  0.0033 \\
      $0.2$--$0.4$      &  23.01 &  0.2844 &          0.4126 & \textbf{0.3930} &          0.4188 &  0.0052 \\
      $0.4$--$0.8$      &  19.56 &  0.5490 &          0.7101 & \textbf{0.6691} &          0.7068 &  0.0091 \\
      $0.8$--$1.6$      &  12.97 &  1.0751 &          1.3068 & \textbf{1.2166} &          1.2635 &  0.0238 \\
      $1.6$--$3.2$      &   7.28 &  2.1513 &          2.7178 & \textbf{2.3912} &          2.5565 &  0.0479 \\
      $3.2$--$6.4$      &   3.99 &  4.2927 &          5.6004 & \textbf{5.0155} &          5.3776 &  0.1678 \\
      $6.4$--$12.8$     &   2.05 &  8.5006 &         11.3277 & \textbf{9.8601} &         10.9161 &  0.2168 \\
      $\geq 12.8$       &   1.50 & 20.6905 &         27.6714 & \textbf{26.3950} &         27.9687 &  0.2774 \\
      \bottomrule
    \end{tabular}
    \label{tab:performance-matpes-rel}
  \end{table*}

\subsection{Near-equilibrium properties: MatCalc}
The MatCalc benchmark suite~\cite{kaplan2025foundational} assesses formation
energy, bulk and shear moduli, the harmonic heat capacity, force softening
and the fingerprint distance between the MLFF-relaxed and the DFT reference
structure. Its reference data follow the same PBE/PBE+$U$ convention as
OMat24. We deviate from the MatCalc defaults in two respects. First, for the
elastic moduli the atomic positions of each strained cell are re-relaxed,
yielding relaxed-ion rather than clamped-ion constants, and the strain set is
extended from four to six values per mode, $(\pm0.01,\pm0.00667,\pm0.00333)$
for the normal and $(\pm0.06,\pm0.04,\pm0.02)$ for the shear deformations,
for a better-conditioned strain--stress fit. Second, formation energies are
referenced to tabulated PBE elemental energies rather than to elemental
structures relaxed with the MLFF itself. Phonon supercells follow the MatCalc
implementation, which enforces a minimum lattice-vector length of $20\,$\AA{}
rather than the $2\times2\times2$ supercell quoted in
Ref.~\cite{kaplan2025foundational}.

\begin{figure*}[ht]
    \centering
    \captionsetup[subfigure]{position=top, justification=raggedright,
                             singlelinecheck=false, skip=0pt}
    \begin{subfigure}{0.3\textwidth}
        \caption{}
        \includegraphics[width=\textwidth]{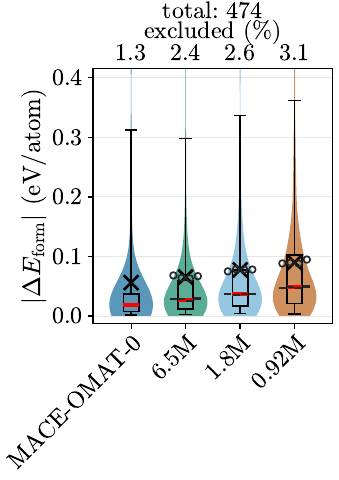}
    \end{subfigure}
    \hfill
    \begin{subfigure}{0.3\textwidth}
        \caption{}
        \includegraphics[width=\textwidth]{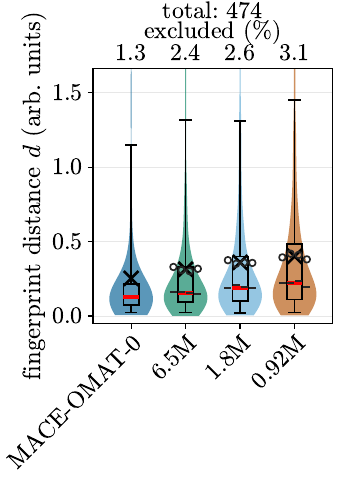}
    \end{subfigure}
    \hfill
    \begin{subfigure}{0.3\textwidth}
        \caption{}
        \includegraphics[width=\textwidth]{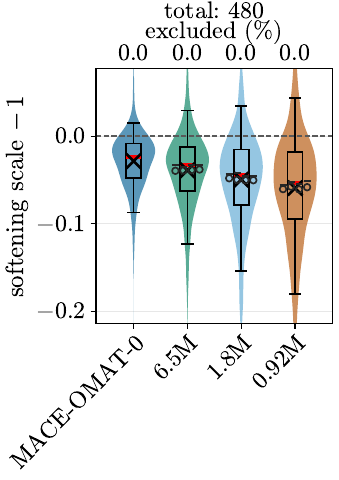}
    \end{subfigure}

    \vskip\baselineskip

    \begin{subfigure}{0.3\textwidth}
        \caption{}
        \includegraphics[width=\textwidth]{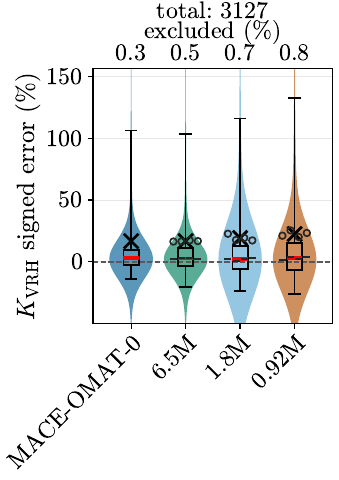}
    \end{subfigure}
    \hfill
    \begin{subfigure}{0.3\textwidth}
        \caption{}
        \includegraphics[width=\textwidth]{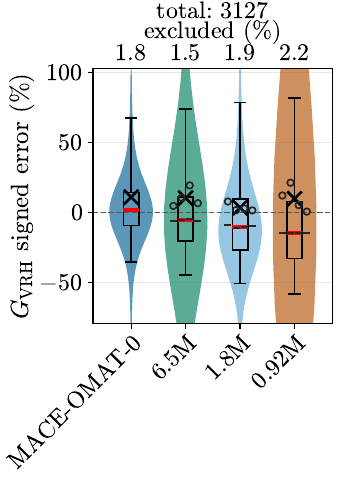}
    \end{subfigure}
    \hfill
    \begin{subfigure}{0.3\textwidth}
        \caption{}
        \label{fig:matcalc-cv-degrade}
        \includegraphics[width=\textwidth]{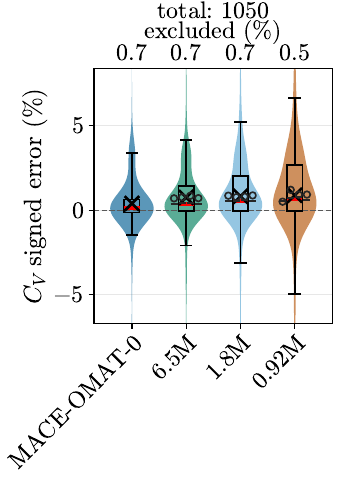}
    \end{subfigure}
    \caption{MatCalc benchmark of MACE-OMAT-0 and MACE MLFFs trained on randomly drawn OMat24 subsets of decreasing size (6.5 million, 1.8 million and 0.92 million atoms as specified on the horizontal axis): (a) formation energies, (b) fingerprint distance between MLFF- and DFT-relaxed structures, (c) systematic force softening, (d) bulk modulus, (e) shear modulus and (f) constant-volume heat capacity in the harmonic approximation. Violin plots: cross and red bar, mean and median; box edges, 25th and 75th percentiles; whiskers, 5th and 95th percentiles. Black circles and black bars
    are the mean and median values of the 4 individual seeds of the random subsets.
    Above every violin, the percentage of structures that were excluded due to diverged relaxations
    or unphysical property predictions (i.e non-positive elastic moduli) is given
    and at the top the total number of data points for the respective benchmark property are given}
    \label{fig:matcalc-bench-degradation}
\end{figure*}

To test how the predicted properties degrade if the training data are reduced
even further, two additional sets of random OMat24 subsets were drawn, with
$1.8$ and $0.92$ million atoms (four seeds each), and trained in the same way
as the $6.5$ million-atom random models. Fig.~\ref{fig:matcalc-bench-degradation}
shows that the magnitude and spread of the errors increase gradually as the
subset shrinks; most clearly, the shear modulus is increasingly underestimated
(median $-5.6\%$, $-10.2\%$ and $-14.7\%$), together with stronger force
softening. Even at these more extreme compression ratios, however, no
qualitative breakdown of the near-equilibrium properties is observed.
For formation energy, force softening and shear modulus, the per-seed medians
scatter much less than they shift between subset sizes. For the fingerprint
distance and $C_V$, in contrast, the scatter between seeds grows as the subset
shrinks, so that the per-seed medians of the $1.8$ and $0.92$ million-atom
models partly overlap.

\subsection{MD stability: MLIP Arena}
The MLIP Arena stability benchmark~\cite{chiang2026mlip} probes how long an
MLFF can drive MD before the dynamics break down. Each of 80 initial
structures, random mixtures of stable materials restricted here to
non-$f$-block elements, is propagated for $10\,$ps in the NPT ensemble while
the temperature is ramped linearly from $300$ to $3000\,$K and the pressure
from $0$ to $500\,$GPa. Apart from the element restriction, the settings of
the reference implementation were used: thermostat and barostat parameters
$t_{\mathrm{time}}=25\,$fs and $p_{\mathrm{factor}}=(75\,\mathrm{fs})^2\times100\,$GPa,
a time step of $0.5$ ($2.0$)\,fs for cells with (without) hydrogen, a
sampling interval of 10 steps, and a wall-clock budget of $600\,$s per
attempt with two restarts ($30\,$min in total per structure).

\begin{figure}[ht]
    \centering
    \includegraphics[width=\linewidth]{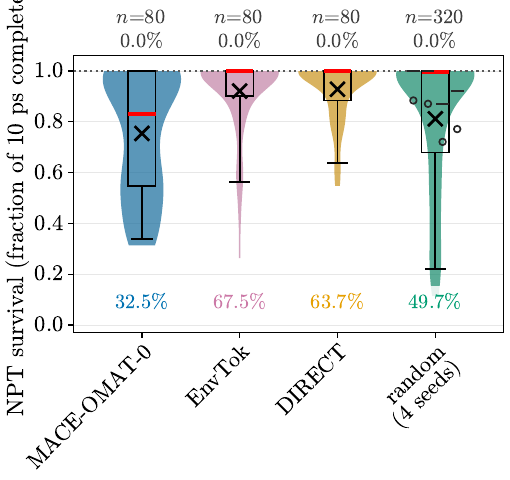}
    \caption{Distribution of MD steps completed per run in the MLIP Arena
    stability benchmark for MACE-OMAT-0 and the subset-trained models 
    (violin plots; red bar and black cross: median and mean; box edges: 
    25th and 75th percentiles; whiskers: 5th and 95th percentiles)}
    \label{fig:stability-survival}
\end{figure}

Runs are classified as \emph{completed}, \emph{truncated} (still
integrating normally when the wall-clock budget expired) or
\emph{physically diverged}. The last category comprises collapse or
explosion of the simulation cell, divergence of atomic forces or of the
temperature, and out-of-memory failures, which occur when the collapse of
the cell causes an extreme growth of the neighbour list; such runs are
already lost and would not recover with more memory. The survival curves of
Fig.~3a of the main text are complemented by the completed fraction of MD steps
in Fig.~\ref{fig:stability-survival}.

\subsection{Li migration barriers: nebDFT2k}
The nebDFT2k benchmark of the LiTraj suite~\cite{dembitskiy2025benchmarking}
comprises 1681 Li-vacancy hops in Materials-Project-derived crystals; for
each hop the dataset provides a BVSE-NEB preconditioned band of five or seven
images as initial guess and a DFT climbing-image NEB relaxed band as
reference. After the element cut, 1656 hops remain, 621 of which contain no
Hubbard-$U$ element in the Materials Project convention. Following the
published protocol, the two end-points and then the band are relaxed with
the MLFF using the FIRE optimizer and a climbing-image NEB with spring
constant $5\,$eV/\AA$^2$ in the improved-tangent formulation; the migration
barrier is the difference between the highest and lowest image energy along
the relaxed band. Errors are evaluated with the scoring functions shipped
with the dataset.

\begin{figure}[ht]
    \centering
    \includegraphics[width=\linewidth]{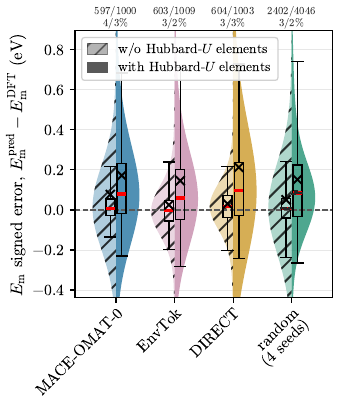}
    \caption{Distributions of Li$^{+}$ migration-barrier errors on the
    nebDFT2k benchmark (violin plots; red bar and cross: median and mean; box
    edges: 25th and 75th percentiles; whiskers: 5th and 95th percentiles),
    split by whether the host contains one of the Materials Project Hubbard-$U$ elements (Co, Cr, Fe, Mn, Mo, Ni, V, W) used for OMat24. Above
    every violin, the percentage of hops excluded due to non-converged
    relaxations and the number of valid data points are given. The median
    signed error of every model shifts by $62$--$80\,$meV between the two
    blocks, reflecting the PBE vs.\ PBE+$U$ mismatch between the benchmark
    reference and the training data.}
    \label{fig:nebdft2k-violins}
\end{figure}

The reference barriers were computed with non-spin-polarised PBE without a Hubbard $U$, whereas the models are trained on spin-polarised PBE/PBE+$U$ data.
Results are therefore reported separately for hops whose chemistry
carries no Hubbard $U$ and for those that do, as in the original benchmark.
We deviate from the published protocol in two respects. First, both
end-points are relaxed to the same force threshold of $0.05\,$eV/\AA,
whereas the reference implementation relaxes the source to $0.05$ and the
target to $0.1\,$eV/\AA. Second, the step budget of the end-point and band
relaxations is raised from 100 to $10\,000$, so that convergence is limited
by the force criterion rather than the step cap. The convergence criterion
of the published protocol, a maximum force component below $0.1\,$eV/\AA{}
across the band and both end-points, is retained; 
non-converged hops are recorded rather than discarded, and the reported numbers are those of the converged subset; two hops that exceeded the wall-clock limit of $2\,$h (one each for the EnvTok model and one of the random models) are omitted. 
The signed error distributions are shown in
Fig.~\ref{fig:nebdft2k-violins}.

\section{EnvTok descriptors in MLFF latent space}
\label{sec:si-latent}
\subsection{Protocol}
MACE-OMAT-0 (medium) node features were computed for all atoms of the
OMat24 validation set. 
As feature vectors we used the invariant ($l=0$) node features of both
message-passing layers, concatenated to length $2\times128=256$. 
After exclusion of $f$-block elements the set
comprises over $657$ thousand structures and $12$ million atoms. The
feature vectors were reduced via PCA to a 50-dimensional space retaining over $94\%$ of the variance 
and, for each EnvTok descriptor
represented by at least 30 atoms ($18\,445$ descriptors), three quantities
were evaluated:
\begin{itemize}
    \item \textbf{Modality}: the number of connected components of the
    symmetric $k$-nearest-neighbour graph ($k=15$) restricted to the atoms of
    that descriptor.
    \item \textbf{Compactness}: 
  how tightly the atoms of a descriptor cluster
    around their common centre, compared with randomly chosen atoms of the
    same element. For the $n$ atoms of a descriptor we compute the mean
    squared distance from their centroid, $R^2$ (the squared radius of
    gyration), and compare it with the values obtained for many random sets
    of $n$ atoms of the same element, yielding a $z$-score:
    \begin{equation*}
        z = \frac{\langle R^2\rangle_\mathrm{rand} - R^2}{\sigma_\mathrm{rand}},
    \end{equation*}
    where $\langle R^2\rangle_\mathrm{rand}$ and $\sigma_\mathrm{rand}$ are
    the mean and standard deviation of $R^2$ over the random sets. Positive
    values mean the descriptor is tighter than such a random set; negative
    values mean it is more spread out, and these descriptors are termed
    \emph{diffuse}.
    \item \textbf{$k$-nearest-neighbour label purity}: the fraction of an
    atom's $k$ nearest neighbours  ($k=30$) in feature space that carry the same
    descriptor, averaged over the atoms of the class.
\end{itemize}

Over the validation set, $98.4\%$ of the analysed descriptors form a single connected
region, $96.9\%$ are compact, and the purity spans the full range from
$0.003$ to $1.0$ with a mean of $0.4$. 
For $40\,000$ atoms drawn uniformly at random from the atoms of these sufficiently frequent 
EnvTok descriptors, 
we determined the $k=30$ nearest neighbors among all atoms of the validation set.
Of the neighbours carrying a different
descriptor, $99.9\%$ share the central element, $69.7\%$ the same set of
ligand elements and $56.9\%$ the same coordination number. 
The EnvTok
descriptor therefore tends to over-partition the latent space into chemically
adjacent classes; selecting for maximal EnvTok diversity samples the latent
space essentially without gaps, at the cost of some redundancy relative to a
hypothetical minimal subset. The exceptions are the multimodal and diffuse
minorities discussed below, where one label maps to several latent regions
and finer sampling would be required to retain that diversity.

\subsection{Descriptors with multiple disjoint regions}
The $1.6\%$ (292) of descriptors that split into two or three regions of
latent space differ weakly at most from those forming a
single region for every EnvTok descriptor field including the element of the central atom.
For most of them the splitting is material-based: atoms
sharing a label separate into one region or another according to the host
structure in which they occur. This is not simply a matter of chemical
diversity, since many single-region descriptors occur in at least as many
chemically distinct materials as those that split. We further tested
whether splitting arises from exposure of MACE-OMAT-0 to too few distinct
structures containing these environments: 
compared with single-region descriptors of the same size,
split descriptors appear about as often in the training set, but in fewer
different materials. Descriptors that never appear in the training set are
the least likely to split.
We
therefore attribute the splitting neither primarily to the locality of the descriptor
nor to the training data alone but also to the latent-space representation of
MACE-OMAT-0 itself, possibly a consequence of its comparatively short
training of only a few epochs~\cite{batatia2025cross}. 
This interpretation is, however, difficult to verify directly.

\subsection{Diffuse descriptors}
$3.1\%$ of the descriptors are diffuse. Among tabulated element and
coordination features, diffuseness is driven primarily by coordination
to highly electronegative anions: descriptors with O or F in the
coordination shell constitute $11.9\%$ of the total but account for $67\%$
of all diffuse ones. The central cation's own electronegativity is not
predictive on its own, nor is transition-metal character as such: 
a typical $d$-block element is not more often diffuse than other elements, 
and Y, Zr and Hf as well as Ru, Ir and Pt are among the most compact
environments observed. 
Beyond ligand chemistry, the identity of the central
element is, however, the strongest single predictor. 
After controlling for
ligand chemistry, coordination number and descriptor abundance, 
the odds of being diffuse are about two to seven times higher for 
V, Cr, Mn, Fe, Cu, Mo and Ag and for the 
heavy $p$-block cations Tl, Pb, Sn, In and Sb than for the
other elements, whereas alkali and alkaline-earth centres are diffuse an
order of magnitude less often.
 
A physically motivated interpretation is that for highly polar bonding the same
nominal coordination polyhedron collapses a variety of second-shell
chemistries, bond lengths and local distortions into a single label, which
MACE, with its larger receptive field, resolves.
We verified that the spread is not caused by the exchange--correlation
treatment: 
atoms of the same descriptor occupy nearly the same region of
latent space whether their host structure was computed with pure PBE or
with PBE+$U$, and the functional accounts for only a small fraction of a
descriptor's spread compared with the chemistry of the host structure.
The spread therefore reflects chemical heterogeneity rather than an
artefact of the mixed-functional reference data.

\bibliography{references}